\documentclass[traditabstract]{aa}
\usepackage{graphicx}
\usepackage{rotating,subfigure,amssymb,afterpage}
\usepackage{txfonts}
\usepackage{natbib}

\begin{document}
   \title{The Cosmic Large-Scale Structure in X-rays (CLASSIX) \\
   cluster survey IV: Superclusters in the local Universe at $z \le 0.03$
\thanks{
   Based on observations at the European Southern Observatory La Silla,
   Chile and the German-Spanish Observatory at Calar Alto}}

   \author{Hans B\"ohringer\inst{1,2} \& Gayoung Chon\inst{1}} 

   \offprints{H. B\"ohringer, hxb@mpe.mpg.de}

   \institute{$^1$ Universit\"ats-Sternwarte M\"unchen, Fakult\"at f\"ur Physik,
                  Ludwig-Maximilians-Universit\"at M\"unchen,
                  Scheinerstr. 1, 81679 M\"unchen, Germany.\\
              $^2$ Max-Planck-Institut f\"ur extraterrestrische Physik,
                   D-85748 Garching, Germany.
}

   \date{Submitted 18/5/21}

\abstract{It is important to map the large-scale matter distribution in the local Universe
for cosmological studies, such as the tracing of the large-scale peculiar velocity flow, the
characterisation of the  environment for different astronomical objects, and 
for precision measurements of cosmological parameters. 
We used X-ray luminous clusters to map this matter distribution 
and find that about 51\% of the groups and clusters are members of 
superclusters which occupy only a few percent of the volume. In this paper we
provide a detailed description of these large-scale structures. With a friends-to-friends algorithm,
we find eight superclusters with a cluster overdensity ratio of at least two with five or more galaxy group 
and cluster members in the cosmic volume out to $z = 0.03$. The four most
prominent ones are the Perseus-Pisces, the Centaurus, the Coma, and the Hercules
supercluster, with lengths from about 40 to over 100 Mpc and estimated masses of
$0.6 - 2.2 \times 10^{16}$ M$_{\odot}$. The largest of these structures is the Perseus-Pisces
supercluster. The four smaller superclusters include the Local and the Abell 400 supercluster
and two superclusters in the constellations Sagittarius and Lacerta.
We provide detailed maps, member catalogues, and physical descriptions of the
eight superclusters. By constructing superclusters with a range of 
cluster sub-samples with different lower X-ray luminosity limits, we show that the main structures
are always reliably recovered.}

 \keywords{galaxies: clusters, cosmology: observations, 
   cosmology: large-scale structure of the Universe,
   X-rays: galaxies: clusters} 

\authorrunning{B\"ohringer et al.}
\titlerunning{Superclusters in the local Universe}
   \maketitle
%

\section{Introduction}

Galaxy clusters are good tracers of the large-scale matter distribution in the
Universe (e.g. \cite{Bar1986,Kai1986,Boe2020}).
The most frequent way to characterise the large-scale structure traced by
galaxy clusters is the two point correlation
function or its Fourier counterpart, the power spectrum (e.g. \cite{Hau1973,
Bah1983}). This does not capture, however, larger non-linear structures in
the cluster distribution called superclusters. Such structures have become
interesting astrophysical study objects and they can also be used to test
cosmological and structure formation models (e.g. \cite{Bas2001,Ein2021}). 

Superclusters were detected and characterised
as soon as large galaxy and photographic surveys such as the Shapley-Ames survey of
bright galaxies and the National Geographic-Palomar Observatory Sky Survey became available
(e.g. \cite{Sha1932, Abe1961}). \citet{Oor1983} provided a comprehensive review 
of the status of supercluster research at the time. He already described the Local, 
the Perseus, the Coma, and the Hydra-Centaurus superclusters as the major nearby
large-scale structures, together with a few smaller superclusters (at distances $< 50$ Mpc)
and the more distant Corona-Borealis Supercluster. He noted sizes of about 40 to 90 Mpc
and possibly larger and points out the interesting finding by \citet{Gio1983} 
(see also  \cite{Gio1986}) that the
filamentary structure of the Perseus supercluster is traced much sharper by 
early-type compared to late-type galaxies. The latter discovery shows that superclusters
as a whole have their own distinct astrophysical properties. The astrophysical
interest in superclusters as laboratories for the study of galaxy evolution is
now well established by more detailed investigations of the galaxy population
in supercluster environments (e.g. \cite{Par2007,Lie2007,Alp2015,Ein2020}).

With the availability of redshifts for galaxy clusters more detailed supercluster
studies were conducted \citep{Bah1984,Zuc1993,Ein1997,Ein2003a,Ein2003b,Lii2012}, 
mostly on Abell's catalogue and its southern extension \citep{Abe1958,Abe1989}.
The superclusters were found in these studies mostly by a friends-of-friends technique.
\citet{Ein2001} also included two small samples of X-ray detected clusters. 
When large galaxy redshift surveys were published, superclusters were also 
constructed from the galaxy distribution, mainly by detecting overdense regions above a certain
threshold in galaxy or luminosity density maps from the 2dFGRS \citep{Ein2007a,Ein2007b}
and the Sloan Digital Sky Survey (SDSS) (e.g. \citep{Ein2006,Cos2011,Lup2011}). The similarity of the
structures found with clusters and with galaxies was, for example, discussed
by \citet{Lup2011}. Radio observations of HI in galaxies were used to extend
the optical studies of large-scale structures also into the regions of high
extinction (e.g. \citep{Hau1987,Cha1990,Ram2016,Kra2018}).

Except for \citet{Ein2001} and the HI observations, these studies have been
based on optical data. For galaxy clusters as tracer objects, X-ray detections
have the advantage that the X-ray emission ensures that the systems are gravitationally
bound, the X-ray luminosity is closely related to the cluster mass, for example, 
\citep{Pra2009},
and projection effects are minimised. In addition the heavily used galaxy 
cluster catalogue of Abell has no clear selection function. 
Therefore it is worth to make a make a fresh approach based on our large and
highly complete sample of X-ray luminous galaxy
clusters from the {\sf CLASSIX} (Cosmic Large-Scale Structure in X-rays) galaxy 
cluster survey, which has a well defined selection function. With its flux-limit, the survey 
provides an X-ray luminosity-limited (and closely mass-limited) cluster sample
in each redshift shell. We have shown not only with various tests
applied to observations, but also with cosmological simulations,
that the X-ray luminous clusters provide true probes of the large-scale matter distribution,
as further described in section 2.

\begin{figure}[h]
   \includegraphics[width=\columnwidth]{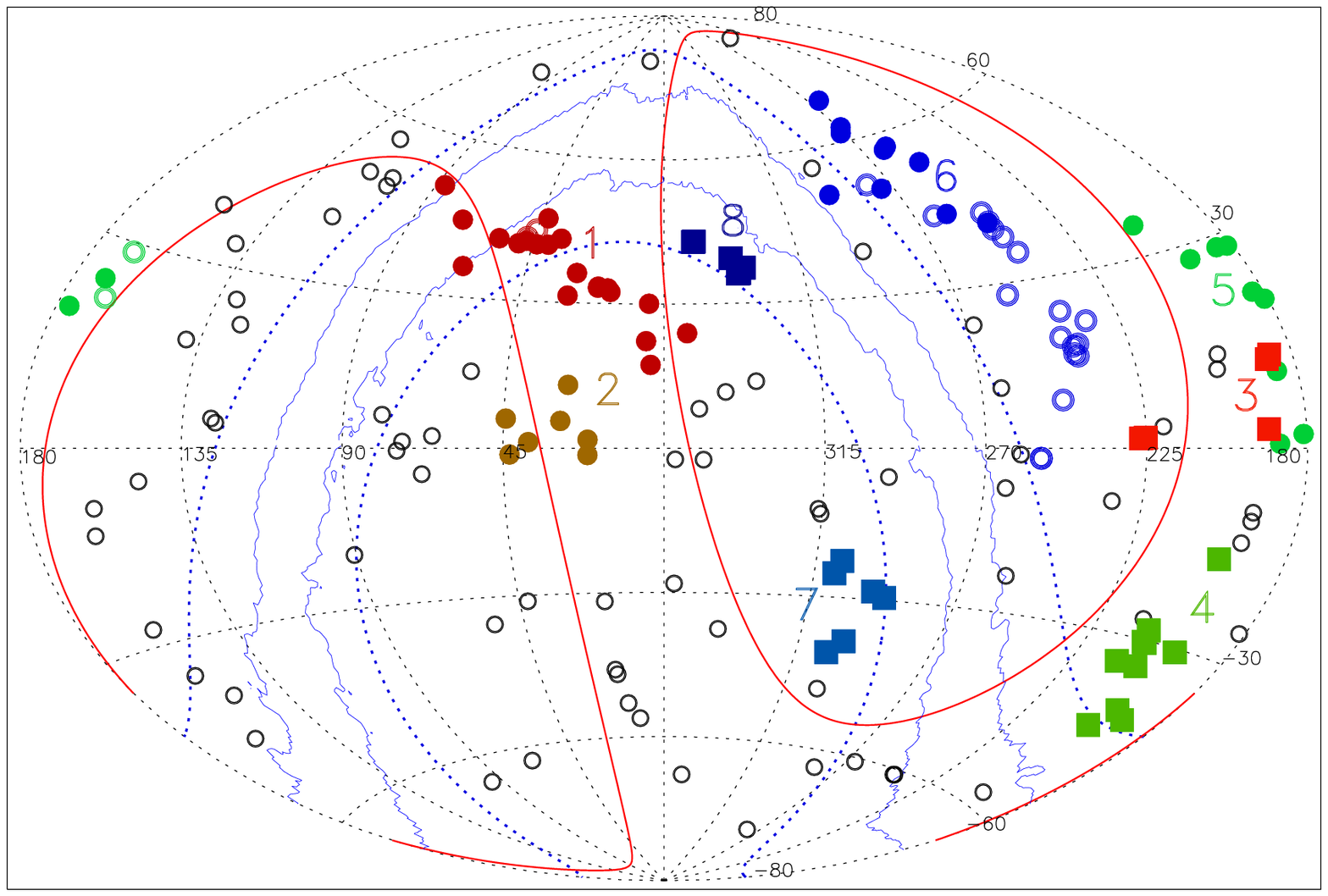}
   \includegraphics[width=\columnwidth]{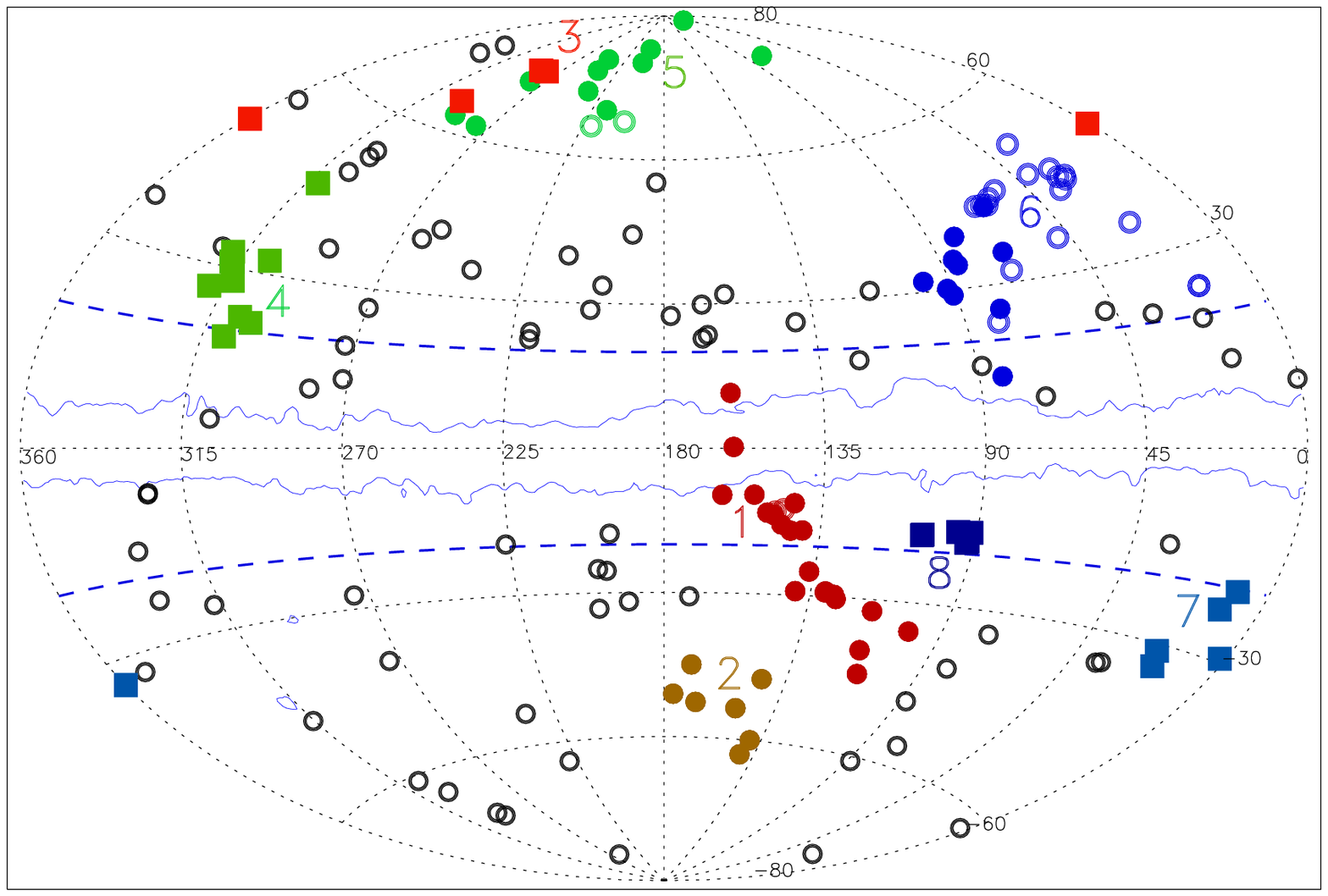}
\caption{Sky distribution of the superclusters at $z \le 0.03$ in 
equatorial coordinates (upper panel) and galactic coordinates (lower panel) . 
The superclusters are marked by numbers: 1 = Perseus-Pisces Supercluster,
2 = Abell 400 supercluster, 3 = Local Supercluster, 4 = Centaurus Supercluster, 5 = Coma Supercluster 
6 = Hercules supercluster, 7 = Sagittarius Supercluster, 8 = Lacerta Supercluster.
The full coloured symbols are {\sf CLASSIX} supercluster members, with $z \le 0.03$, while more
distant members are shown by coloured open symbols. All other {\sf CLASSIX} clusters 
at $z \le 0.03$ are shown as black open circles. The galactic band
($b_{II} = \pm 20^{o}$) is shown by blue dotted (dashed) lines, the region with high hydrogen 
column density ($n_H \ge 2.5 \times 10^{21}$ cm$^{-2}$) is indicated by the solid blue lines,
and the supergalactic band ($SGZ = \pm 20^{o}$) by red lines. 
}\label{fig1}
\end{figure}

In a series of papers we used the sample of {\sf CLASSIX} galaxy clusters to explore
the cosmography of the local Universe at $z \le 0.03$. In \citet{Boe2020}
we have shown that our cosmic neighbourhood has a lower matter density
by about 15- 30\% out to $\sim 100$ Mpc in the northern sky and $\sim 140$ Mpc in the southern
sky. In \citet{Boe2021a} we found that in a region out to about 100 Mpc 
the matter is strongly segregated towards the Supergalactic plane and in 
\citet{Boe2021b} we explored the structure of the Perseus-Pisces 
and the A400 Supercluster (also known as Southern Great Wall). 
In this paper we continue to study the 
nearby large-scale structures by characterising the six superclusters 
found at $z \le 0.03$ in addition to the Perseus-Pisces
and A400 superclusters.

In this study we construct superclusters which feature typical overdensity ratios in the
matter distribution, $R_{DM} = \rho_{DM} / < \rho_{DM}>$, 
of about a factor of 1.5 - 2.5 by means of a friends-of-friends (FoF) technique and
characterise their properties. In the literature such structures are generally called 
superclusters. Since this term is used for a large range of structures, we have recently
introduced a new physical characterisation, by distinguishing
those mass concentrations which will collapse in the future, which we call `superstes-clusters'
\citep{Cho2015} and structures that will not survive as a whole 
due to the accelerated expansion
of a $\Lambda$CDM universe. Superstes-clusters require a present day overdensity ratio
greater than 7.8 for a future collapse. Here we deal with
larger structures at lower oversensity, and we refer to them as 
superclusters (SC) in the following.

The paper has the following structure. In section 2 we describe 
the {\sf CLASSIX} galaxy cluster survey and its applications to large-scale
structure studies and section 3 deals with methodological aspects. 
The results of our analysis is presented in section 4 with a detailed
description of the SC. Section 5 provides a discussion and section 6
the summary and conclusion.
In the Appendix we provide X-ray/optical images of the members of six of 
the eight superclusters (the Perseus-Pisces and the A400 superclusters
have already been described in detail in  \citet{Boe2021b}).
For physical properties which depend on distance we adopt the following
cosmological parameters: a Hubble constant of $H_0 = 70$ km s$^{-1}$ Mpc$^{-1}$,
$\Omega_m = 0.3$, and a spatially flat metric. For the cosmographical analysis we
use Supergalactic coordinates, defined by the location of the
Supergalactic North Pole at $l_{II} = 47.3700^o$ and $b_{II} = 6.3200^o$,
as established by De Vaucouleurs et al. in the 3rd Catalog 
of Bright Galaxies (1991, see also \citet{Lah2000}). X-ray luminosities
are determined in the ROSAT band, $0.1 - 2.4$ keV.

   \begin{table*}
      \caption{Properties of the superclusters in the local Universe at $z \ge 0.03$
constructed with a minimal linking length, $l_0 = 19$ Mpc. The superclusters are: PP = Perseus-Pisces SC,
A400 = A400 SC, Coma = Coma SC, Vir = Local SC, Cen = Centaurus SC, Her = Hercules SC,
Sag = Sagittarius SC, Lac = Lacerta SC. The suffix $_{in}$ designates that part
of the supercluster which lies inside $\le 0.03$, otherwise the redshift constraint was relaxed in the SC
construction. $N_{CL}$ is the number of SC members and $wN_{CL}$ the weighted number of members. The SC volume
is in units of $10^5$ Mpc$^3$, $M_{CL~tot}$ is the sum of the $m_{200}$ cluster masses in units of $10^{14}$ M$_{\odot}$,
$M_{est}$ is the estimated SC mass in units of  $10^{15}$ M$_{\odot}$ and the length is in units of Mpc. The
cluster density, $n_{CL}$ is in units of $10^{-4}$ Mpc$^{-3}$ and the overdensity ratios $R_{CL}$ and
$R_{DM}$ are defined in the text.
}
         \label{T1}
      \[
         \begin{array}{lrrrrrrrrrrrr}
            \hline
            \noalign{\smallskip}
{\rm name}& N_{CL}& wN_{CL} &{\rm volume}& M_{CL~tot} & M_{est}   & {\rm length}   
        & <z> & z_{min} & z_{max} & n_{CL} & R_{CL} & R_{DM} \\
            \noalign{\smallskip}
            \hline
            \noalign{\smallskip}
{\rm PP}      &  22 &  58.3 &    2.9 &   35.8 &    24.9 &   115.7 & 0.0205 & 0.0147 & 0.0314 &  2.0 &   2.8 &   2.1\\
{\rm PP}_{in} &   20 &  49.0 &   2.8  &  31.6 &    21.5 &   115.7 & 0.0195 & 0.0147 & 0.0300 &  1.8 &   2.4 &   1.9\\
{\rm A400}     &   7 &  14.6 &   1.1 &    5.0 &     6.8 &    45.8 & 0.0202 & 0.0171 & 0.0243 &  1.4 &   1.9 &   1.6\\
{\rm Vir}     &   5 &   5.0 &    0.5 &    2.4 &     2.5 &    18.4 & 0.0044 & 0.0031 & 0.0062 &  1.0 &   1.4 &   1.2\\
{\rm Cen}     &  10 &  11.7 &    1.0 &    7.1 &     5.6 &    37.1 & 0.0131 & 0.0087 & 0.0160 &  1.2 &   1.6 &   1.4\\
{\rm Coma}      &  13 &  37.8 &    2.3 &   18.9 &    16.8 &    78.0 & 0.0254 & 0.0202 & 0.0322 &  1.7 &   2.3 &   1.8\\
{\rm Coma}_{in} &  11 &  28.8 &    1.9 &   16.8 &    13.0 &    64.8 & 0.0242 & 0.0202 & 0.0280 &   1.5 &   2.1 &   1.7\\
{\rm Her}     &  24 &  94.0 &    3.7 &   26.0 &    38.7 &   141.1 & 0.0297 & 0.0236 & 0.0349 &   2.6 &   3.6 &   2.6\\
{\rm Her}_{in} &  10 &  30.6 &   1.7 &   6.7 &     13.4 &    71.1 & 0.0270 & 0.0236 & 0.0297 &   1.8 &   2.5 &   1.9\\
{\rm Sag}    &   6 &  11.4 &    0.9 &    2.7 &     5.5 &    33.8 & 0.0204 & 0.0191 & 0.0246 &   1.2 &   1.7 &   1.4\\
{\rm Lac}    &   6 &   9.7 &    0.6 &    4.4 &     4.3 &    19.9 & 0.0178 & 0.0169 & 0.0192 &   1.7 &   2.3 &   1.8\\
            \noalign{\smallskip}
            \hline
            \noalign{\smallskip}
         \end{array}
      \]
\label{tab1}
   \end{table*}

\section{The CLASSIX galaxy cluster survey}
The {\sf CLASSIX} galaxy cluster survey, which comprises the {\sf REFLEX II} 
survey in the southern sky \citep{Boe2013} and the {\sf NORAS II} survey
in the northern hemisphere \citep{Boe2017}, provides
a total sky coverage of 8.26 ster
at galactic latitudes $|b_{II}| \ge 20^o$. An extension of {\sf CLASSIX}
towards lower galactic latitudes includes part of the `zone of avoidance' (ZoA), 
covering that region where the interstellar Hydrogen column 
density~\footnote{The values for the
interstellar hydrogen column density are taken from the 21cm survey of \citet{Dic1990}.
We have compared the
interstellar hydrogen column density compilation by \citet{Dic1990}
with the more recent data set of the 
Bonn-Leiden-Argentine 21cm survey \citep{Kal2005}
and found that the differences relevant for us are of the order of at most one percent.
Because our survey has been constructed with a flux cut based
on the Dickey \& Lockman results, we keep the older hydrogen column density
values for consistency reasons.} $n_H \le 2.5 \times 10^{21}$ cm$^{-2}$,
adding another 2.56 ster. In this region the completeness of the cluster detection 
is not as high as for {\sf REFLEX} and {\sf NORAS} and also follow-up observations 
to obtain redshifts are still incomplete.
The statistical properties of the cluster distribution in the ZoA is 
therefore somewhat qualitative.  We also added three known X-ray luminous
clusters in the ZoA at higher $n_H$, also detected in the RASS.

The {\sf CLASSIX} galaxy cluster survey is compiled from
the X-ray detection of galaxy clusters in the ROSAT All-Sky Survey
(RASS, \citep{Tru1993,Vog1999}). The survey construction,  
selection function, and tests of the completeness
are described in \citet{Boe2013,Boe2017}. In brief, the 
nominal unabsorbed flux limit for the galaxy cluster detection in the RASS is
$1.8 \times 10^{-12}$ erg s$^{-1}$ cm$^{-2}$ at 0.1 - 2.4 keV.
For the present study we use a minimum source photon count limit of 20. 
Under these conditions the nominal flux limit quoted above is reached in about
80\% of the survey. In regions with lower exposure and higher interstellar
absorption the flux limit is accordingly higher 
(see Fig.\ 11 in \citet{Boe2013} and Fig.\ 5 in \citet{Boe2017}. 
This effect is well modelled and taken into account in the survey selection function.
The survey selection function as a function of sky position and redshift is described
in  \citet{Boe2013}  for {\sf REFLEX II} and \citet{Boe2017}
for {\sf NORAS II}, where numerical data are provided in the on-line material.

In total we found 146 groups and clusters of galaxies with 
these selection criteria and with $L_X \ge 10^{42}$ erg s$^{-1}$ in the study region 
at $z \le 0.03$, which provides us with a sufficiently high cluster density for the
mapping of the large-scale structure. The average distance between clusters
in this region is about 36.8 Mpc. We have shown in \citet{Boe2020}
that the cluster density is a robust biased measure of the matter density with an
accuracy corresponding roughly to the Poisson uncertainty of the cluster counting
statistics. This is based on an analysis of the correlation of cluster and 
matter density in the Millennium Simulations \citep{Spr2005}.
The bias found in this study is consistent with the theoretical predictions
(e.g. \citep{Tin2010,Bal2011}).

We have used the {\sf REFLEX I}  
\citep{Boe2004} and  {\sf REFLEX II} surveys
to study the cosmic large-scale matter distribution
through, for example, the correlation function \citep{Col2000}, 
the power spectrum \citep{Sch2001,Sch2002,Sch2003a,Sch2003b,Bal2011,Bal2012}, 
Minkowski functionals, \citep{Ker2001},
and for the study of superstes-clusters \citep{Cho2013,Cho2014}.
We found the results consistent with theoretical 
expectations, which helped to establish the use of clusters for
cosmographical investigations.

The X-ray luminosity and mass of clusters are important parameters
in this study. The X-ray luminosity in the 0.1 to
2.4 keV energy band was derived within a cluster radius of 
$r_{500}$ \footnote{$r_{500}$ is the radius where the average
mass density inside reaches a value of 500 times the critical density
of the Universe at the epoch of observation.}. To estimate the cluster
mass from the observed X-ray luminosity, we use the 
scaling relation from {\citet{Pra2009} as described in
\citet{Boe2014,Boe2021b}.

\section{Methods}

A FoF method was used to construct the superclusters.
Because the flux limited survey features an increasing luminosity limit with 
redshift (as shown in Fig. A1 in the appendix), the mean distance between clusters
is also a function of redshift. As a consequence the linking length of the
FOF algorithm has to be adjusted to this luminosity limit which we achieve
by means of a weighting scheme. The weights were calculated
from an integration of the luminosity function, $\phi(L_X)$, as follows:

\begin{equation}
w_i = {\int_{L_{X_0}}^{\infty} \phi(L) dL \over \int_{L_{X_i}}^{\infty} \phi(L) dL} ~~~~~~~~~~~, 
\end{equation}
\vspace{0.3cm}

where $L_{X_0}$ is the nominal lower X-ray luminosity limit of the sample 
and $L_{X_i}$ is the lower luminosity limit that can be reached at the 
sky location and redshift of the cluster to be weighed. In the
FoF algorithm we adopt a minimal linking length, $l_0$, and we adjust
the linking length $l_i = l_0 \times (w_i)^{1/3}$ if $L_{X_i}$ is higher 
than $L_{X_0}$. Since the linking length is calculated for each cluster at its
location, we take the average $l_i$ of both clusters in the percolation process
by means of the formula $<l> = l_0 \times (2/(1/w_1 + 1/w_2))^{1/3}$.

For this study we adopted a lower X-ray luminosity limit of $10^{42}$ erg  s$^{-1}$.
With this value for $l_0$, the survey is volume limited in most of the sky out
to a redshift of $z \sim 0.016$.
For the linking length we used a minimal value of $l_0 = 19$ Mpc. 
This corresponds approximately to an overdensity ratio, $R_{Cl} = n_{Cl}/<n_{Cl}>$, 
of about a factor of 2 compared to the mean density of clusters in the nearby Universe.
This factor is obtained as follows. From the X-ray luminosity
function obtained by \citet{Boe2014} we determined the mean density of
CLASSIX clusters with $L_X \ge 10^{42}$ erg s$^{-1}$ to be $7.2 \times 10^{-5}$ Mpc$^{-3}$.
The linking length for an overdensity ratio, $R$, of a factor of 2 is then 
given by: $l = R^{-1/3} \cdot n^{-1/3} \sim 19$ Mpc.

The adopted luminosity limit corresponds to a cluster mass limit of about 
$m_{200} = 2.1 \times 10^{13}$ M$_{\odot}$. We are therefore including
less massive galaxy groups in our study. They are definitely gravitationally bound
entities, as shown by their extended X-ray emission, but are optically often characterised
by a giant elliptical galaxy surrounded by a few smaller galaxies, which in its extreme
is called a `fossil group'.

\section{Results}

   \begin{table}
      \caption{Properties of the same superclusters as shown in Table~\ref{tab1},
but here the volume and mass estimates are based on volumes determined with a radius
of 10 Mpc around each cluster. For an explanation of the labels of rows and columns
see Table~\ref{tab1}.}
         \label{T2}
      \[
         \begin{array}{lrrrrr}
            \hline
            \noalign{\smallskip}
{\rm name}&{\rm volume}& M_{est} &  n_{CL}   & R_{CL} & R_{DM} \\
            \noalign{\smallskip}
            \hline
            \noalign{\smallskip}
{\rm PP}     &     0.7 &   21.5 &     8.9 &  12.4 &   8.1\\
{\rm PP}_{in} &    0.6 &    18.2 &     7.8 &  10.9 &   7.2\\
{\rm A400}    &    0.2 &      5.5 &     6.3 &   8.8 &   5.8\\
{\rm Vir}    &     0.1 &   1.9 &     5.3 &   7.4 &   5.0\\
{\rm Cen}    &     0.2 &      4.5 &      4.8 &   6.7 &   4.5\\
{\rm Coma}     &    0.5 &     14.1 &     7.8 &  10.8 &   7.1\\
{\rm Coma}_{in} &    0.4 &   10.8 &     7.1 &   9.9 &   6.6\\
{\rm Her}    &      0.8 &     34.3 &     12.3 &  17.1 &  11.0\\
{\rm Her}_{in}&        0.4 &      11.3 &    8.4 &  11.7 &   7.7\\
{\rm Sag}   &      0.2 &      4.3 &      5.8 &   8.1 &   5.4\\
{\rm Lac}   &      0.1 &      3.6 &      8.0 &  11.0 &   7.3\\
            \noalign{\smallskip}
            \hline
            \noalign{\smallskip}
         \end{array}
      \]
\label{tab2}
   \end{table}

\subsection{The superclusters}
 
\begin{figure}[h]
   \includegraphics[width=\columnwidth]{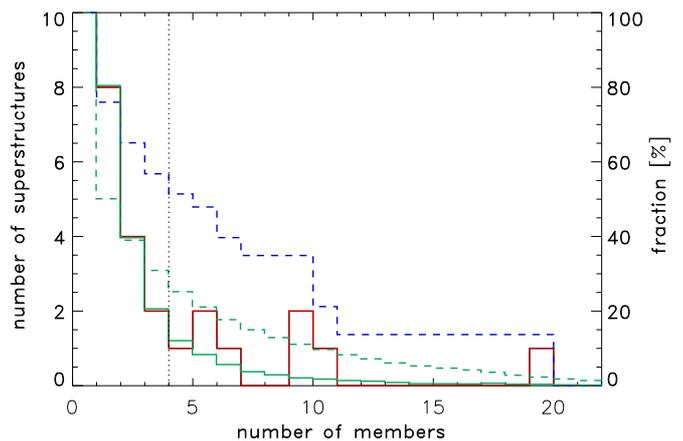}
\caption{Supercluster multiplicity function found from observations (red). 
The cumulative fraction of {\sf CLASSIX} clusters in superclusters
as a function of the number of members is shown in blue (with 
corresponding axis on the right).
The multiplicity function for a simulated random distribution of clusters
is shown in green and the corresponding fraction of clusters in superclusters 
as green dashed line.
}\label{figA1}
\end{figure}

The goal to study structures at low matter overdensities, but larger in size than
smaller superclusters, led us select the requirement that the SC contain at least
five members. A study of the multiplicity function, that is the number distribution of 
SC as a function of the number of members, discussed in the next subsection,
provides a justification for this choice. Fig.~\ref{figA1} shows the multiplicity function
derived for a minimal linking length of 19 Mpc. We note that the number of SC with
four members or less increases fast towards low richness, very similar to the distribution
we find for a simulated random cluster distribution with same volume and spatial density.
However, SC with more than five members are much more frequent in the data than in
the random simulations. Thus, these larger SC are special, while the smaller ones
can also be produced by shot noise. Being generous in the lower number cut, we included 
the bin of SC with five members in the transition region in our study.

With this requirement and the adopted linking length we found eight SC.
They are listed together with some of their main properties in Tables~\ref{tab1} and \ref{tab2}
and shown in an Aithoff projection in Fig~\ref{fig1}. We show the sky distribution in equatorial 
and in galactic coordinates for an easier comparison to different sky surveys.
Most of the SC are well-known superclusters including the Perseus-Pisces SC, the A400 SC
(also referred to as the Southern Great Wall),
the Coma SC (also known as Great Wall) with the Coma cluster, 
the Local SC with the Virgo cluster, the Centaurus
SC, and a large structure that also extends well beyond our study volume, which includes the 
core of the Hercules SC. We also refer to this structure inside the study volume as Hercules SC.
Two additional structures are not so well known. There is a group of six objects at a mean redshift
of $z = 0.0204$ in the constellation Sagittarius, which contains the Abell cluster A3698 and several
galaxy groups. The last SC in the constellation Lacerta is completely contained 
in the ZoA at $b_{II} = -17.9~{\rm to}~-15.9$ and has therefore not caught much attention 
in optical surveys.

The properties of the superclusters we list in Table~\ref{tab1} were determined as follows. 
To compare the `richness' among the different SC objectively, taking 
the selection function into account, we determine the weighted number of members by summing 
the weights derived with Eq.~1. The volume of the systems was determined by taking for 
each member cluster a sphere with a radius of 19 Mpc, equal to the minimal linking length. 
The volume of the SC is then 
the sum of the spheres, taking the overlap into account. The mass of the clusters in the 
SC is the sum of the $m_{200}$ values. The SC mass is estimated from the volume times 
the cosmic mean density times the matter overdensity ratio, $R_{DM}$, 
of the SC. The length of the SC is determined from the largest separation 
of two SC members. The cluster density, $n_{CL}$, is determined by the weighed 
number of clusters divided by the SC volume. The overdensity ratio of 
clusters in the SC, $R_{CL} = n_{CL} / <n_{CL}>$ is determined with respect 
to the mean cluster density in section 3. The dark matter
overdensity is derived by means of the relation $R_{DM} = (R_{CL} - 1) /b + 1$. 
We used a bias factor $b = 1.6$, which is the average value determined for the 
relevant cluster sample with given lower mass limit.

\begin{figure}[h]
   \includegraphics[width=\columnwidth]{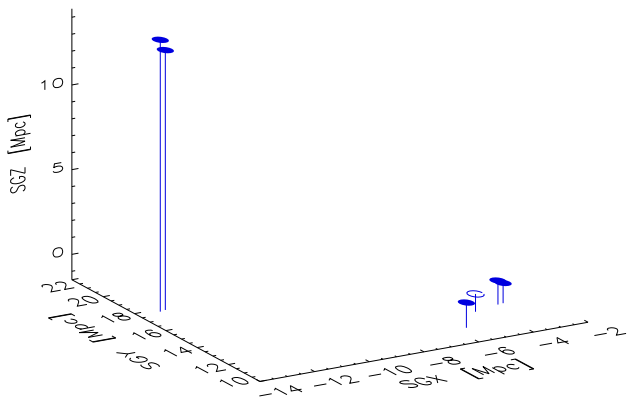}
\caption{Three-dimensional representation of the Local SC.
The four objects in the lower right are the Virgo cluster, with M87, M86
and M49 as well as NGC4636. M49 is marked with an open symbol because
its X-ray luminosity is below the adopted luminosity limit.
}\label{fig2}
\end{figure}

The volume calculation described above is quite generous. 
We adopted this as the most obvious 
choice, because it yields an overdensity ratio of about 2, consistent with 
the above considerations concerning the linking length.
One could in principle also argue that one should
only take half the radius to the next nearest neighbour. We provide the results
from such calculations with an alternative sphere radius of 
10 Mpc in Table~\ref{tab2}. We note that this results in a decrease of the estimated
SC mass by only 13 - 23\% (27\% in the case of the Local SC).

   \begin{table*}
      \caption{Groups and clusters which are members of the Local SC. The flux, $F_X$, is in units of $10^{-12}$ 
erg s$^{-1}$ cm$^{-2}$ in the 0.1 - 2.4 keV band and the error in the following column is in per cent. The X-ray 
luminosity, $L_X$, is in units of $10^{44}$ erg s$^{-1}$ at 0.1 to 2.4 keV within $r_{500}$, $m_{200}$ is the cluster
mass estimated from the $L_X$-mass relation within $r_{200}$, $r_{out}$ is the radius in arcmin out to
which the X-ray luminosity is detected in the RASS, and $n_H$ is the interstellar column density 
in the line-of-sight in units of $10^{20}$ cm$^{-2}$.
In the last column we designate groups by the name of the central dominant galaxy.}
         \label{Tab1}
      \[
         \begin{array}{lrrlrrrrrrl}
            \hline
            \noalign{\smallskip}
{\rm name}&{\rm RA}&{\rm DEC}&{\rm redshift}& F_X & {\rm error}&L_X&m_{200}&r_{out}&n_H & {\rm alt. name} \\
            \noalign{\smallskip}
            \hline
            \noalign{\smallskip}
{\rm RXCJ1226.2+1257}& 186.5540 &  12.9577 & 0.0033^{a)} & 34.2216 &   3.80 &   0.0166 &   0.295 &  17.5 &   2.6&{\rm M86}\\
{\rm RXCJ1230.7+1223}& 187.6838 &  12.3915 & 0.0033 & 371.7451 &   1.20 &   0.1792 &   1.291 &  17.5 &   2.5&{\rm M87 ~(Virgo)}\\
{\rm RXCJ1242.8+0241}& 190.7098 &   2.6884 & 0.0031 &  16.6059 &   8.50 &   0.0112 &   0.231 &  10.5 &   1.8&{\rm NGC~4636}\\
{\rm RXCJ1501.1+0141}& 225.2994 &   1.6981 & 0.0061 &  13.1537 &   6.30 &   0.0267 &   0.396 &   8.5 &   4.2&{\rm NGC~5813}\\
{\rm RXCJ1506.4+0136}& 226.6237 &   1.6022 & 0.0062 &   8.2915 &  13.80 &   0.0111 &   0.230 &  13.0 &   4.3&{\rm NGC~5846}\\
            \noalign{\smallskip}
            \hline
            \noalign{\smallskip}
         \end{array}
      \]
{\bf Notes:}$^{a)}$M86 has a blueshift in a heliocentric reference system of about 223 km s$^{-1}$. Here 
we give as a distance measure the redshift of the Virgo cluster, to which M86 belongs. 
\label{tab3}
   \end{table*}

   \begin{table*}
      \caption{Groups and clusters which are members of the Centaurus SC. The meaning of the columns is the same as in Table~\ref{tab3}}
         \label{Tab2}
      \[
         \begin{array}{lrrrrrrrrrl}
            \hline
            \noalign{\smallskip}
{\rm name}&{\rm RA}&{\rm DEC}&{\rm redshift}& F_X & {\rm error}&L_X&m_{200}&r_{out}&n_H & {\rm alt. name} \\
            \noalign{\smallskip}
            \hline
            \noalign{\smallskip}
{\rm RXCJ1248.7-4118}& 192.1997 & -41.3078 & 0.0114 & 251.0173 &   2.40 &   0.7665 &   3.165 &  80.0 &   8.3&{\rm A~3526~(Centaurus)}\\
{\rm RXCJ1304.2-3030}& 196.0696 & -30.5154 & 0.0117 &   8.8235 &  13.60 &   0.0299 &   0.424 &  20.0 &   6.2&{\rm NGC~4936}\\
{\rm RXCJ1307.2-4023}& 196.8136 & -40.3950 & 0.0159 &   2.3100 &  18.70 &   0.0154 &   0.280 &  11.0 &   6.7&{\rm ESO-323 0.0159}\\
{\rm RXCJ1315.3-1623}& 198.8499 & -16.3897 & 0.0087 &  72.5643 &   4.00 &   0.1357 &   1.083 &  36.0 &   4.9&{\rm NGC~5044}\\
{\rm RXCJ1321.2-4342}& 200.3137 & -43.7128 & 0.0118 &   3.1000 &  21.80 &   0.0136 &   0.260 &   9.0 &   8.5&{\rm NGC5090/5091}\\
{\rm RXCJ1336.6-3357}& 204.1616 & -33.9584 & 0.0123 &   2.4243 &  18.00 &   0.0107 &   0.224 &   9.5 &   4.1&{\rm A~3565}\\
{\rm RXCJ1347.2-3025}& 206.8014 & -30.4194 & 0.0145 &   2.5884 &  50.60 &   0.0152 &   0.278 &  10.0 &   4.4&{\rm A~3574W}\\
{\rm RXCJ1349.3-3018}& 207.3304 & -30.3094 & 0.0160 &   8.2988 &  25.00 &   0.0504 &   0.584 &  20.0 &   4.4&{\rm A~3574E}\\
{\rm RXCJ1352.8-2829}& 208.2140 & -28.4977 & 0.0159 &   2.0869 &  30.80 &   0.0149 &   0.274 &   9.0 &   4.4&{\rm NGC~5328}\\
{\rm RXCJ1403.5-3359}& 210.8995 & -33.9879 & 0.0132 &   9.0368 &  14.40 &   0.0402 &   0.508 &  17.0 &   5.6&{\rm AS~0753}\\
            \noalign{\smallskip}
            \hline
            \noalign{\smallskip}
         \end{array}
      \]
\label{tab4}
   \end{table*}

\subsection{Supercluster multiplicity function}

The multiplicity function of SC, that is their richness distribution, is an 
interesting statistical tool to gain some understanding of the nature of
the structures. We show in Fig.~\ref{figA1} the multiplicity function resulting from
the SC construction with a minimal linking length of 19 Mpc, relaxing
the requirement of a minimum number of members. We note that the distribution
shows a steep increase in the number of SC for less than five members towards
low richness, while we see a long tail in the distribution with more than 
five members. To better understand these results, we performed simulations of
spatially random distributions of clusters with the same study volume and total 
number and subjected them to the same SC construction process. We repeated
these simulations 1000 times and compare the mean results of these to the 
observations in Fig.~\ref{figA1}. At the low richness end ($N_{CL} < 5 $)
the random distribution reproduces the observations quite well. This implies
that the steep increase towards low richness can be produced by shot noise.
However, in the high richness tail the simulated SC are much less
frequent than the  observed ones, underlining the conclusion that these 
structures are especially interesting.

\begin{figure}[ht]
   \includegraphics[width=\columnwidth]{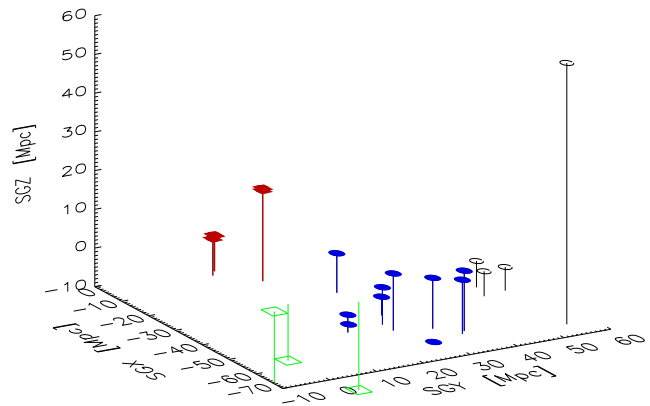}
\caption{Three-dimensional representation of the Centaurus SC 
(blue circles) and the Local SC (red squares). The open green 
squares show (from left to right) the Norma, Antlia and Hydra clusters.
Other {\sf CLASSIX}
clusters in this volume not associated with the two SC are
shown as open black circles. 
}\label{fig3}
\end{figure}

\begin{figure}[h]
   \includegraphics[width=\columnwidth]{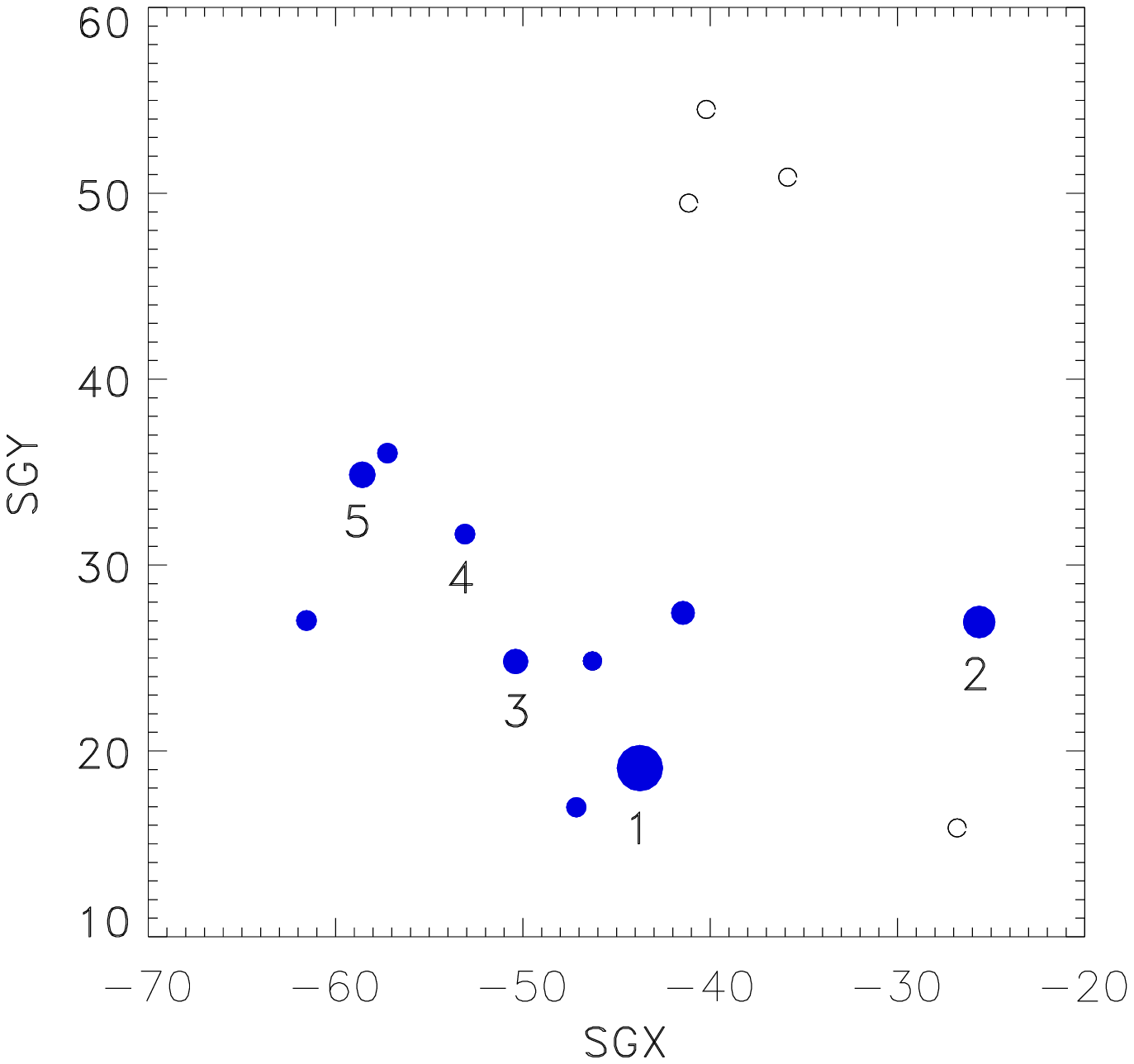}
   \includegraphics[width=\columnwidth]{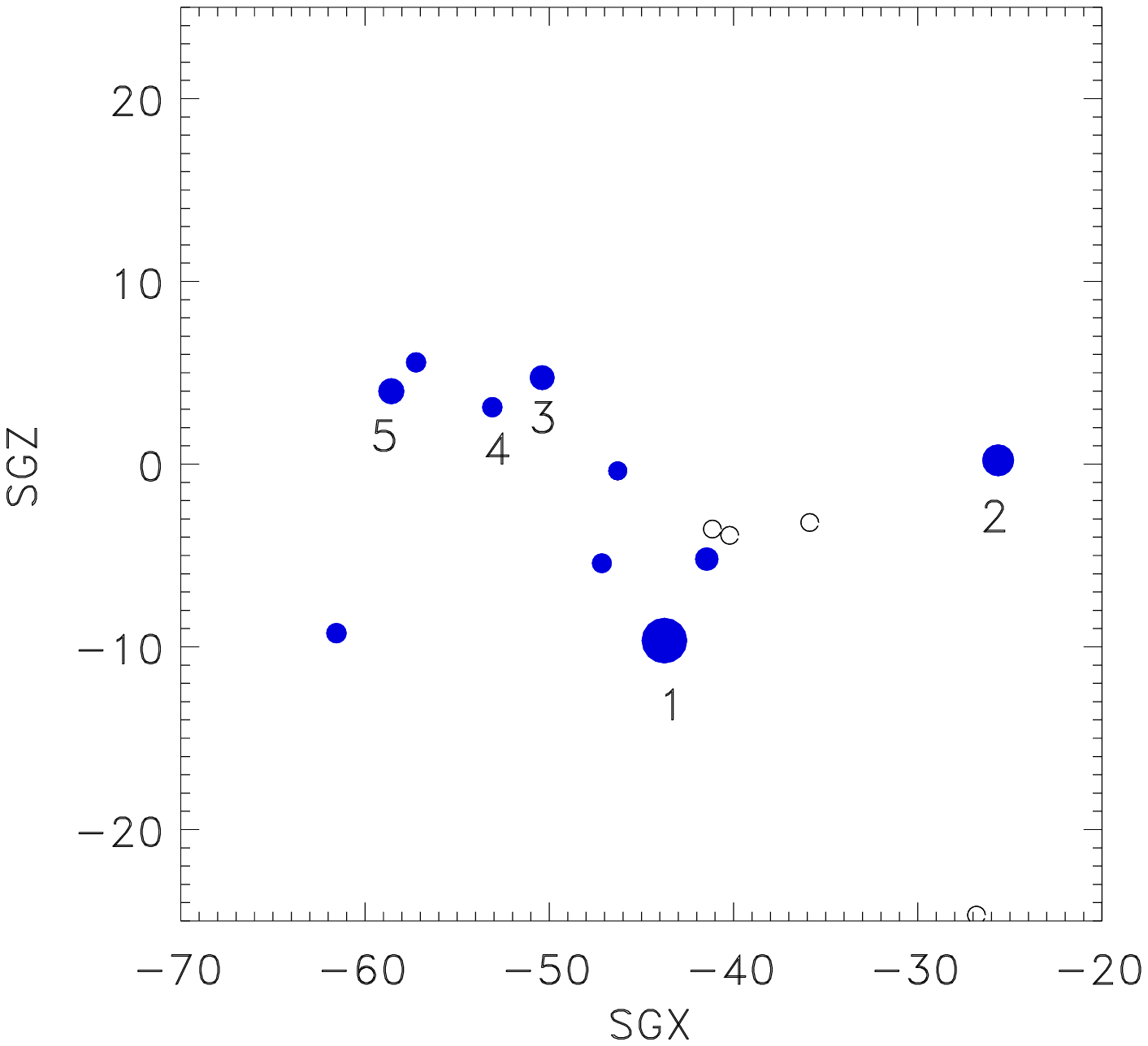}
   \includegraphics[width=\columnwidth]{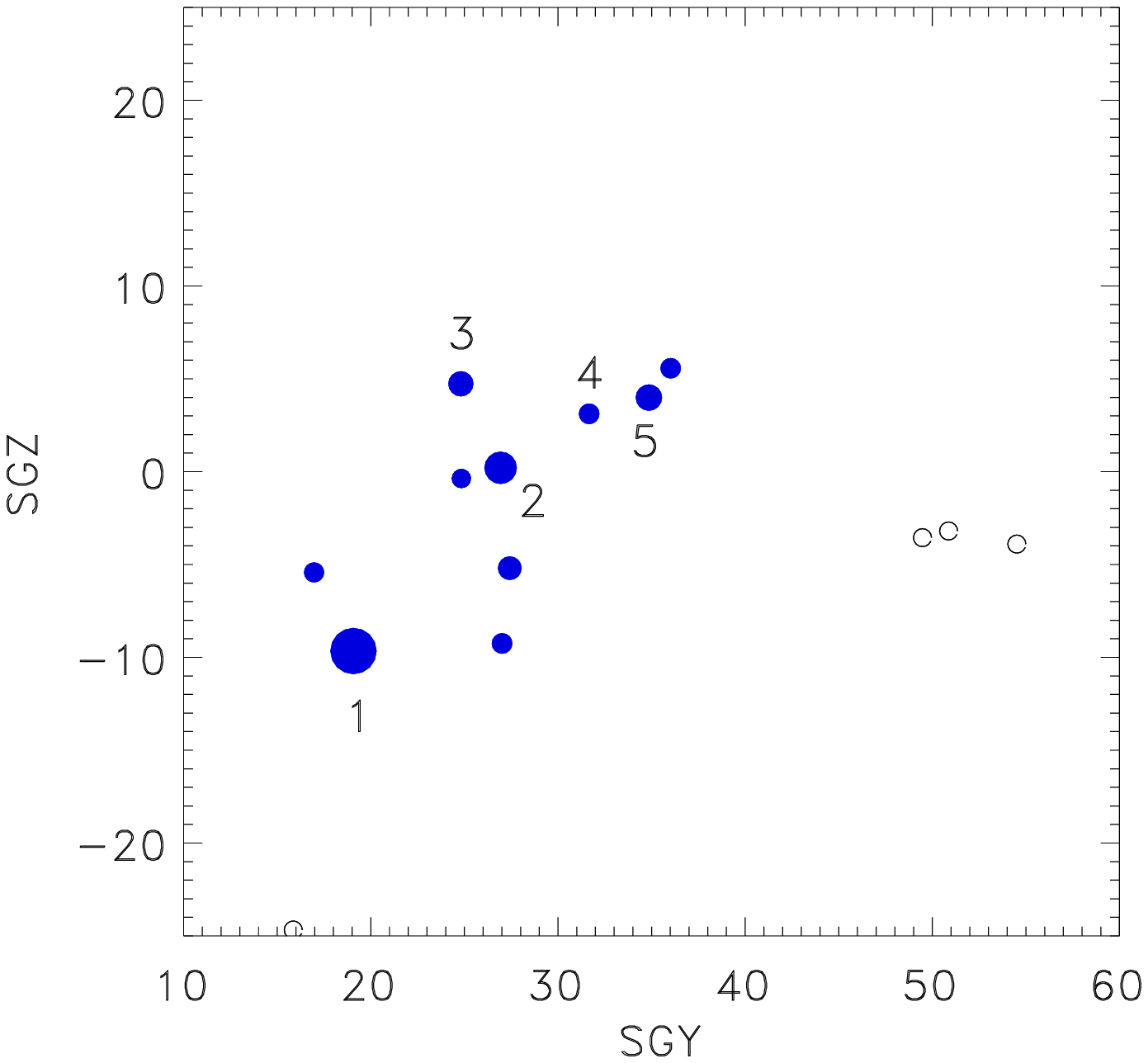}
\caption{Centaurus supercluster in three projections in Supergalactic
coordinates. The SC members are shown as filled blue circles, 
while other {\sf CLASSIX} clusters in this volume are marked by 
open circles. The size of the symbols is proportional to the 
cube root of the mass of the
clusters as explained in the text.
Several prominent clusters are marked. 1: Centaurus cluster, A3526,
2: NGC 5044 group, 3: AS 753, 4: A 3574 West, 5: A 3574 East.
}\label{fig4}
\end{figure}

\subsection{The Local Supercluster with the Virgo cluster}

The structure found including the Virgo cluster, listed in Table~\ref{tab3}, is the Local
SC \citep{Dev1953,Dev1956,Dev1958}. It contains only one galaxy
cluster, Virgo, and several smaller galaxy groups. If we would have applied the same strict 
criteria, that we applied to the other superclusters, the Local SC
would not have been included. We considered the Virgo cluster as three separate
dynamical units, the X-ray halos around the giant elliptical galaxies
M87, M86 and M49, which can be distinguished in the RASS. Among these X-ray halos, 
M49 falls (with its luminosity of $L_X \sim 3 \times 10^{41}$ erg s$^{-1}$) below the sample
luminosity limit and is thus excluded. The halos of M87 and M86 overlap on
the sky, but one can model the X-ray surface brightness distribution with two distinct 
halos \citep{Boe1994}. Also in redshift space the two halos can be distinguished
(e.g. \citet{Bin1987}). For other systems we considered different 
components only if they do not overlap on the sky in the RASS. 
Allowing for this exception for the 
nearby Virgo cluster provided us with five X-ray halo members for the Local SC, making it
part of the SC sample. 

In  addition, inspecting the X-ray 
luminosities of the five objects of the Local SC, we find that 
three of them have a low luminosity within a factor of two of the luminosity
limit and the fourth one is only slightly more luminous. Therefore the only massive 
object in this structure is the main body of the Virgo cluster formed by the halo of M87.
If we would have set the lower luminosity limit to $\ge 2 \times 10^{42}$ erg s$^{-1}$, 
the only two members left would have been M87 and NGC4636. We show in Fig.~\ref{figA2} 
the mean luminosity limit as a function of redshift. One notes
that a luminosity limit $L_X > 2 \times 10^{42}$ erg s$^{-1}$ effectively applies
for all systems with redshifts $z \ge 0.0203$. Thus again the Local SC
would not have been included if it would not be so close. But we had a strong
interest to include this well known SC in the description of the 
local cosmography of our Universe.

Fig~\ref{fig2} shows the three-dimensional configuration of the group of clusters. The two components
of the Virgo cluster and NGC4636 form a tight group as well as the pair consisting of NGC5813 and NGC5846,
while the two associations have a separation of about 17 Mpc, close to the linking length. Further properties
of this SC are given in Table~\ref{tab1} and \ref{tab2}, with an estimated mass of about  
$1.9 -2.5 \times 10^{15}$ M$_{\odot}$.

\begin{figure}[ht]
   \includegraphics[width=\columnwidth]{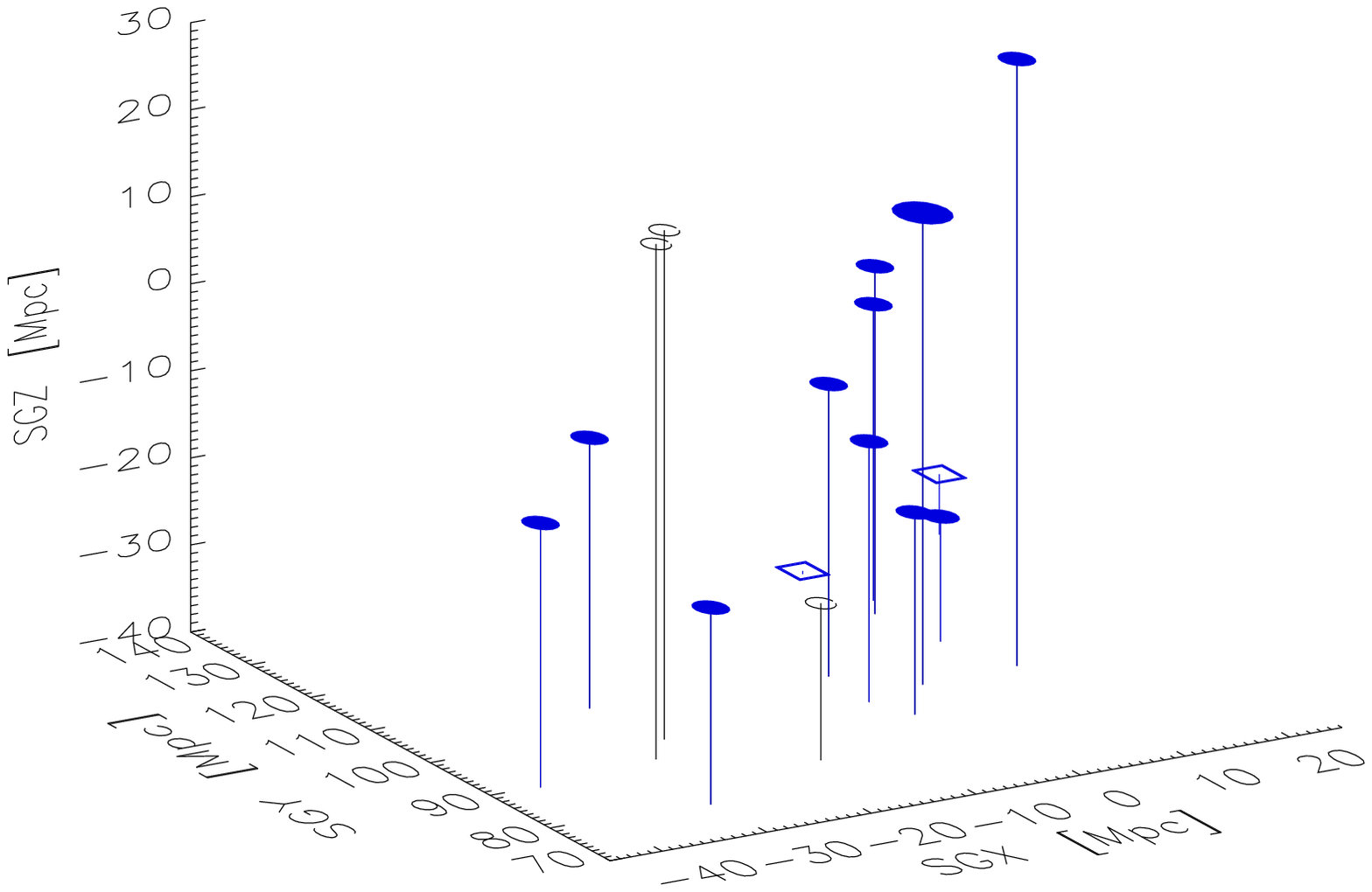}
\caption{Three-dimensional representation of the Coma SC in Supergalactic coordinates.
The members of the Coma SC are shown as full blue circles, the Coma cluster is marked by
a larger symbol and the two clusters at $z > 0.03$ are shown as open squares. All
other {\sf CLASSIX} clusters in the volume are shown as black open circles.
}\label{fig5}
\end{figure}

\subsection{The Centaurus Supercluster}

The Centaurus SC is found with ten members whose properties are listed in Table~\ref{tab4}.
The Centaurus cluster, A3526, is by far the most massive object. One galaxy group,
NGC 5090/5091 is located in the ZoA at $b_{II} \sim 18.1^o$.
Fig.~\ref{fig3} shows a three-dimensional representation of the
SC and its location with respect to the Local SC. The Centaurus SC is mostly oriented along
the Supergalactic plane. Its has a length of about 37.1 Mpc with an extension in the SGZ direction of 
only about 15.2 Mpc. The estimated mass is about $4.5 - 5.6 \times 10^{15}$ M$_{\odot}$.
It is located close to the Local SC. A linking length of 20.3 instead of 19 Mpc  would merge the
two superclusters through the systems RXCJ1315.3-1623 and RXCJ1501.1+0141.

In  Fig.~\ref{fig4} we show three projections of the Centaurus SC in supergalactic coordinates.
In this and similarly for the following figures we indicate the estimated cluster masses by the size
of the symbols, with a diameter scaling with the cube root of the estimated mass. Five of
the more massive members are marked in the image. The dominant Centaurus cluster sits on one side of
the SC. In Fig.~\ref{fig3} it appears on the left (low values of SGY) together with the group
NGC5090/5091 (which has the lowest SGY coordinate). The second most massive system in the SC is
the group NGC 5044, with an estimated mass, $m_{200}$, of about $1.08 \times 10^{14}$ M$_{\odot}$.
The other SC members have estimated masses below $6 \times 10^{13}$ M$_{\odot}$. RXCJ1349.3-3018,
A3574E, includes an X-ray bright AGN, the X-ray emission of which was subtracted in this analysis, as
further explained in the Appendix. 

In the literature this structure is often described as part of the Hydra-Centaurus
SC. With our recipe to construct the nearby SC, the Hydra cluster, A1060, fails
to be merged with the Centaurus SC by a large margin. Similarly, the other two prominent clusters
in this region, Antlia and Norma (A3627), are too distant to be linked. For the Hydra, Norma and Antlia 
clusters we find a distance to the nearest Centaurus SC member of 27.2, 35.1 and 22.9 Mpc, respectively,
while the linking length with the weighting for the specific location turns out to be, 19, 20.3 and 19 Mpc.
We discuss this further below, when we compare different linking schemes.

\subsection{The Coma Supercluster}

The Coma Supercluster, often also referred to as the Great Wall,
is found with 11 group and cluster members in the volume out to $z = 0.03$. If we relax
the boundary constraint, two more clusters are associated to this SC at redhifts $z = 0.03 - 0.0322$
as shown in Table ~\ref{tab5}. The two most prominent members of the Coma SC are the 
Coma cluster and A1367. All other groups and clusters have estimated masses below $1.5 \times 10^{14}$ M$_{\odot}$.

Fig~\ref{fig5} displays a three-dimensional representation of the Coma SC. This structure has a slightly larger
extent in the SGZ direction of 55.4 (65.3) Mpc compared to the SGX and SGY directions with an extent of
50.2 (61.8) and 38.0 (52.4) Mpc, respectively, where the number in brackets refer to the structure 
including the two clusters at $z > 0.03$. Compared to the Perseus-Pisces and Centaurus SC it is oriented much more
in a perpendicular direction to the Supergalatic plane. The total length of the Coma SC is 64.8 (78) Mpc. 
It is thus the third largest supercluster, also in mass, next to the Perseus-Pisces and Hercules SC. 

   \begin{table*}
      \caption{Groups and clusters which are members of the Coma SC. The meaning of the columns is the same as in Table~\ref{tab3}}
         \label{Tab3}
      \[
         \begin{array}{lrrrrrrrrrl}
            \hline
            \noalign{\smallskip}
{\rm name}&{\rm RA}&{\rm DEC}&{\rm redshift}& F_X & {\rm error}&L_X&m_{200}&r_{out}&n_H & {\rm alt. name} \\
            \noalign{\smallskip}
            \hline
            \noalign{\smallskip}
{\rm RXCJ1109.7+2146}& 167.4291 &  21.7682 & 0.0315 &   4.8267 &  25.00 &   0.1142 &   0.964 &  13.5 &   1.5&{\rm A~1177} \\
{\rm RXCJ1110.5+2843}& 167.6429 &  28.7208 & 0.0322 &   6.4190 &  12.10 &   0.1590 &   1.183 &  14.0 &   1.8&{\rm A~1185}\\
{\rm RXCJ1122.3+2419}& 170.5937 &  24.3191 & 0.0258 &   2.2962 &  18.40 &   0.0417 &   0.517 &   8.0 &   1.4&{\rm HCG~51}\\
{\rm RXCJ1145.0+1936}& 176.2631 &  19.6166 & 0.0217 &  42.4193 &   3.50 &   0.5161 &   2.466 &  17.5 &   2.5&{\rm A~1367}\\
{\rm RXCJ1204.1+2020}& 181.0487 &  20.3468 & 0.0226 &   3.6735 &  18.00 &   0.0438 &   0.534 &  16.0 &   2.4&{\rm NGC~4066} \\
{\rm RXCJ1204.4+0154}& 181.1049 &   1.9005 & 0.0202 &  17.6879 &   5.90 &   0.1786 &   1.278 &  17.0 &   1.9&{\rm MKW~4}\\
{\rm RXCJ1206.6+2810}& 181.6555 &  28.1827 & 0.0280 &   4.8100 &  12.50 &   0.0923 &   0.846 &  12.5 &   1.7&{\rm NGC~410}\\
{\rm RXCJ1213.4+2136}& 183.3511 &  21.6101 & 0.0243 &   1.3294 &  35.00 &   0.0180 &   0.307 &  14.5 &   2.3&{\rm UGC~7224}\\
{\rm RXCJ1219.8+2825}& 184.9643 &  28.4229 & 0.0272 &   1.6120 &  16.20 &   0.0344 &   0.458 &   6.5 &   1.9&{\rm CGCG~185-075} \\
{\rm RXCJ1223.1+1037}& 185.7761 &  10.6230 & 0.0257 &   8.6869 &  11.20 &   0.1649 &   1.213 &   9.0 &   2.2&{\rm NGC~4325}\\
{\rm RXCJ1231.0+0037}& 187.7726 &   0.6283 & 0.0232 &   1.4198 &  25.00 &   0.0194 &   0.322 &   9.5 &   1.9&{\rm NGC~4493}\\
{\rm RXCJ1259.6+2756}& 194.9196 &  27.9337 & 0.0231 & 305.0191 &  10.00 &   3.6314 &   8.259 &  70.0 &   0.9&{\rm A~1656~(Coma)}\\
{\rm RXCJ1334.3+3441}& 203.5976 &  34.6904 & 0.0241 &   3.0242 &  11.90 &   0.0436 &   0.532 &  11.5 &   0.9&{\rm NGC~522}\\

            \noalign{\smallskip}
            \hline
            \noalign{\smallskip}
         \end{array}
      \]
\label{tab5}
   \end{table*}

A display of the Coma SC in three projections in Supergalactic coordinates is given by Fig~\ref{fig6}.
The five most prominent members of the Coma SC are identified. We note that the Coma cluster
and NGC 522 (RXCJ1334.3+3441) are located at positive SGZ coordinates and are 
separated from most other groups and clusters in the SC. NGC 522 (RXCJ1334.3+3441)
is also the object furthest to the east in the sky. The two clusters with $z > 0.03$ have
the most negative SGZ coordinates. 

\begin{figure}[h]
   \includegraphics[width=\columnwidth]{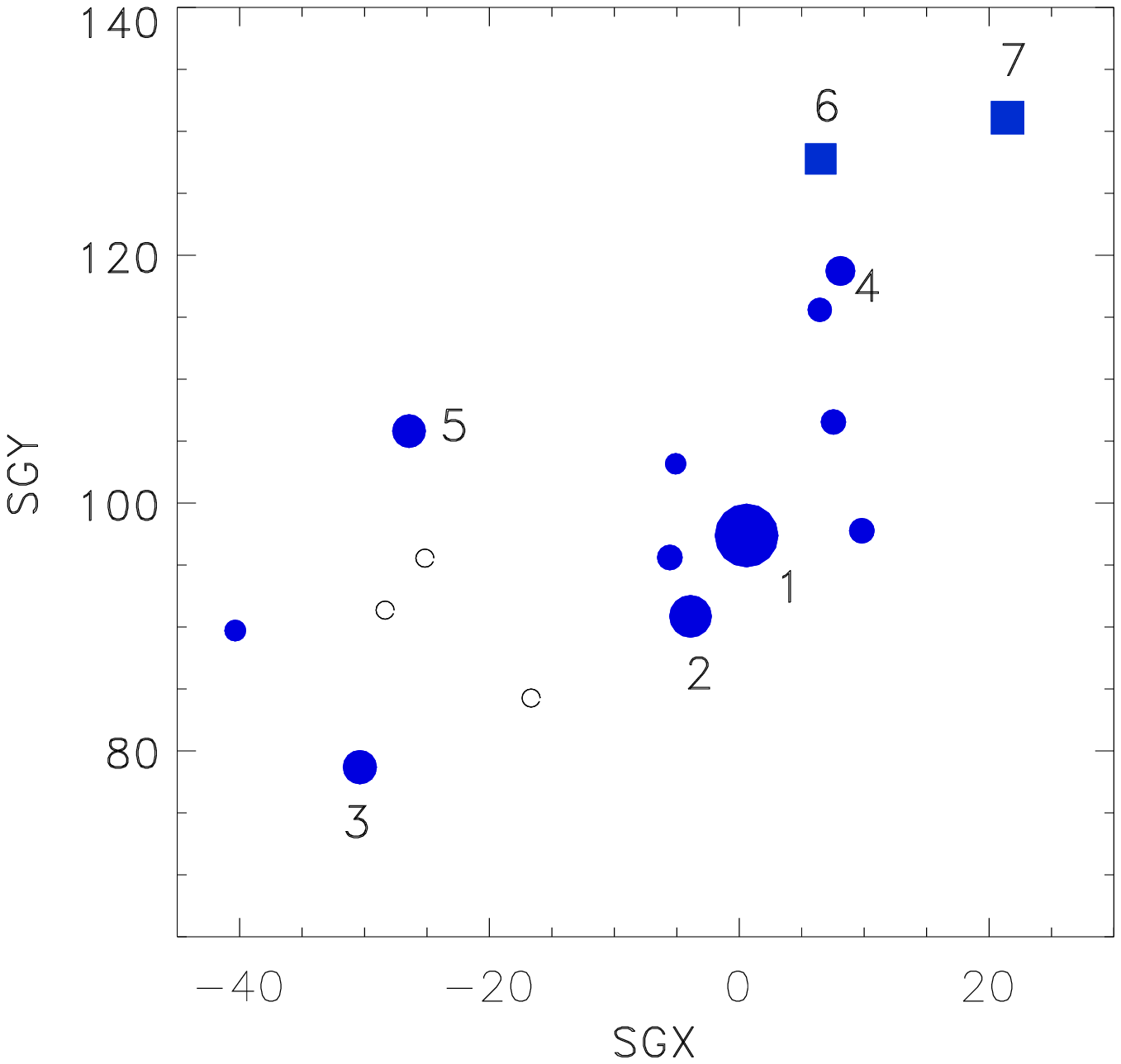}
   \includegraphics[width=\columnwidth]{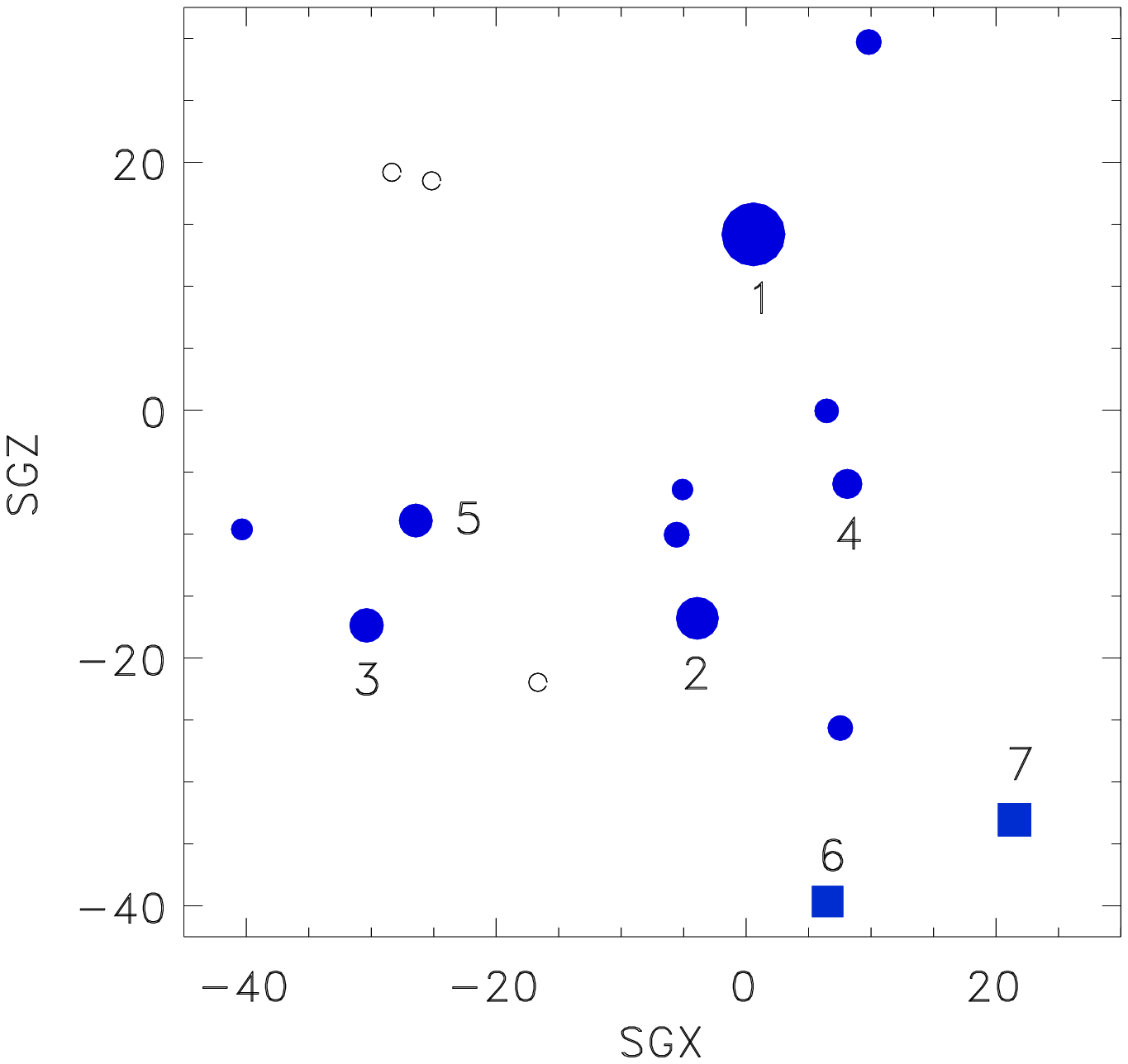}
   \includegraphics[width=\columnwidth]{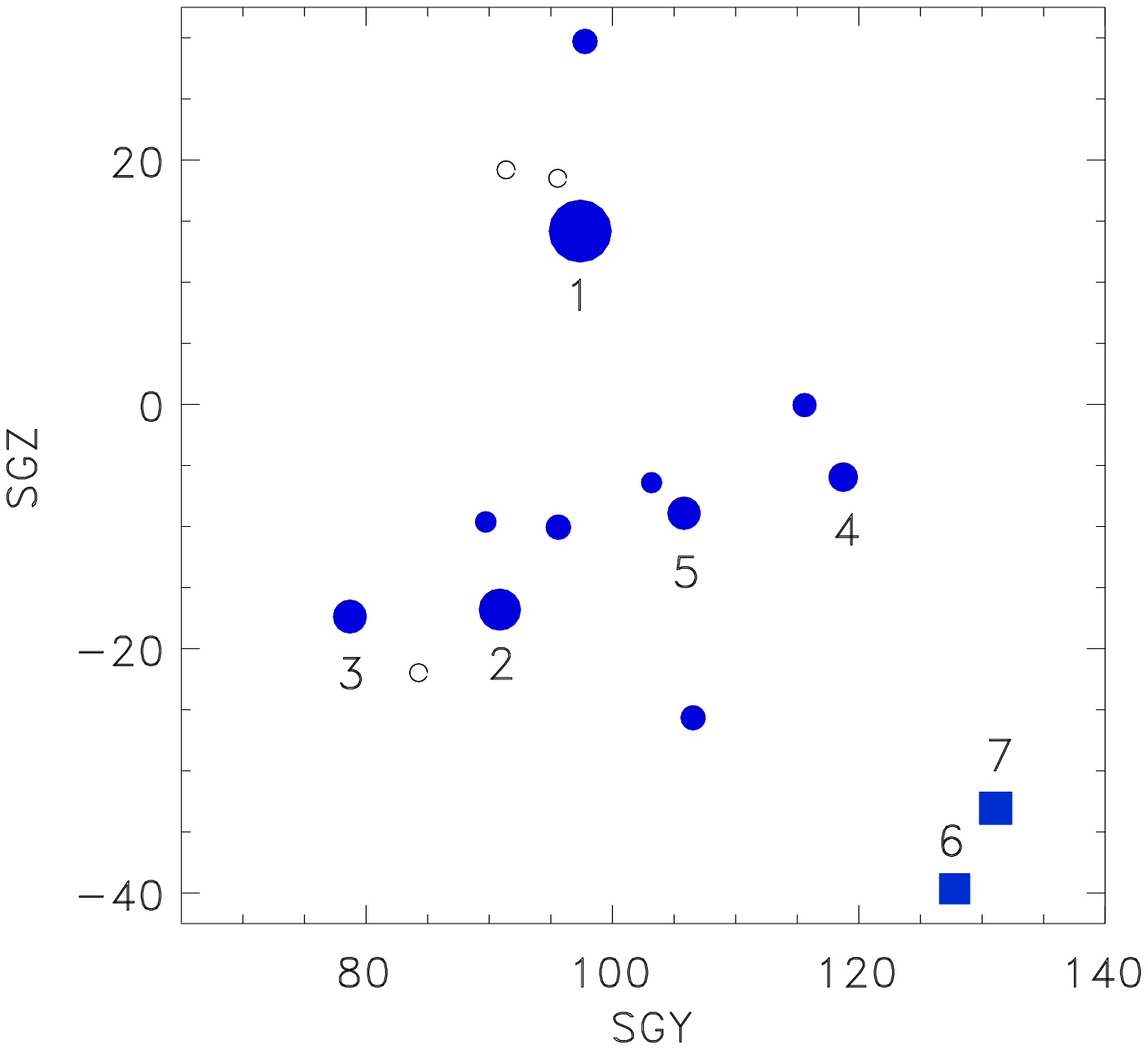}
\caption {Coma Supercluster in Supergalactic coordinates in three projections. Members of the structure are 
shown as filled blue circles, where the size of the symbol reflects the estimated mass
of the cluster. The two clusters at $z > 0.03$ are shown as blue squares. 
All other clusters are marked by black open circles. The most massive members
of the Coma supercluster are identified by numbers: 1: Coma, 2: A 1367, 3: MKW4, 
4: NGC 410, 5: NGC 4325, 6: A1177, 7: A 1185.
}\label{fig6}
\end{figure}

\subsection{The Hercules Supercluster}

The SC with the members shown in Table~\ref{tab6} consists of ten groups and poor 
clusters. The SC is located more 
than 60 Mpc above the Supergalactic plane (Figs~\ref{fig7}, \ref{fig8}). It is part of 
a larger structure with its major parts outside the radius of $z = 0.03$.
The core of this larger structure is the classical Hercules SC, with the members 
A2147, A2151, A2152 (e.g. \citet{Sha1932,Abe1961,Tar1979,Tar1980,Gio1997}. 
\citet{Abe1961} considered the former three clusters together with A2162,
A2197 and A2199 as one supergalactic system ('second order cluster'). \citet{Bar1998}
included further clusters, A2107, A2063 and A2052 as possible members of the SC. Apart from A2197
which lies at the boundary of our study region all these clusters are located at $z > 0.03$.
What we observe in our study-volume is just the extension of this much more massive structure,
that is linked together if we extend the friends-of-friends analysis with our recipe beyond
$z = 0.03$. We describe the entire structure in more detail in a subsequent publication
and concentrate here on our study volume. 

The part of the Hercules SC inside the study region contains mostly less massive
systems with masses below $10^{14}$ M$_{\odot}$, except for RXCJ1715.3+5724 (NGC 6338)
with an estimated mass of about $m_{200} = 1.7 \times 10^{14}$ M$_{\odot}$. 
The next massive object is 
RXCJ1629.6+4049 (A2197E) through which this structure connects to the classical Hercules
SC. Fig.~\ref{fig7} shows the locations of the Hercules SC
members at $z \le 0.03$ and a few clusters at higher redshift including the 
concentration A2197E, 2197W and A2199.

NGC 6338 has been observed with Chandra and XMM-Newton
and studied in detail by \citet{Pan2012} and \citet{Osu2019}.
It is found to be an interesting merger of a smaller group with the
main system. NGC 6338 also hosts interesting radio lobe cavities.
The temperature outside the core is about 2 - 3 keV \citep{Osu2019}.

   \begin{table*}
      \caption{Groups and clusters which are members of the Hercules SC at $z \le 0.03$. 
               The meaning of the columns is the same as in Table~\ref{tab3}.}
         \label{Tab4}
      \[
         \begin{array}{lrrrrrrrrrl}
            \hline
            \noalign{\smallskip}
{\rm name}&{\rm RA}&{\rm DEC}&{\rm redshift}& F_X & {\rm error}&L_X&m_{200}&r_{out}&n_H & {\rm alt. name} \\
            \noalign{\smallskip}
            \hline
            \noalign{\smallskip}
{\rm RXCJ1629.6+4049}& 247.4245 &  40.8231 & 0.0297 &   4.2600 &  18.00 &   0.0876 &   0.818 &  15.0 &   1.0&{\rm A~2197E}\\
{\rm RXCJ1649.3+5325}& 252.3283 &  53.4230 & 0.0298 &   2.8180 &  13.00 &   0.0579 &   0.632 &  13.5 &   2.8&{\rm Arp~330}\\
{\rm RXCJ1714.3+4341}& 258.5802 &  43.6882 & 0.0276 &   3.0830 &   9.90 &   0.0570 &   0.627 &  12.0 &   2.2&{\rm NGC~6329 }  \\
{\rm RXCJ1715.3+5724}& 258.8401 &  57.4082 & 0.0293 &  13.9714 &   3.50 &   0.2858 &   1.704 &  17.0 &   2.8&{\rm NGC~6338 }  \\
{\rm RXCJ1723.4+5658}& 260.8504 &  56.9785 & 0.0271 &   1.6518 &  10.60 &   0.0328 &   0.446 &   8.5 &   3.2&{\rm NGC~6370}  \\
{\rm RXCJ1736.3+6803}& 264.0919 &  68.0569 & 0.0256 &   2.0965 &  10.00 &   0.0316 &   0.435 &  15.0 &   4.4&{\rm NGC~6420^{a)}}  \\
{\rm RXCJ1755.8+6236}& 268.9557 &  62.6124 & 0.0259 &   3.9252 &   5.00 &   0.0610 &   0.655 &  16.0 &   3.4&{\rm ^{a)}}  \\
{\rm RXCJ1806.5+6135}& 271.6432 &  61.5974 & 0.0236 &   1.5521 &  20.00 &   0.0239 &   0.368 &   7.5 &   3.5&{\rm VII~Zw~767}\\
{\rm RXCJ1818.7+5017}& 274.6882 &  50.2837 & 0.0262 &   1.4078 &  12.00 &   0.0233 &   0.260 &  11.0 &   4.1&{\rm UGC~11202}  \\
{\rm RXCJ1941.7+5037}& 295.4422 &  50.6201 & 0.0243 &   3.3785 &  12.30 &   0.0475 &   0.561 &  14.0 &  12.8&{\rm UGC~11465}  \\
            \noalign{\smallskip}
            \hline
            \noalign{\smallskip}
         \end{array}
      \]
{\bf Notes:}$^{a)}$ These two clusters were previously identified in the RASS North Ecliptic Pole survey
by \citet{Hen1995} and are listed there by their RASS source names: RX J1736.4+6804 and RX J1755.8+6236. 
\label{tab6}
   \end{table*}

\begin{figure}[h]
   \includegraphics[width=\columnwidth]{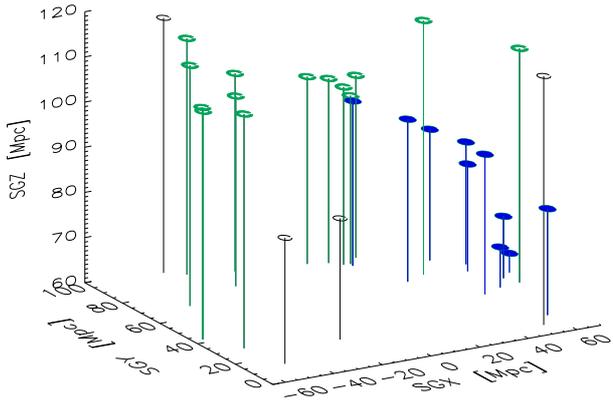}
\caption{Three-dimensional representation of Hercules supercluster
in Supergalactic coordinates. The members of the SC at $z \le 0.03$
are shown as solid blue circles, some of the structure beyond 
with open green circles. Other {\sf CLASSIX} clusters in the study
volume are marked by open black circles.
}\label{fig7}
\end{figure}

\begin{figure}[h]
   \includegraphics[width=\columnwidth]{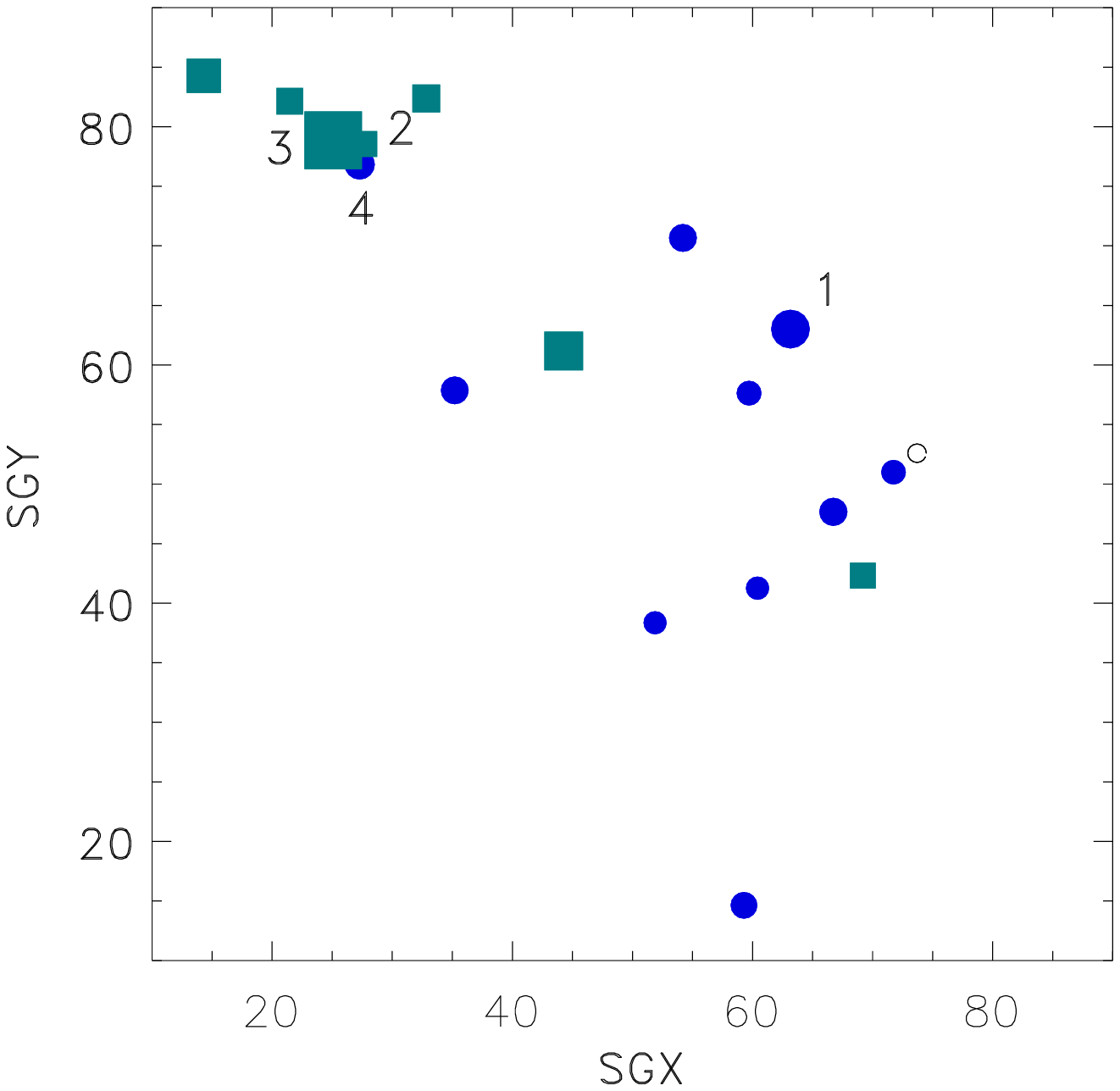}
   \includegraphics[width=\columnwidth]{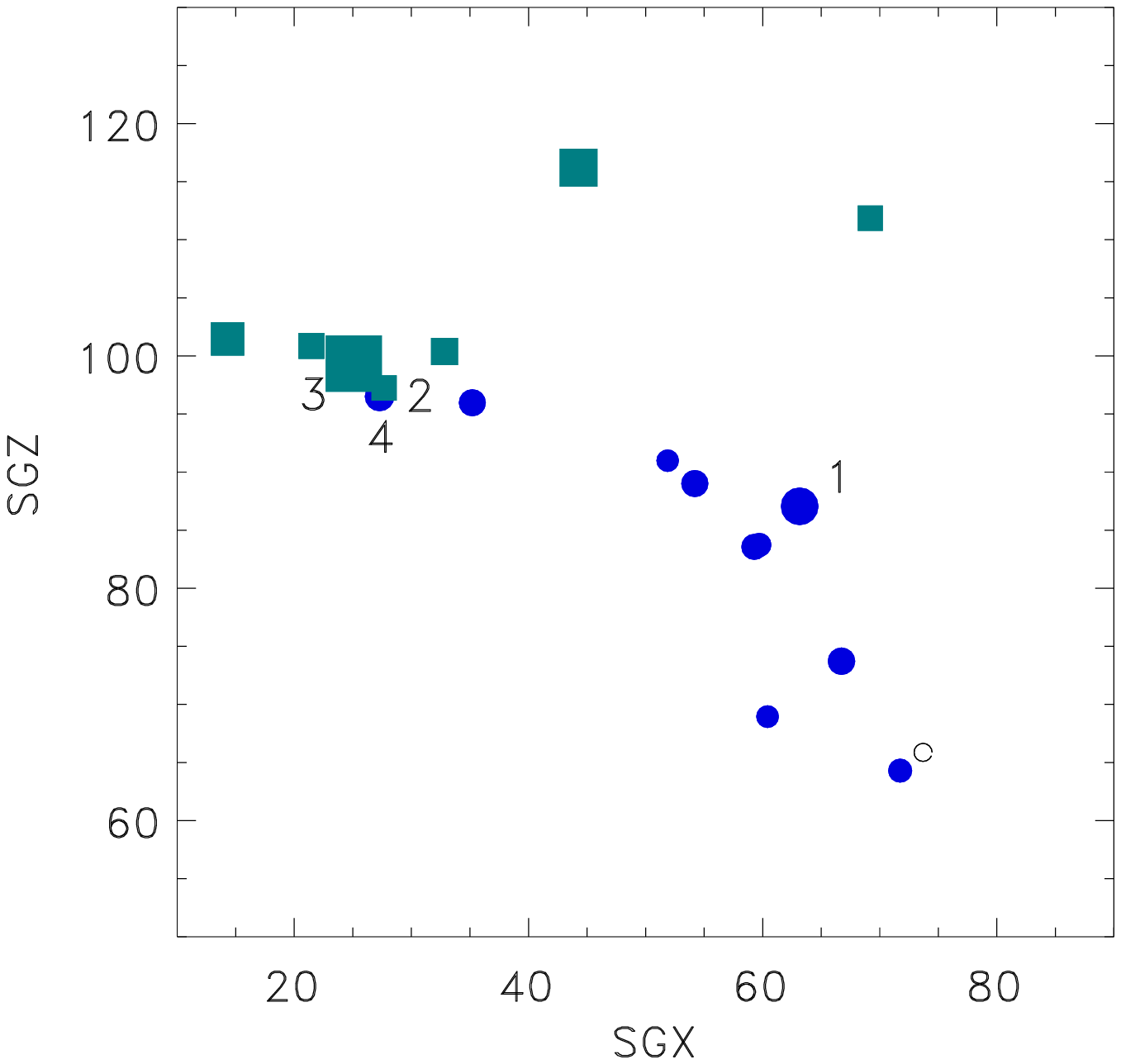}
   \includegraphics[width=\columnwidth]{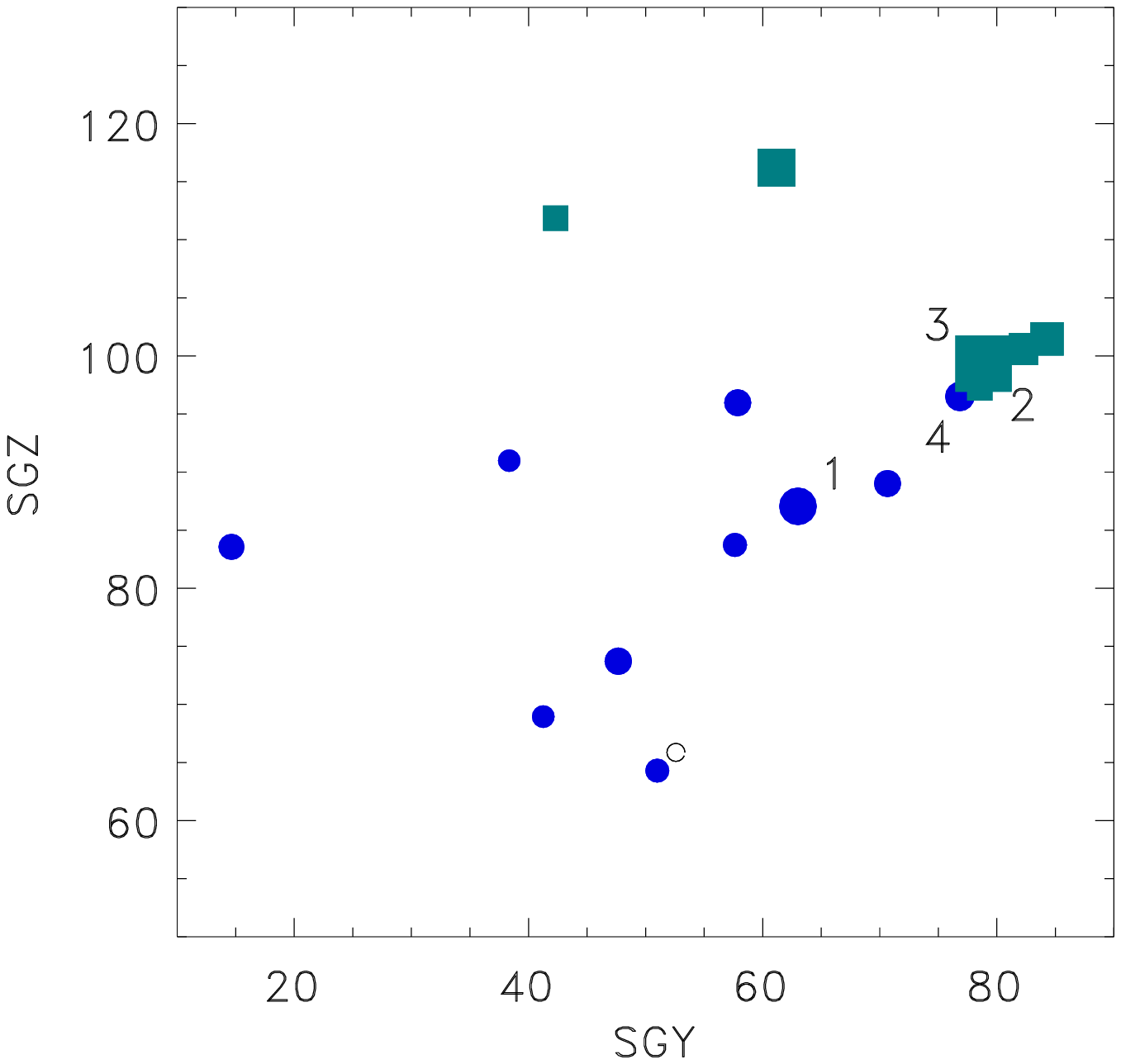}
\caption{Hercules supercluster in Supergalactic coordinates in three projections.
The size of the symbols indicates their mass. Supercluster members
with $z \le 0.03$ are shown as solid blue circles, those at $z > 0.03$ with
blue squares, and other {\sf CLASSIX} clusters in this volume with open black circles. 
The clusters with numbers are: 1 = NGC 6338, 2 = A2197W, 3 = A2199, 4 = A2197E.
}\label{fig8}
\end{figure}

\subsection{The Sagittarius Supercluster}

Six objects in the southern sky, as listed in Table~\ref{tab7}, are linked together 
to a supercluster in the constellation of Sagittarius. We have not 
found a previous description of this structure and 
thus refer to it as the Sagittarius SC.
All members have estimated masses below $m_{200}= 7 \times 10^{13}$ M$_{\odot}$.
The most massive one is the group around the galaxy ESO 460 - G004
($m_{200} = 6.5 \times 10^{13}$ M$_{\odot}$).
Fig.~\ref{fig9} provides a three-dimensional representation of the structure.
The SC has a length of 33.8 Mpc and an estimated mass of 
$4.3 - 5.5 \times 10^{15}$ M$_{\odot}$.

   \begin{table*}
      \caption{Groups and clusters which are members of the Sagittarius SC. The meaning of 
       the columns is the same as in Table~\ref{tab3}.}
         \label{Tab5}
      \[
         \begin{array}{lrrrrrrrrrl}
            \hline
            \noalign{\smallskip}
{\rm name}&{\rm RA}&{\rm DEC}&{\rm redshift}& F_X & {\rm error}&L_X&m_{200}&r_{out}&n_H & {\rm alt. name} \\
            \noalign{\smallskip}
            \hline
            \noalign{\smallskip}
{\rm RXCJ1928.2-2930}& 292.0661 & -29.5002 & 0.0246 &   2.9196 &  19.30 &   0.0602 &   0.649 &   5.5 &   8.5&{\rm ESO 460-G004}\\ 
{\rm RXCJ1944.0-2824}& 296.0096 & -28.4007 & 0.0200 &   3.1075 &  22.60 &   0.0325 &   0.445 &  10.0 &  10.6&{\rm NGC 6816}  \\
{\rm RXCJ2000.6-3837}& 300.1505 & -38.6231 & 0.0191 &   3.4630 &  20.00 &   0.0283 &   0.425 &  14.5 &   6.6& - ^{a)}  \\
{\rm RXCJ2018.4-4102}& 304.6065 & -41.0466 & 0.0192 &   4.4508 &  16.60 &   0.0461 &   0.552 &   9.5 &   4.7&{\rm IC 4991}^{b)}  \\
{\rm RXCJ2029.2-2240}& 307.3020 & -22.6717 & 0.0196 &   1.6227 &  22.00 &   0.0203 &   0.331 &   6.0 &   5.2&{\rm ESO 528-G008}  \\
{\rm RXCJ2035.7-2513}& 308.9348 & -25.2178 & 0.0200 &   2.5383 &  28.10 &   0.0250 &   0.378 &  12.5 &   4.5&{\rm A 3698}  \\
            \noalign{\smallskip}
            \hline
            \noalign{\smallskip}
         \end{array}
      \]
{\bf Notes:}$^{a)}$ The central galaxy of the group is WISEA J00035.61-383736.4 with z = 0.01946 and a brightness of around
15th magnitude. Another member of this group somewhat offset from the centre is IC4931 at
$z = 0.02004$. $^{b)}$ This group was also identified in the search for galaxy groups in the 2MASS redshift
survey as object no. 12252 by  \citet{Cro2007}.
\label{tab7}
\end{table*}

\begin{figure}[h]
   \includegraphics[width=\columnwidth]{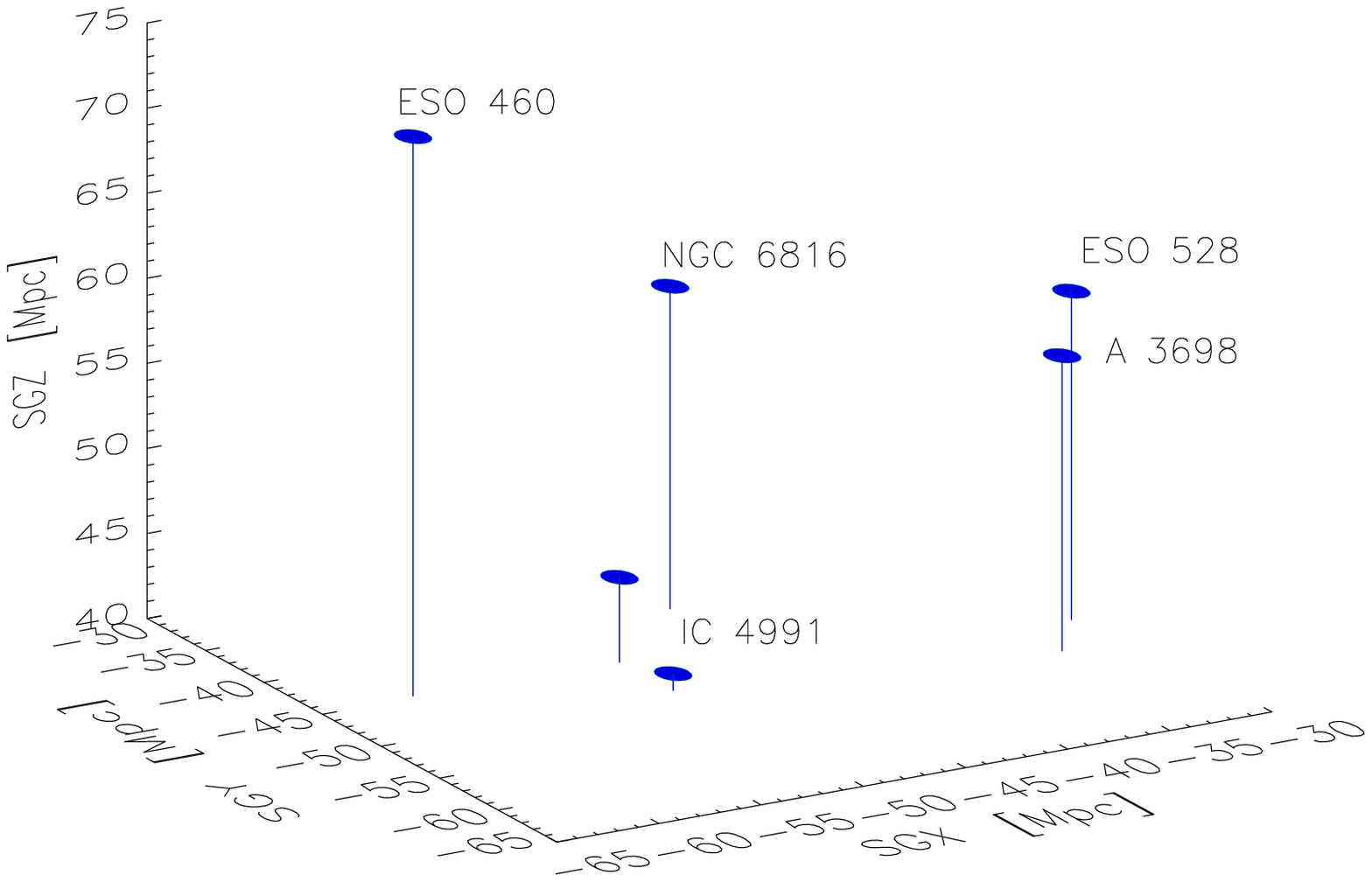}
\caption{Three-dimensional representation of distribution of
the members of the Sagittarius SC in Supergalactic coordinates.
The galaxy groups are designated by the names 
of their central galaxies.
}\label{fig9}
\end{figure}

\subsection{The Lacerta Supercluster} 

In Supergalactic coordinates the Lacerta SC is located about 40
to 50 Mpc above the Perseus-Pisces SC. It contains six members
all of which have estimated masses bellow $m_{200} = 1.1 \times 10^{14}$ M$_{\odot}$,
as listed in Table~\ref{tab8}. The SC is entirely located in the ZoA.
Fig.~\ref{fig10} provides a three-dimensional
view on the SC. One notes the close pair of the two groups,
UGC 12491 (CIZA2318.6+4257) and NGC 7618, which have almost the same 
redshift. They both have been found in Chandra observations to be  
interesting interacting systems with sloshing cold fronts \citep{Kra2006}.

The SC has a length of 19.9 Mpc and an estimated mass of 
$3.6 - 4.3 \times 10^{15}$ M$_{\odot}$. Apart from the Local SC
it is the smallest of the SC in the study volume.
We show below that it merges with the Perseus-Pisces SC with 
linking schemes at a higher X-ray luminosity limit.

   \begin{table*}
      \caption{Groups and clusters which are members of the Lacerta SC. The meaning of the columns is the same as in Table 3.}
         \label{Tab6}
      \[
         \begin{array}{lrrrrrrrrrl}
            \hline
            \noalign{\smallskip}
{\rm name}&{\rm RA}&{\rm DEC}&{\rm redshift}& F_X & {\rm error}&L_X&m_{200}&r_{out}&n_H & {\rm alt. name} \\
            \noalign{\smallskip}
            \hline
            \noalign{\smallskip}
{\rm RXCJ2215.6+3717}& 333.9159 &  37.2908 & 0.0192 &  13.2201 &   7.20 &   0.1196 &   0.997 &  17.5 &  14.6&{\rm NGC~7242}^{a)}\\
{\rm RXCJ2222.4+3612}& 335.6112 &  36.2141 & 0.0169 &   1.4443 &  18.10 &   0.0146 &   0.271 &   5.5 &  12.3&{\rm NGC~7265}\\
{\rm RXCJ2224.2+3608}& 336.0558 &  36.1336 & 0.0186 &   6.4139 &  10.30 &   0.0549 &   0.616 &  15.0 &  12.3&{\rm NGC~7274^{b)}}\\
{\rm RXCJ2231.0+3920}& 337.7690 &  39.3336 & 0.0171 &   4.2679 &  11.50 &   0.0337 &   0.455 &  11.0 &  12.0&{\rm -^{c)} }\\
{\rm RXCJ2318.6+4257}& 349.6694 &  42.9616 & 0.0174 &  16.8028 &   6.80 &   0.1271 &   1.036 &  18.0 &  11.8&{\rm CIZAJ2318.6+4257^{d)}}\\
{\rm RXCJ2319.7+4251}& 349.9398 &  42.8611 & 0.0173 &  16.5387 &   6.70 &   0.1236 &   1.018 &  18.0 &  11.8&{\rm NGC~7618}\\
            \noalign{\smallskip}
            \hline
            \noalign{\smallskip}
         \end{array}
      \]
{\bf Notes:}$^{a)}$ also identified as WBL group 679 \citep{Whi1999}.$^{b)}$ also identified as WBL group 681, the redshift above
is that of the group. $^{c)}$  The group features two bright elliptical galaxies, UC 12064  with a distance of about 3.4 arcmin to 
the reference coordinate and MCG+06-49-027 with a distance of about 5.2 arcmin, both at the redshift of the group. 
$^{d)}$ Central dominant galaxy is  UGC 12491.
\label{tab8}
   \end{table*}

\begin{figure}[h]
   \includegraphics[width=\columnwidth]{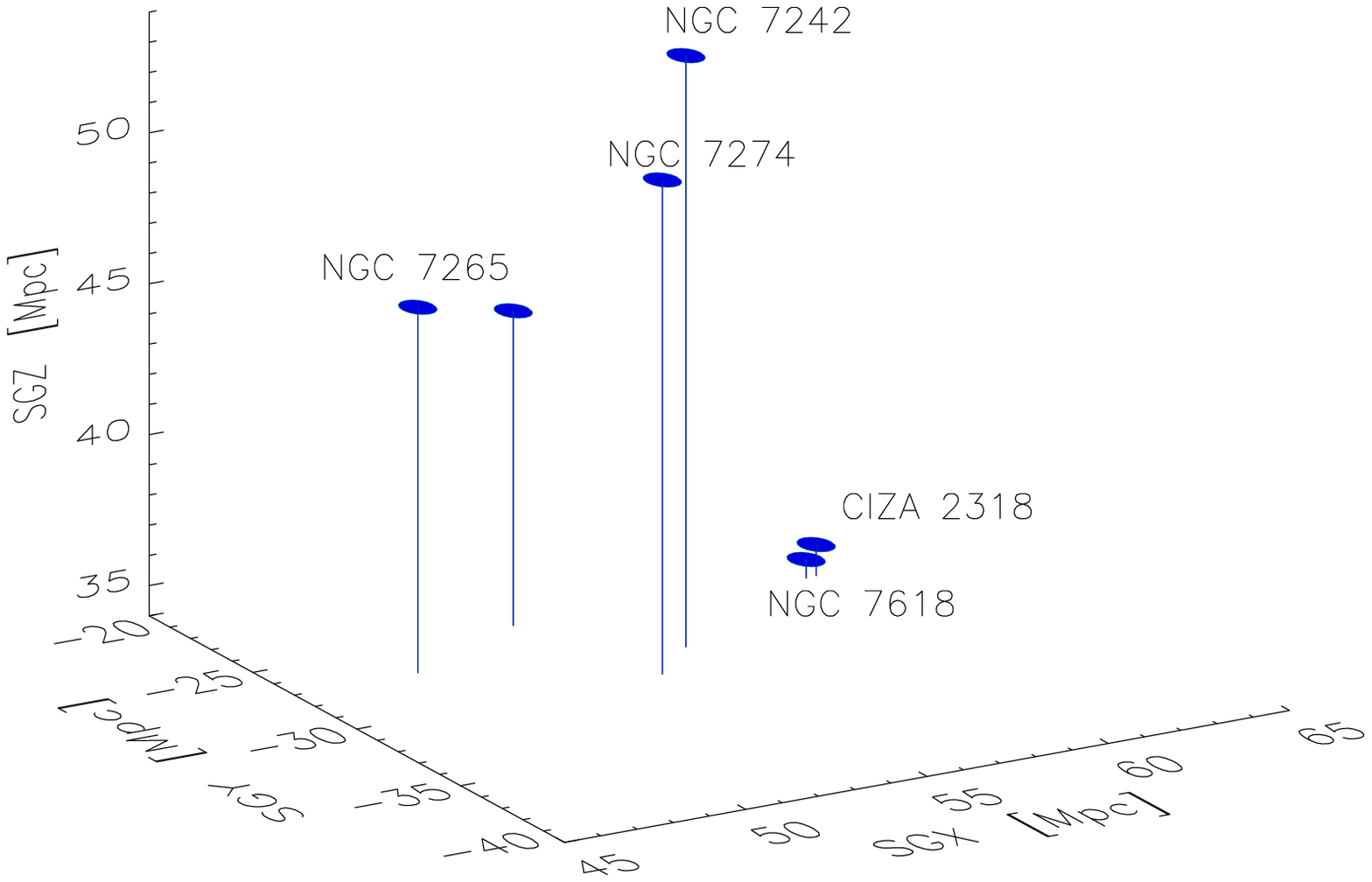}
\caption{Three-dimensional representation
the Lacerta SC in Supergalactic coordinates.
The galaxy groups are designated by the names 
of their central galaxies, except for one unmarked object, which contains
the giant elliptical UC 12064 slightly off-centre.
}\label{fig10}
\end{figure}

\subsection{Overview}

Fig.~\ref{fig11} provides an overview of all the structures in the study volume in
a three-dimensional representation. Among the two panels, the lower one provides 
a slightly better separation of the SC. 
While in total we found 146 groups and clusters with $L_x \ge 10^{42}$ erg s$^{-1}$ at
$z \le 0.03$; 75 of these are part of superclusters (a fraction of 51\%).
This result is similar to that found for superclusters in
the  entire {\sf REFLEX} survey by \citet{Cho2013}. If we compare the volumes based on
the values in Table~\ref{tab1}, however, we find
that the SC occupy only $\sim 14\%$ of the study volume at $z \le 0.03$.
Using the alternative volume calculation with 10 Mpc radius of Table~\ref{tab2},
the volume fraction is only about 1.8\%.

Inspecting the cluster and SC distribution in the plot, we note two clear
asymmetries. In equatorial coordinates we find 86 clusters in the northern
compared to 60 in the southern sky, which is an overabundance by about 
1.5$\sigma$ in the north. But looking at the number of cluster which are members
of SC we find a more than 6$\sigma$ difference, 57
compared to 18, respectively. Thus there are clearly more SC
in the northern sky, including the Perseus-Pisces SC,
most of the A400, the Local, the Coma,
the Hercules, and the Lacerta SC, compared to Centaurus and Sagittarius SC 
in the southern sky (see also Fig.~\ref{fig1}). The other 
inhomogeneity concerns the SC distribution with respect
to the Supergalactic SGZ coordinate. There is no significant  
difference in the number above and below SGZ = 0: 
67/79 for all clusters, 39/36 for clusters in 
SC. But there is a striking difference if we compare the number
of clusters at $SGZ \ge 50$ Mpc with those at $SGZ \le -50$ Mpc, which is
29/18 for all clusters and 16/0 for clusters in SC. Thus we
find no SC in the volume of $z \le 0.03$ and $SGZ \le -50$ Mpc. 
We also note the strong segregation towards the Supergalactic plane,
that we studied in detail in \citet{Boe2021a}.
Among the eight supercluster, the four major ones, Perseus-Pisces,
Centaurus, Coma, and Hercules SC, are clearly recognised as
the largest structures.

\begin{figure*}[h]
   \includegraphics[width=16.0cm]{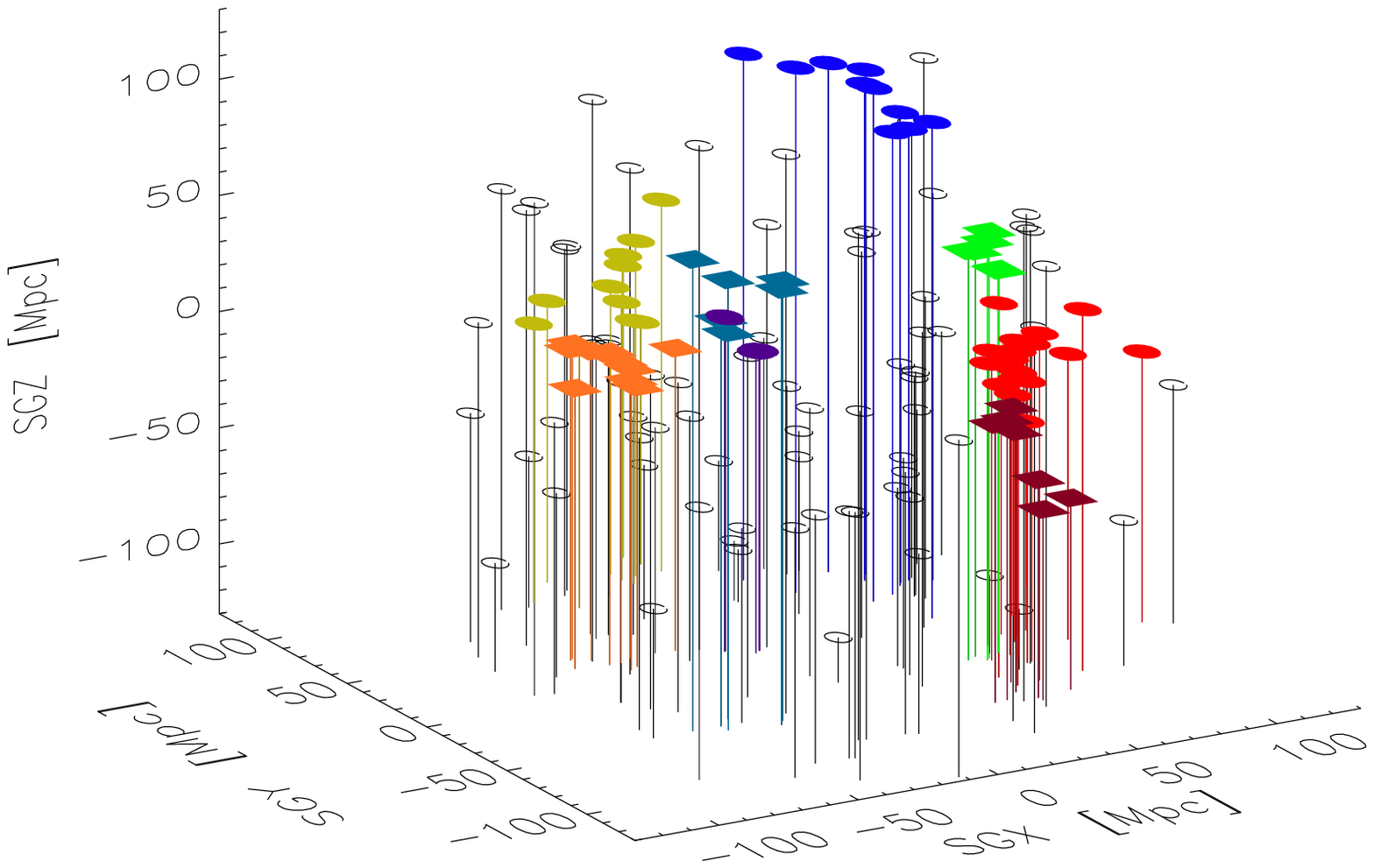}
   \includegraphics[width=16.0cm]{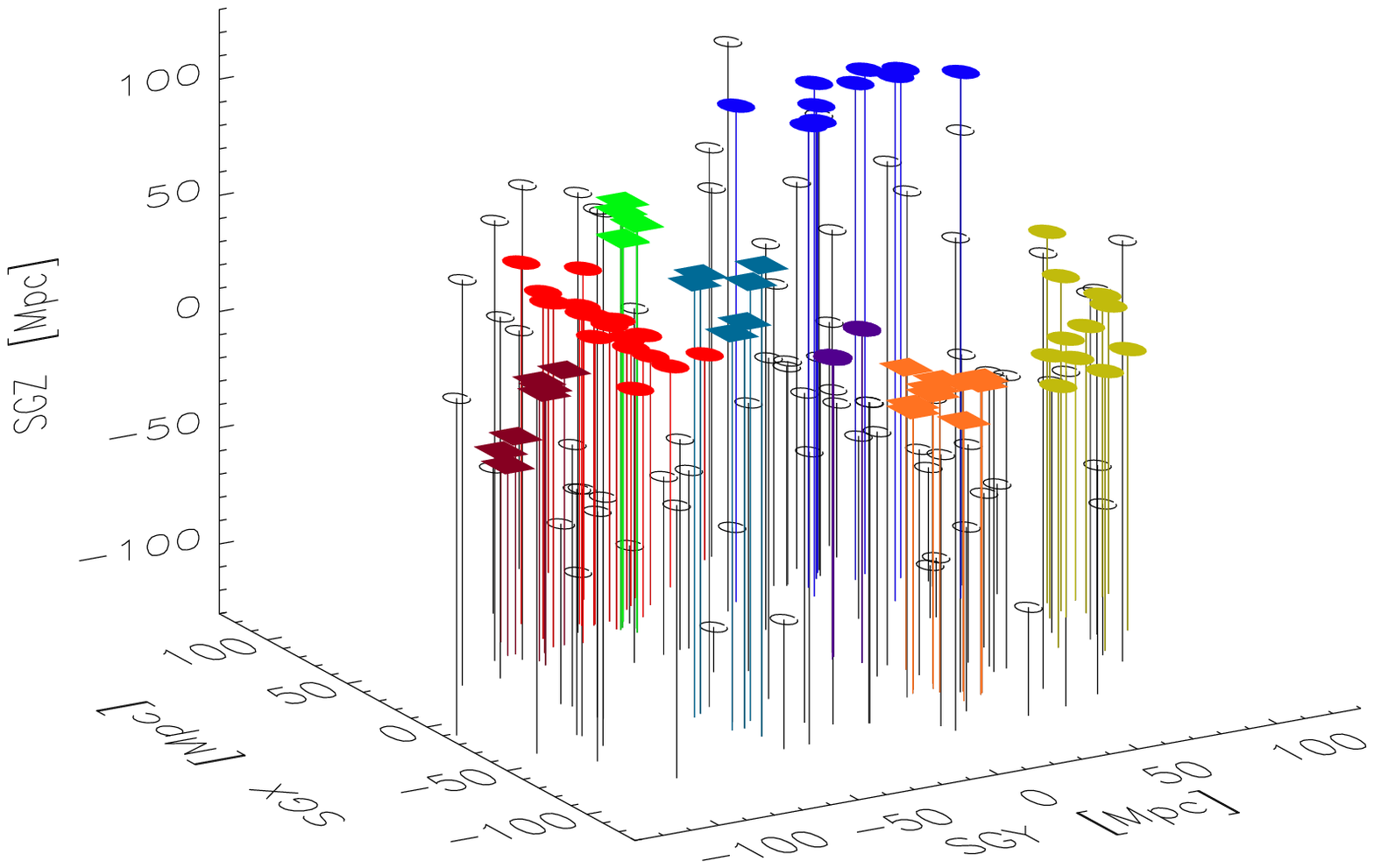}
\caption{Three-dimensional representation of all 
supercluster found within $z \le 0.03$ with at least five members and a minimum linking
length of $l_0 = 19$ Mpc. The structures marked in colour are:
red = Perseus-Pisces SC, red-brown = A400 SC, violet = Local SC, orange = Centaurus SC,
siena = Coma SC, dark blue = Hercules SC, turquoise = Sagittarius SC, light green = Lacerta SC.
Non-supercluster members are shown as black open circles.  
}\label{fig11}
\end{figure*}

\section{Discussion}

\subsection{Effect of the lower luminosity limit for member clusters}

We have constructed the SC described above with groups and clusters
of galaxies with a low X-ray luminosity limit, much lower than what was typically 
used in the past. The advantage of this procedure is that we can base the cluster 
density mapping on a sufficiently high cluster density, which is less subject to shot noise.
But it also raises the interesting question, in how much the SC found
depend on this lower luminosity limit. To test this, we conducted
the following study. We repeated the SC construction with a series of
schemes involving an increasing X-ray luminosity limit. By increasing this limit,
we are thinning out the cluster density. To compensate for this, we
increased the linking length accordingly. This was done using the following equation,

\begin{equation}
\tilde l_0 = l_0 \left({ n_{CL}(L_{X_0})\over n_{CL}(\tilde L_{X_0})}\right)^{1/3} ~~~~~~~~~~~~,
\end{equation}
\vspace{0.3cm}   

where $l_0$ is the nominal (minimal) linking length defined in section 3, $\tilde l_0$ is
the new nominal linking length. $n_{CL} (L_{X_0})$ is the cluster density as a function of
lower luminosity limit, where $L_{X_0}$ and $\tilde L_{X_0}$ are the nominal and new lower X-ray 
luminosity limits.
For this study we varied $\tilde L_{X_0}$ over an order of 
magnitude in eight steps including the linking schemes listed in Table~\ref{tab9},
which gives the values of the lower luminosity limit, $\tilde L_{X_0}$, and
the nominal linking length, $\tilde l_0$, for each step. With each step the density mapping
gets noisier. To provide a feeling for this effect, we give in the forth row, labelled 
$N_{CL}$, the number of clusters at $z \le 0.03$ above $\tilde L_{X_0}$. 
Up to step four we only consider SC with at least five members,
as done above. From step 5 on we also consider structures with four members, since the statistics
gets considerably poorer. In this exercise we are including only groups and clusters 
at redshifts $z \le 0.03$.

   \begin{table*}
      \caption{Supercluster membership as a function of the lower X-ray luminosity limit, $\tilde L_{X_0}$.
 The table provides also the minimum linking length, $\tilde l_0$, and the number of clusters in the study
 volume, $N_{CL}$, as a function of the lower luminosity limit. For each SC we show in the second column
 the number of clusters linking in the first linking step and in subsequent columns the number of members lost and
 gained by the next linking step. If less than five members are found in steps 1 to 4 and less than four members
 in steps 5 to 8, the structure is no longer considered and marked as lost.}
         \label{Tab7}
      \[
         \begin{array}{lrrrrrrrrr}
            \hline
            \noalign{\smallskip}
{\rm scheme}  & 1      & 2     & 3     &  4    & 5     & 6     & 7     & 8     & 8    \\
\tilde L_{X_0} & 0.01  & 0.02    & 0.03  & 0.04 & 0.05  & 0.06  & 0.07  & 0.1   & 0.1  \\
\tilde l_0    & ~~~~~19.0   & ~~~~23.59  & ~~~~26.84 & ~~~~29.40 & ~~~~31.59 & ~~~~33.52 & ~~~~35.26 & ~~~~39.72 & ~~~~39.72\\
N_{CL}        & 146         &    121     &   105      &     87   &      69    &    55    &     49    &      41   &   -  \\ 
            \noalign{\smallskip}
            \hline
            \noalign{\smallskip}
{\rm Perseus-Pisces~SC}   & 20     & -4     &  -    &  -4   &-3, +4  & -2    &  -   & -1    & 10   \\
{\rm A400 SC} & 7      & -1     & {\rm lost}  &  &        &       &      &       & -    \\
{\rm Local~Superlcuster}  & 5      & {\rm lost} &   &       &        &       &      &       & -    \\
{\rm Centaurus~SC}        & 10     & -5, +5 & -2 +1 &   -   & -2, +1 & -3    &  -   & -1, +2& 6    \\
{\rm Coma SC}          & 11     & -3     & +2    & -1, +1&   -    & -4    &  -   & -2    & 4    \\
{\rm Hercules~SC}       & 10     &  -3    & -2    & {\rm lost}&    &       &      &       & -    \\
{\rm Sagittarius~SC}       & 6      &  -     &   -   & {\rm lost}&    &       &      &       & -     \\
{\rm Lacerta~SC}          & 6      & -1     &   -   & -1    & {\rm lost}&    &      &       & -    \\   

            \noalign{\smallskip}
            \hline
            \noalign{\smallskip}
         \end{array}
      \]
\label{tab9}
   \end{table*}

The smaller structures do not survive all the steps. The Local SC gets merged with the Centaurus SC already
in step 2, similar to the behaviour we saw in section 4.3. The A400 SC, which does not contain
many massive clusters, is not recovered in step 3. The Sagittarius SC is lost in step 4 and the
Lacerta SC survives with four members up to step 4 and merges in step 5 with the Perseus-Pisces SC.
Only the larger SC survive till the highest luminosity limit, except for the
Hercules supercluster (lost in step 4). The latter has its core outside the considered redshift range 
and it would only survive if we had included also this part. 
The fact that the smaller structures are not traced well if we lower the statistics is not surprising,
since the density mapping is just not fine enough to recognise them.

In Figs.~\ref{fig12} to \ref{fig14} we give an impression how the larger structures survive the increase
of the X-ray luminosity limit. In these figures the solid circles and squares show the structure recovered
at the lowest and the large open circles the structure found at the highest X-ray luminosity limit.
For the Perseus-Pisces SC we note in Fig.~\ref{fig12} that the main chain of clusters of
this SC is found over the complete luminosity limit range. However, while a small ensemble of 
galaxy groups found at the lowest $\tilde L_{X_0}$ in the west of the SC is lost at higher  $\tilde L_{X_0}$,
the Lacerta SC with four clusters is linked to the Perseus-Pisces SC on the western side in step 5.

For the Centaurus cluster (Fig.~\ref{fig13}) the first step brings a significant increase, where 
the Local SC is linked with two members and also Hydra, Antlia and another cluster are added. 
At the highest luminosity limit also the Norma cluster and another luminous cluster in the 
ZoA gets linked. For the Coma SC, shown in Fig.~\ref{fig14}, we note less changes. In step
3 two groups around the galaxies NGC 5129 and NGC 5171 get linked and then the overall structure
remains with fewer members until the highest luminosity limit. 

Thus we find that the construction of SC with the FoF scheme applied here is robust
and not much sensitive to the details of the linking parameters for the given overdensity
selection. There is some change in the linking of different small extensions and an
overall trend that the structures become slightly larger with increasing luminosity limit.
But the main structures stay the same. Another very important fact is that no new 
structure appeared in this process that would have been missed in the
first analysis.

It is interesting to note that the linking of the clusters at higher $\tilde L_{X_0}$
finds the Hydra-Centaurus SC as one unit, which is similar to most descriptions in
the literature. Similarly interesting is the merging of the Perseus-Pisces and Lacerta SC
at higher $\tilde L_{X_0}$. In this case the inclusion of the Lacerta members
indicates an upturn of the Perseus-Pisces SC on the western side. Such an upturn
is prominently seen in the galaxy distribution if radio observations in HI are
included in the redshift surveys. \citet{Kra2018} show in their Fig. 10 the galaxy
distribution around the Perseus-Pisces SC and one can clearly note the 
pronounced filament of the SC. Around $l_{II} \sim 115^o$ and $b_{ii} \sim -30^o$
this galaxy filament turns northward in Galactic coordinates and crosses the
equatorial plane around $l_{ii} = 90^o$. The ensemble of galaxy groups of the
Lacerta SC falls roughly into the middle of this upturning filament and
thus also traces this SC extension.

\subsection{Comparison to previous studies of superclusters of clusters}

Of the eight systems in our sample of SC, the five most prominent
structures have been previously known and are, for example, described in the review
by \citet{Oor1983}. For these SC we can compare the size of the systems quoted
in the review with our findings. The size of the Local SC was given as $\sim 28.6$ Mpc
(18.5 Mpc), for the Perseus SC 54 Mpc (116 Mpc), for the Coma SC 65 - 114 Mpc (78 Mpc),
for Hydra-Centaurus SC 64 Mpc (37 Mpc), and for the Hercules SC 100 Mpc (140 Mpc), where
the numbers in brackets are our results from Table 1. The sizes from
\citet{Oor1983} were converted from a scaling with a Hubble constant of $H_0 = 50$
km s$^{-1}$ Mpc$^{-1}$ to the value of 70 km s$^{-1}$ Mpc$^{-1}$  used here. The largest
differences are for the Perseus SC, where we now include an extension through the
ZoA, and for the Centaurus SC where we have not included the complex around the
Hydra cluster as discussed in the previous section. Already \citet{Joe1978} found that
the Perseus-Pisces SC is the most prominent superstructure in the nearby Universe.
They assigned a mass of about $2\times 10^{16} h^{-1}_{50}$ M$_{\odot}$ to this SC which 
is similar to our result. This earlier mass is a little smaller as the extension
across the ZoA was not included. Overall these major structures have
been recognised in a similar way in many different studies, some of which were already
mentioned above in the corresponding sections of the SC, which shows that their
recognition in the galaxy cluster distribution as distinct superstructures is robust.

\subsection{Comparison to other studies of the large-scale structure}

Our results can also be compared to methods of characterising the cosmic 
large-scale structure other than those using galaxy clusters, such as the cosmic flow analysis
based on galaxy peculiar velocities, the study of the galaxy density distribution, and various
ways of geometrical characterisations of the cosmic web. The use of galaxy peculiar
velocities to trace the matter distribution in the Universe is a 
sensitive method, but restricted to the local Universe in the volume 
in which peculiar velocities can be determined with sufficient precision. That 
our results are also confined to the nearby Universe in only a slightly larger
volume, makes a comparison with the results from a cosmic flow analysis, like
those of the group of Tully, interesting. In  Fig. 1 of  \citet{Tul2019} 
the four major nearby mass concentration labelled as Virgo and Great Attractor, Coma, 
Perseus-Pisces, and Hercules,
are the same five major structures we found here. Starting from our local position we
find that the Local SC is closely linked to the Centaurus or Hydra-Centaurus complex.
This is what \citet{Tul2014} found in their streaming analysis which links the Local
SC, to the Great Attractor in the Hydra-Centaurus region. All of this
together with the Norma cluster is combined into the Laniakea SC. 
In a stricter mathematical approach to segment the major structure in the local
cosmic flows by \citet{Dup2019,Dup2020} the authors isolate the structures into basins of
attraction by following individual streamlines to common destinations. They used the
Constrained Local UniversE Simulations ({\sf CLUES};
\citet{Yep2009,Got2010}) for their study and identified as major attractors the 
Laniakea SC (volume $= 5\times 10^5~ ($Mpc $ h^{-1})^3$), Coma SC 
(volume $= 1\times 10^6~ ($Mpc $ h^{-1})^3$), and Perseus-Pisces SC 
(volume $= 7\times 10^5~ ($Mpc $ h^{-1})^3$). The typical mass of their basins 
of attraction is about  $= 5\times 10^{16}~ {\rm  M}_{\odot}~ h^{-1}$. The volumes 
are about three to five times larger than the values we find and the typical mass
is about two to three times higher than that for the larger structures in our sample.
This difference is due to the different definitions of the structures. While we only
consider the overdense regions of the SC, the basins of attraction include the 
complete surroundings of the SC. Accounting for this fact by considering 
the filling factors of our SC, the results become quite similar.

Most of the more recent studies of SC in the galaxy distribution are based 
on the SDSS or the 2degree Field Galaxy Redshift Survey (2dFGRS) at redshifts 
outside our study volume. Therefore we can only make a statistical comparison. 
\citet{Ein2007a,Ein2007b} have identified SC in the 2dFGRS with a density field method, 
finding large structures in the galaxy density distribution smoothed with an Epanechnikov 
kernel of radius $8 h^{-1}$ Mpc as overdensities. The richest of these SC have typical 
radii of about 50 to slightly over 100 $h^{-1}$ Mpc. They also analyse cosmological
simulations, the Millennium run galaxy catalogue by \citet{Cro2006}, for comparison
and find SC with similar sizes. These structures can well
be identified with the type of SC we find. \citet{Lii2012} identified and 
studied superclusters in a similar way in the SDSS
with the density field method. For a
plausible density threshold  $D \sim 5$ for the selection of SC
they find the largest SC to have diameters of $100 - 200~ h^{-1}$ Mpc.
With the extended percolation analysis \citet{Ein2018,Ein2019,Ein2021} detected
SC in the galaxy distribution of the SDSS with a similar density threshold of
$D = 5$ as used by \citet{Lii2012} and find that the size function is cut off
at diameters slightly larger than  $100~ h^{-1}$ Mpc. \citet{Ein2016} have analysed
the region of the Sloan Great Wall with a similar recipe as \citet{Lii2012}.
This is one of the largest structures found with an estimated length of about
$328~ h_{70}^{-1}$ Mpc. In their analysis the Sloan Great Wall breaks up into
two larger and three smaller SC. The two larger SC have masses in the
range $ 1 - 2 \times 10^{16}~ h_{70}^{-1}$ M$_{\odot}$. We therefore note that these studies
of SC in the galaxy distribution, which cover a much larger cosmic volume, 
do not unvail much larger SC than what we found in the nearby
Universe.  
 
An overview on different geometrical ways to characterise the cosmic web is for example
given by \citet{Cau2014} (see also \citet{Lie2018}). They identify structures with different
morphologies in the cosmic web: clusters, filaments, sheets, and voids. They find that 
filaments dominate the cosmic web at least since $z \sim2$, carrying about 50\% of
the total mass at present. The filaments have a fractal distributions over a 
range of scales and most of the mass is carried by the most massive filaments.
The largest filaments have an extent over $100~ h^{-1}$ Mpc and 
connect several clusters in a linear configuration. Therefore we can  
identify the superclusters we find with the massive filaments of
the geometrical analysis. We thus find good agreement, with the 
minor exception, that we would not stress that these structures 
are generally linear chains of clusters, which we
find in such a pronounced way particularly for the Perseus-Pisces SC.

\begin{figure}[h]
   \includegraphics[width=\columnwidth]{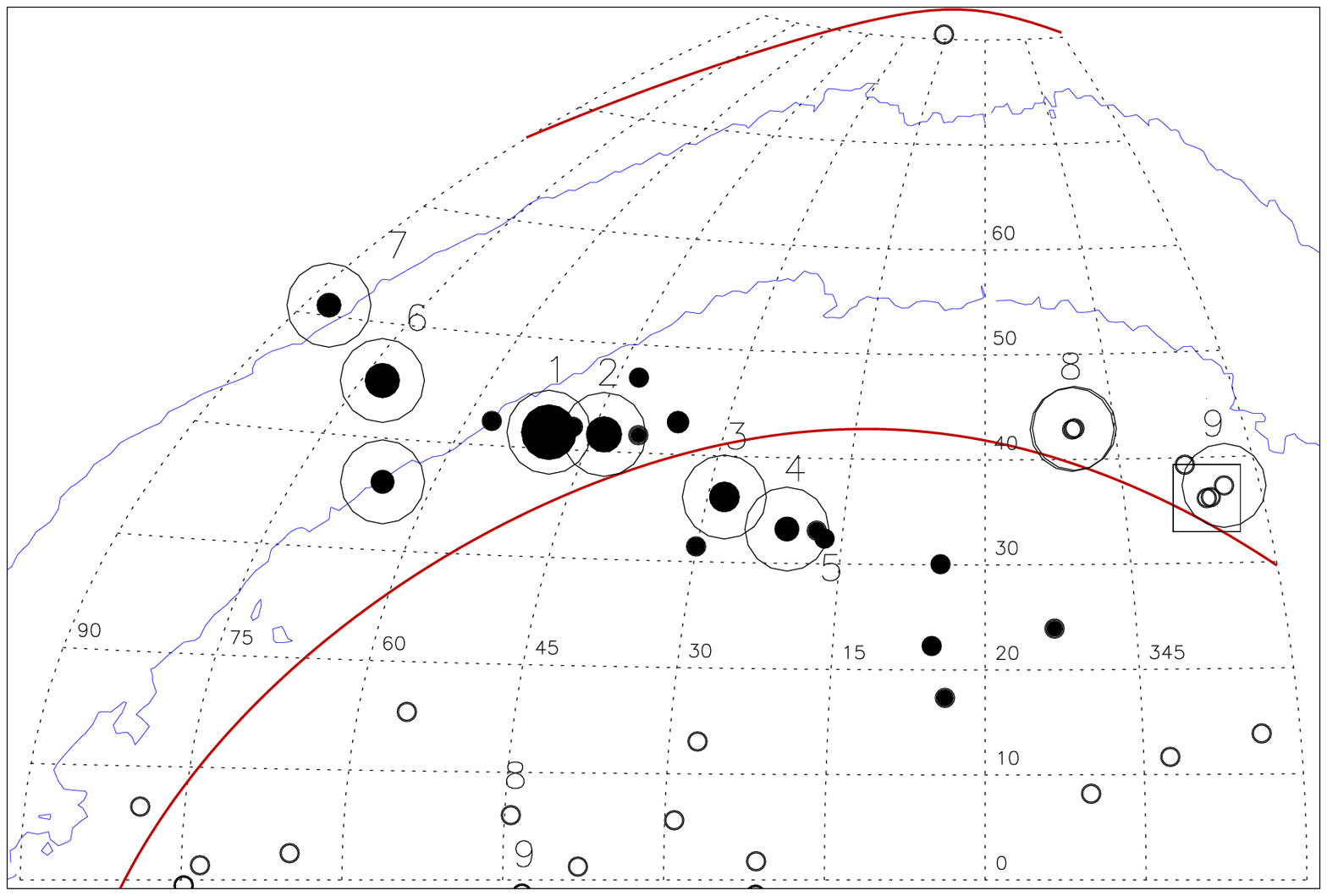}
\caption{Cluster members of the Perseus-Pisces SC for different linking
schemes. The solid black points show the SC members found for scheme 1 and
the large open circles show the members found for scheme 8. The 
object marked by an open square is first found with scheme 5. Some prominent
clusters are marked: 1 = Perseus, 2 = AWM7, 3 = A262, 4 = NGC 507, 5 = NGC 383,
6 = 3C129, 7 = UGC 12655, 8 = double cluster RXCJ2318.6+4257, RXCJ2319.7+4521,
9 = NGC7242. The blue lines constrain the region of
high interstellar absorption, the red lines show galactic latitudes $b_{II} \pm 20^O$. 
}\label{fig12}
\end{figure}

\begin{figure}[h]
   \includegraphics[width=\columnwidth]{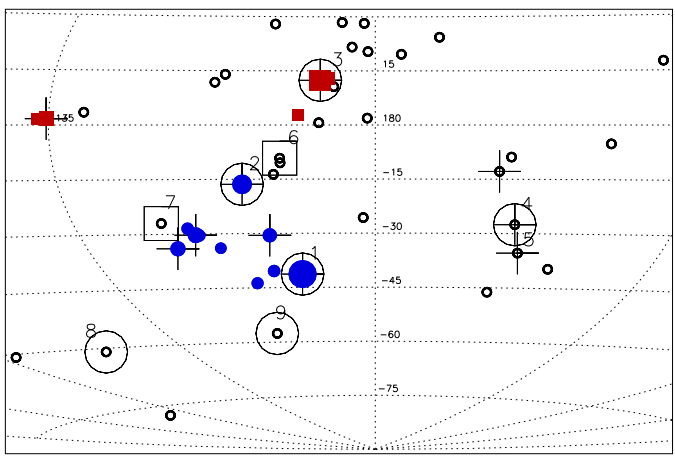}
\caption{Cluster members of the Centaurus SC and the Local SC for 
different linking schemes. The solid circles and the red squares show
the Centaurus and Virgo superclusters found with scheme 1, 
the crosses mark those found with scheme 2 and the large open circles 
mark those found with scheme 8.
The open squares show two clusters linked at intermediate schemes
3 and 5, respectively. Some prominent clusters are marked: 1 = Centaurus,
2 = NGC 5044, 3 = M87, 4 = Hydra (A1060), 5 = Antlia, 6 = HCG 62, 7 = A3581,
8 = Norma (A3627), 9 = RXCJ1324.7-5736. 
}\label{fig13}
\end{figure}

\begin{figure}[h]
   \includegraphics[width=\columnwidth]{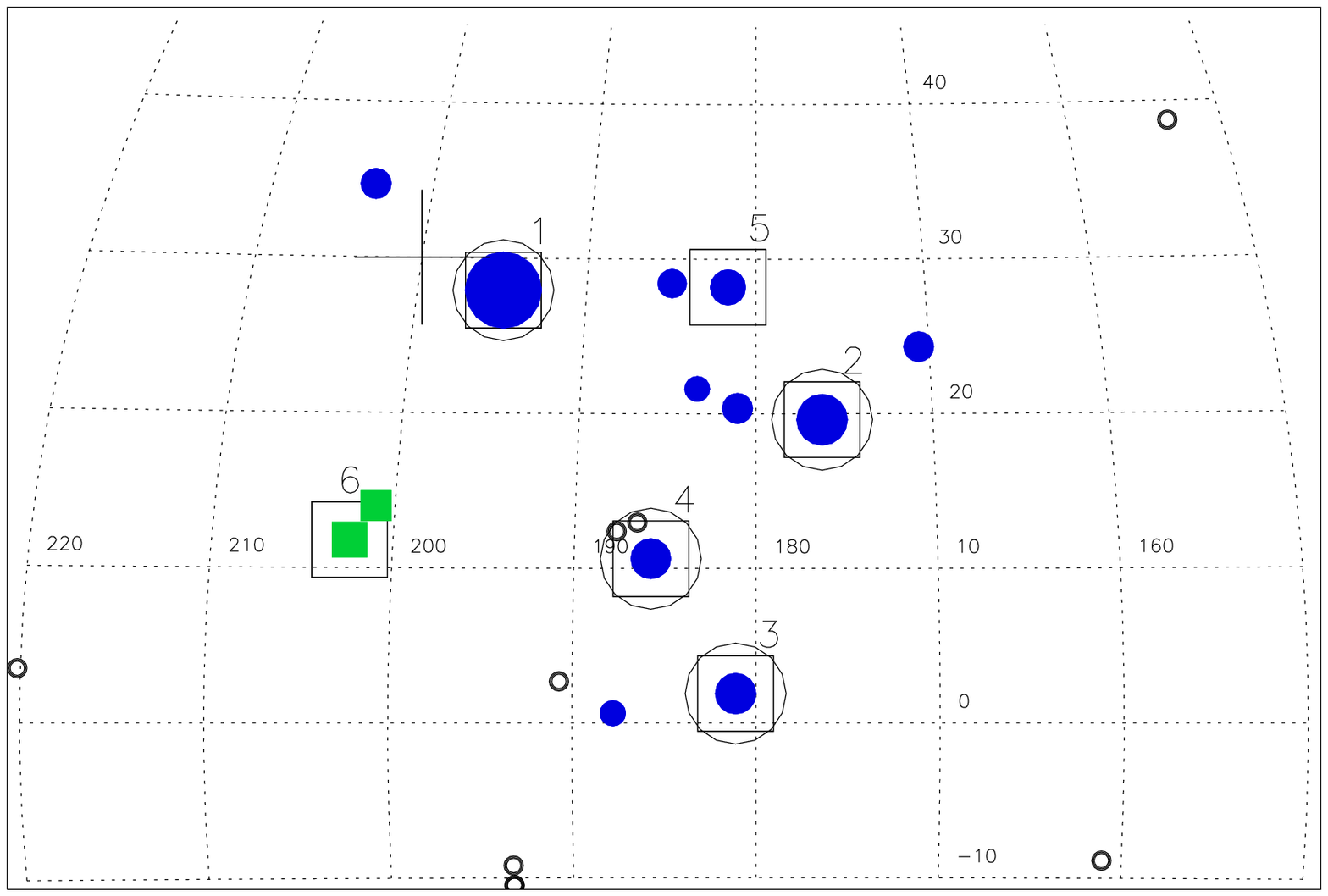}
\caption{Cluster members of the Coma SC for 
different linking schemes. The solid blue circles describe the structure 
found with scheme 1, while the large open squares and open circles mark
the structure found with schemes 7 and 8, respectively. The green squares
show the clusters linked with scheme 3.
Some prominent clusters are marked: 1 = Coma cluster,
2 = A 1367, 3 = MKW4, 4 = NGC 4325, 5 = NGC 410, 6 = NGC 5171.
}\label{fig14}
\end{figure}

\subsection{Luminosity distribution of Supercluster members}

In our previous study on the Perseus-Pisces and A400
SC, we found a hint that the clusters in SC are on average more X-ray luminous
than in the field. This was found at high significance in the study of all
the superstes-clusters in the REFLEX sample and in simulations 
by \citet{Cho2013} and \citet{Cho2014}. Fig.~\ref{figXX} shows the luminosity distribution
for the clusters at $z \le 0.03$ inside and outside the superclusters. 
There are more very luminous clusters in SC, but at the 
low luminosity end we see also more small systems in the SC.
The mean X-ray luminosity of all clusters is $L_X = 2.0 \times 10^{43}$ 
erg s$^{-1}$ for the field and $L_X = 2.5 \times 10^{43}$ erg s$^{-1}$ for
the SC members. The difference is caused by the two most luminous clusters.
We get a somewhat clearer picture of the difference of the luminosity distribution,
if we concentrate on clusters at higher luminosity, with, for example, a lower
luminosity limit of $L_X = 0.5 \times 10^{43}$ erg s$^{-1}$. 
This is shown in Fig.~\ref{figYY} for the comparison of members in all SC
to those in the field (in the left panel) and for members of the four largest
SC compared to the field (in the right panel). The hint to a difference \
becomes clearer for the four largest SC, while the four smaller SC rather dilute
this difference. It is also interesting that the SC contain also a large number
of small systems, which weaken the overall trend.
Kolmogorov-Smirnov statistical tests show, however, that these
results are not highly significant and better statistics is needed to firmly 
establish these findings.

 \begin{figure}[h]
   \includegraphics[width=\columnwidth]{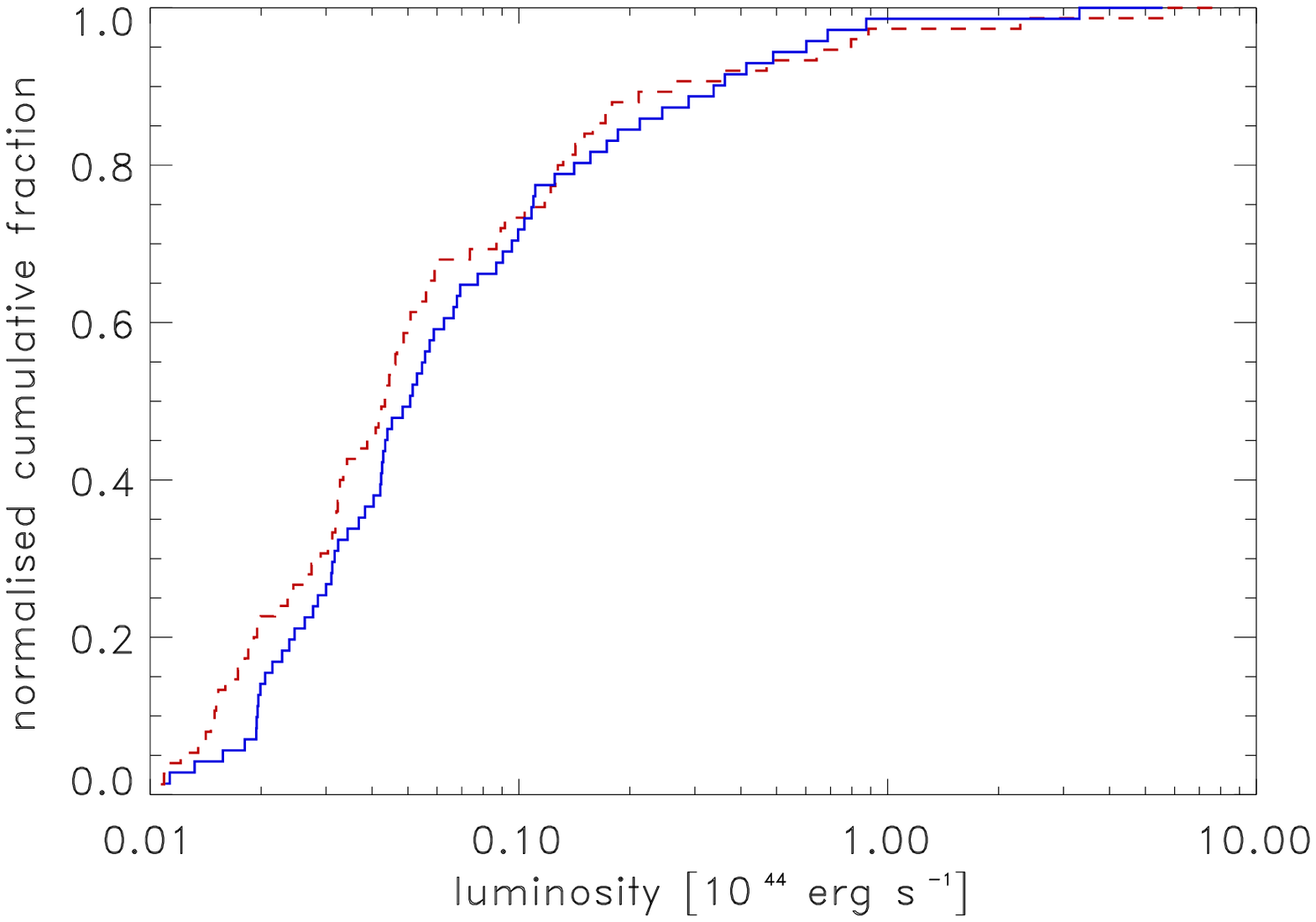}
\caption{Cumulative normalised X-ray luminosity distribution of clusters in
the eight superclusters (red dashed line) compared to the distribution 
for clusters in the field (blue solid line).
}\label{figXX}
\end{figure}

 \begin{figure}[h]
\hbox{
   \includegraphics[width=4.5cm]{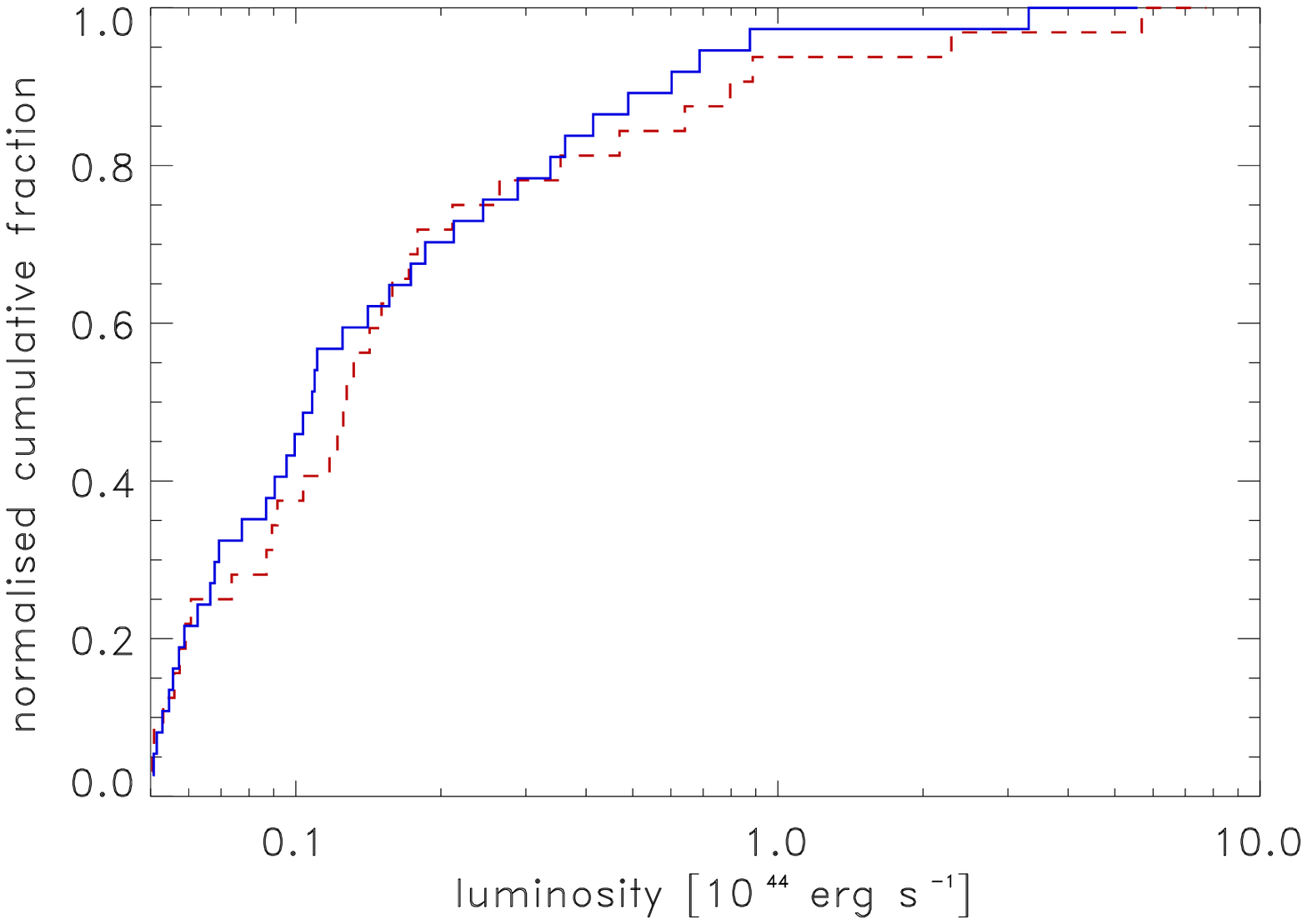}
   \includegraphics[width=4.5cm]{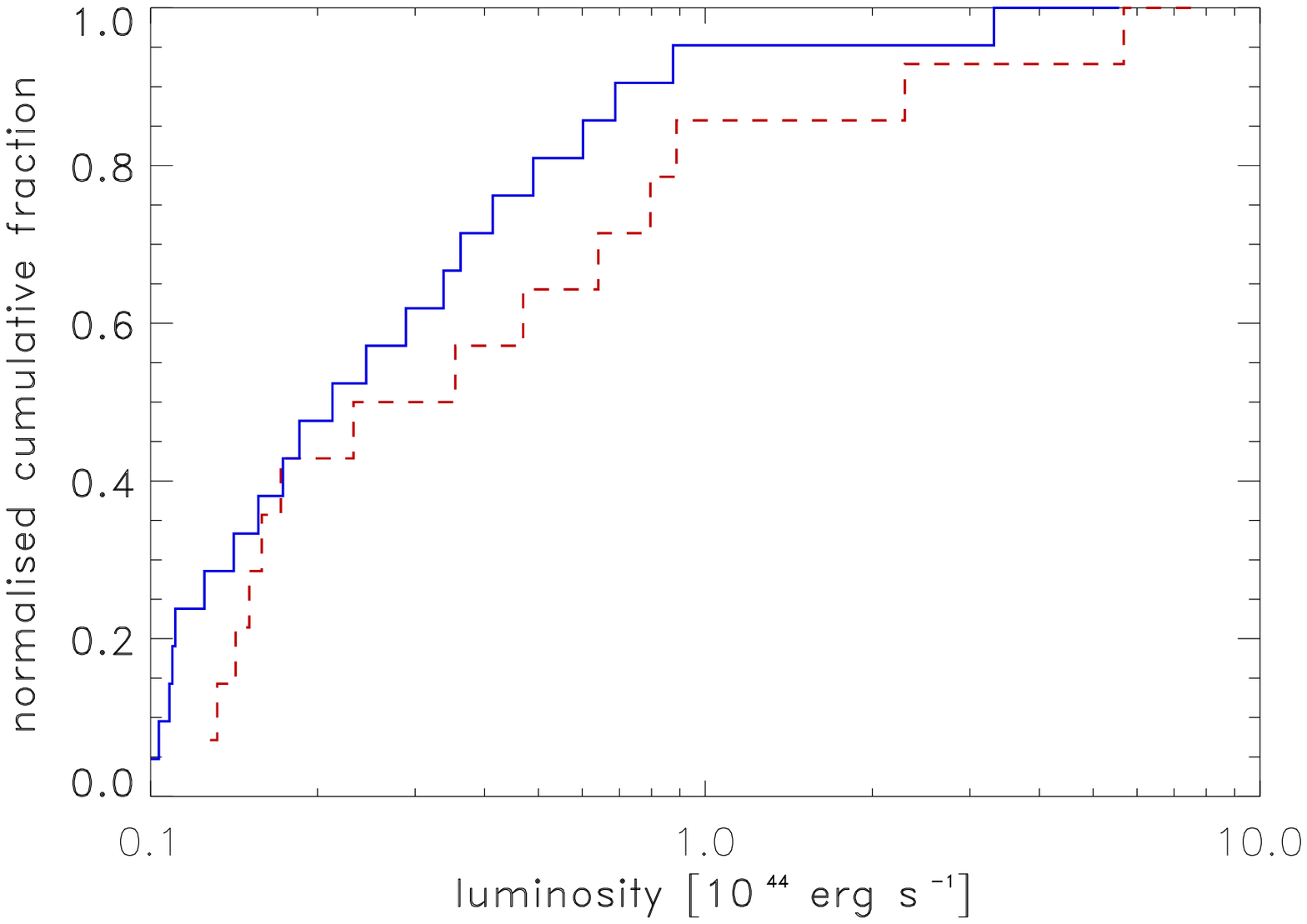}
}
\caption{Cumulative normalised X-ray luminosity distribution of clusters in
the eight superclusters (red dashed line) compared to the distribution 
for clusters in the field (blue solid line) with a lower X-ray luminosity 
limit of $0.5 \times 10^{43}$ erg s$^{-1}$. 
The data in the left panel include all eight superclusters, those in the right
panel only the four large superclusters.
}\label{figYY}
\end{figure}

\section{Summary and Conclusion}

We searched for large-scale structures in the matter distribution in the nearby
Universe at $z \le 0.03$ with overdensity ratios of about two
by means of the distribution of X-ray luminous
galaxy groups and clusters with an estimated mass larger than 
about $m_{200} = 2.1 \times 10^{13}$ M$_{\odot}$.
In total we found eight superclusters with at least five group or cluster members.
Four of these are smaller with estimated masses in the range of 
$2 - 7 \times 10^{15}$ M$_{\odot}$. They are the Local, the A400,
the Sagittarius and the Lacerta SC. The latter two are structures which have
not been described as such. These smaller structures are only found by including
less massive groups of galaxies with low X-ray luminosities of a 
few $10^{42}$ erg s$^{-1}$.

The other four SC are well known, prominent mass concentrations, the Perseus-Pisces SC,
the Centaurus SC (sometimes identified as Great Attractor), the Coma SC with the 
massive Coma cluster, and the low redshift part of the Hercules SC. These have
estimated masses in the range $0.5 - 2.2 \times 10^{16}$ M$_{\odot}$. 
The largest structure is clearly the Perseus-Pisces SC.
We showed that variations of the structure construction schemes robustly recover
approximately the same prominent structures and no other structures appeared for a
wide range of construction parameters. These major structures are consistent
with the results from the analysis of cosmic flows from galaxy peculiar velocities
(e.g. \citet{Tul2019}) and also with other large-scale structure studies.

We provided catalogues and maps of all the member groups and clusters 
in the Appendix and verified that all
of them show significantly extended X-ray emission in the ROSAT All-Sky Survey.
In total 51\% of all the X-ray luminous groups and clusters at $z \le 0.03$ are
members of these SC.
This characterisation of the large-scale environment of superclusters in
the local Universe should be interesting for studies of the environmental 
dependence of the properties of different astronomical objects. It is
also important for a better understanding of our local reference system
from which we perform precision cosmology observations.  

\begin{acknowledgements} We thank the referee, Jaan Einasto, for 
very helpful comments and suggestions.
We acknowledge support of the Deutsche 
Forschungsgemeinschaft through the Munich Excellence Cluster `Universe'. 
G.C. acknowledges support by the DLR under grant no.
50 OR 1905.
\end{acknowledgements}

\appendix
\section{Online Material}

\subsection{Survey luminosity limit as function of redshift}

For our analysis we used a nominal X-ray luminosity limit of 
$10^{42}$ erg s$^{-1}$. This luminosity limit is reached in the RASS 
only at redshifts below about $z \sim 0.0146$. Fig.~\ref{figA2}
shows how the average lower X-ray luminosity limit in the RASS 
varies with redshift. This is taken into account when adjusting
the linking length as described in section 3. While here
we show the average value in the survey, the linking length
adjustment also takes the local variations into account, which are due
to varying exposure time and interstellar absorption.

\begin{figure}[h]
   \includegraphics[width=\columnwidth]{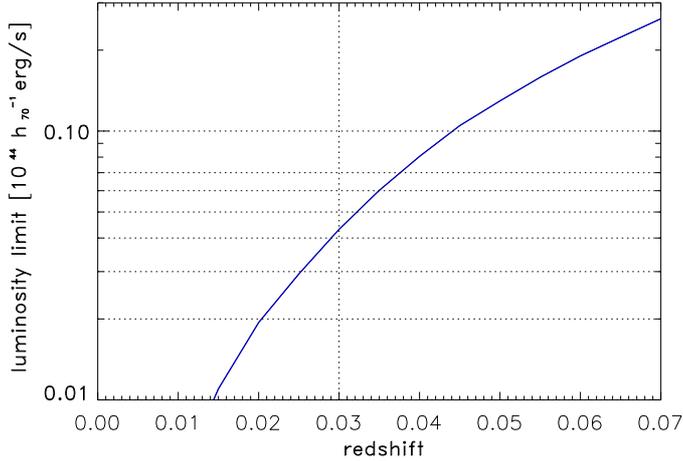}
\caption{Mean survey luminosity limit of the {\sf CLASSIX} Survey
at $|b_{II}| > 20^o$ as a function of redshift. The dashed lines show 
the different X-ray luminosity limits used in the study in the
discussion section and the redshift limit.
}\label{figA2}
\end{figure}

\clearpage

\subsection{Images of the Local Supercluster members}

Fig.~\ref{figA2} provides images of the member groups and clusters of 
the Local Supercluster. The optical images are obtained from the 
Digital Sky Survey (DSS) scans of photographic plates \footnote{available at
http://archive.eso.org/dss/dss}
and the contours show the X-ray surface brightness observed in the RASS. 

All groups and clusters of this structure have significantly extended X-ray
emission in the RASS and no peculiar X-ray spectral properties. The two
groups, NGC 5813 and NGC 5846, have 
been observed in deep Chandra observations, NGC 5813 by \citet{Ran2015}
and NGC 5846 by \citet{Mac2011}, and interesting cavities and interaction effects
of the central AGN with the intragroup medium were found. 

\begin{figure*}[h]
\hbox{
\hspace{1cm}
   \includegraphics[width=7.5cm]{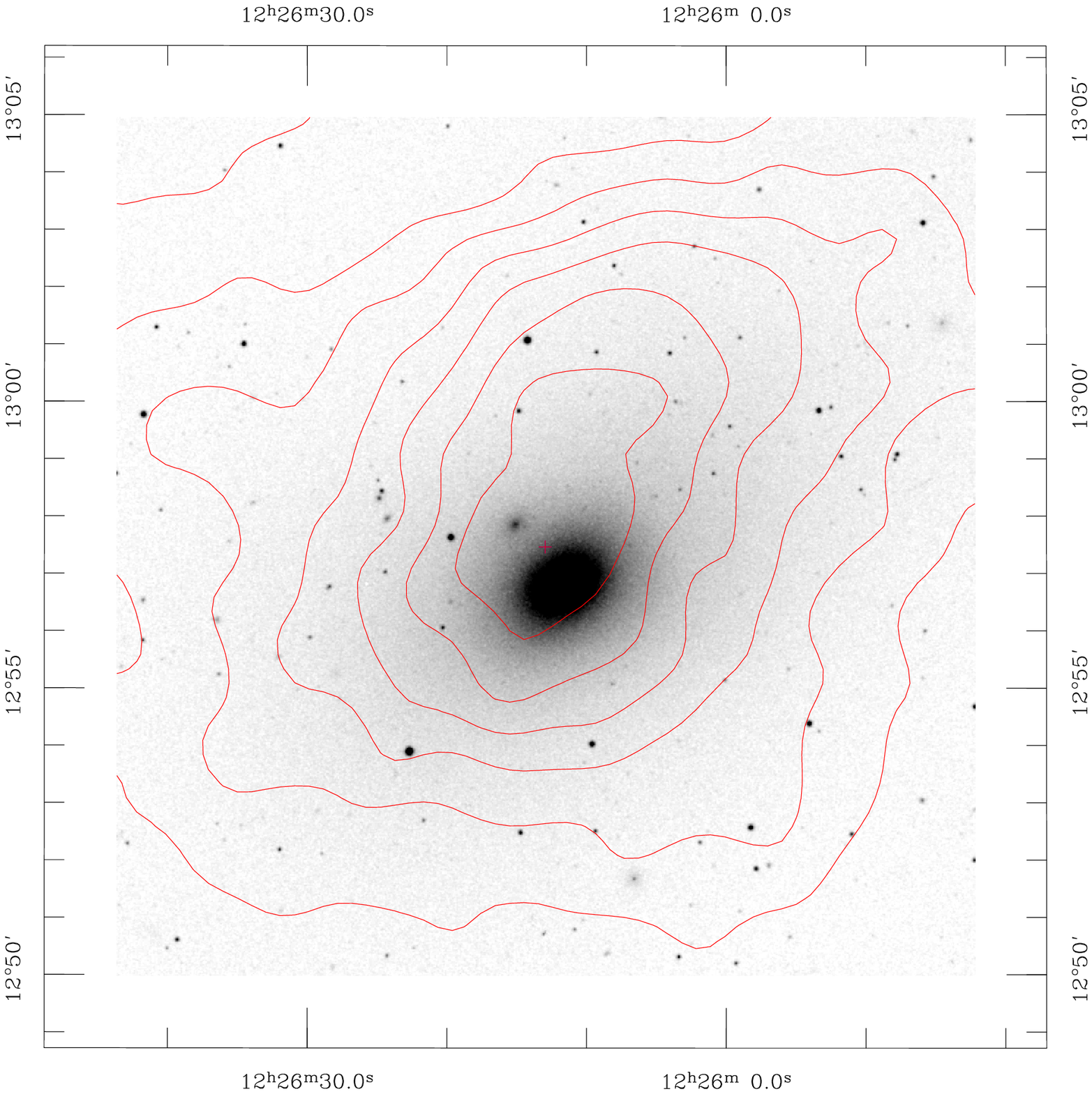}
\hspace{1cm}
   \includegraphics[width=7.5cm]{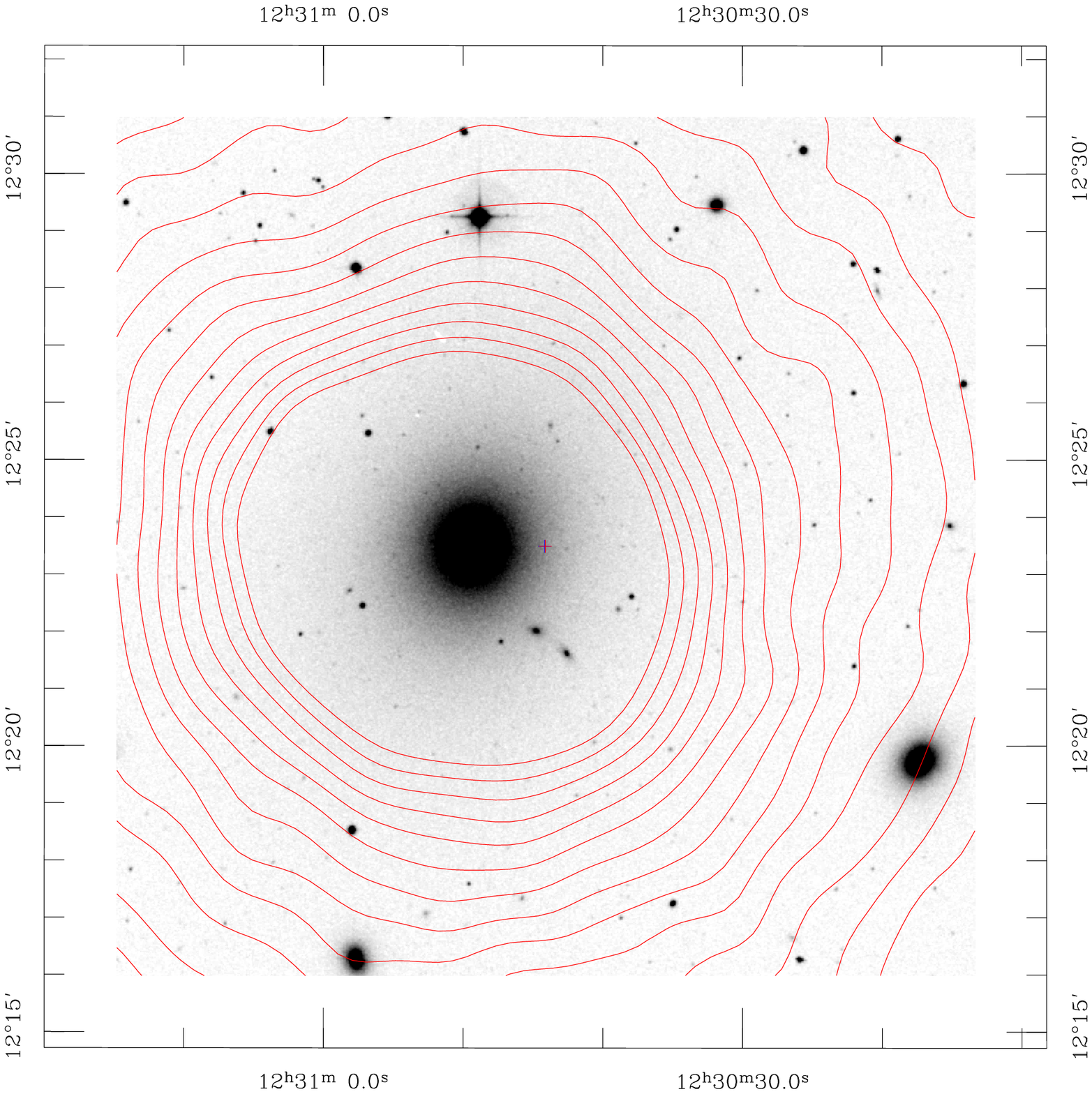}
}
\hbox{
\hspace{1cm}
   \includegraphics[width=7.5cm]{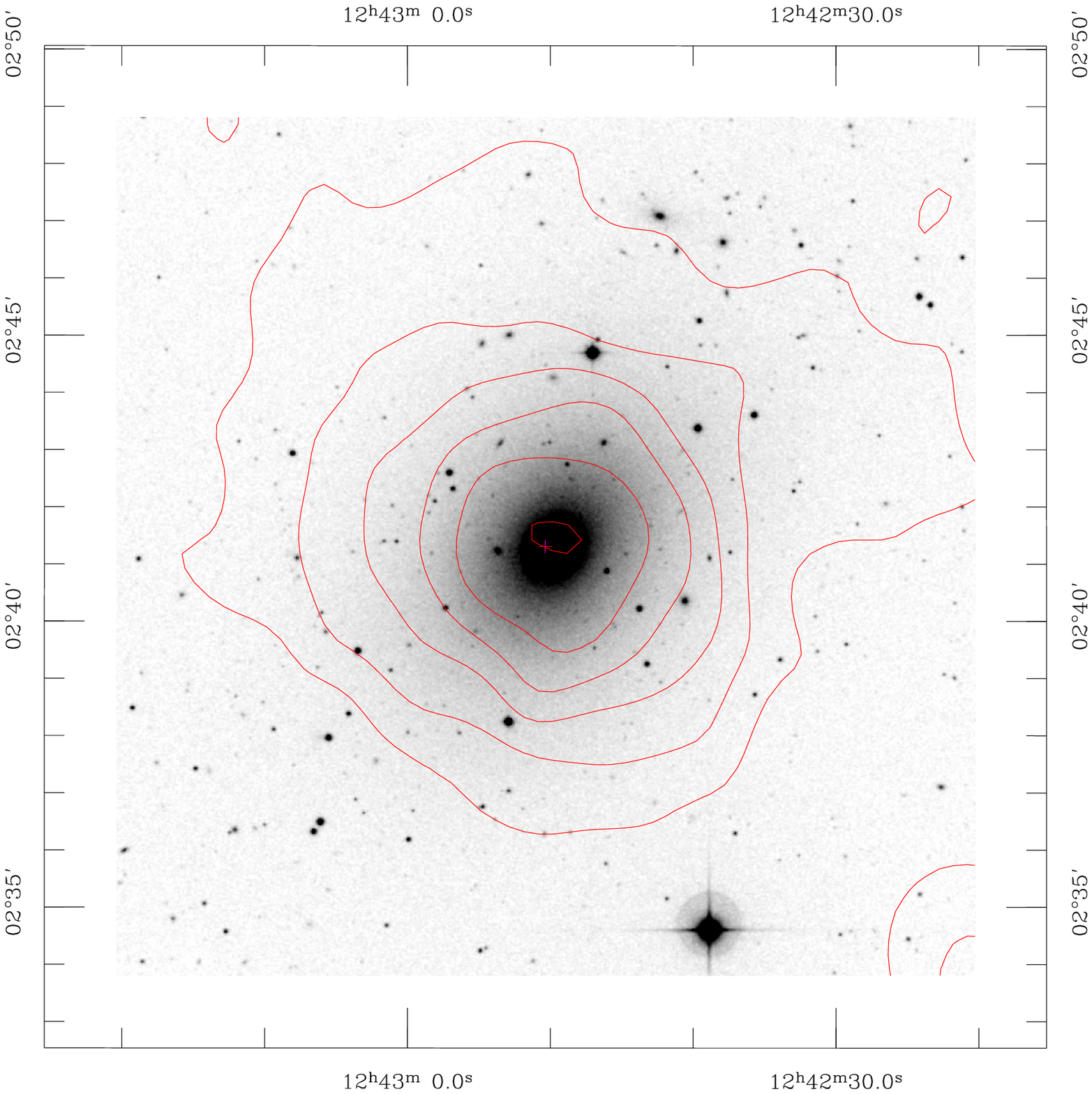}
\hspace{1cm}
   \includegraphics[width=7.5cm]{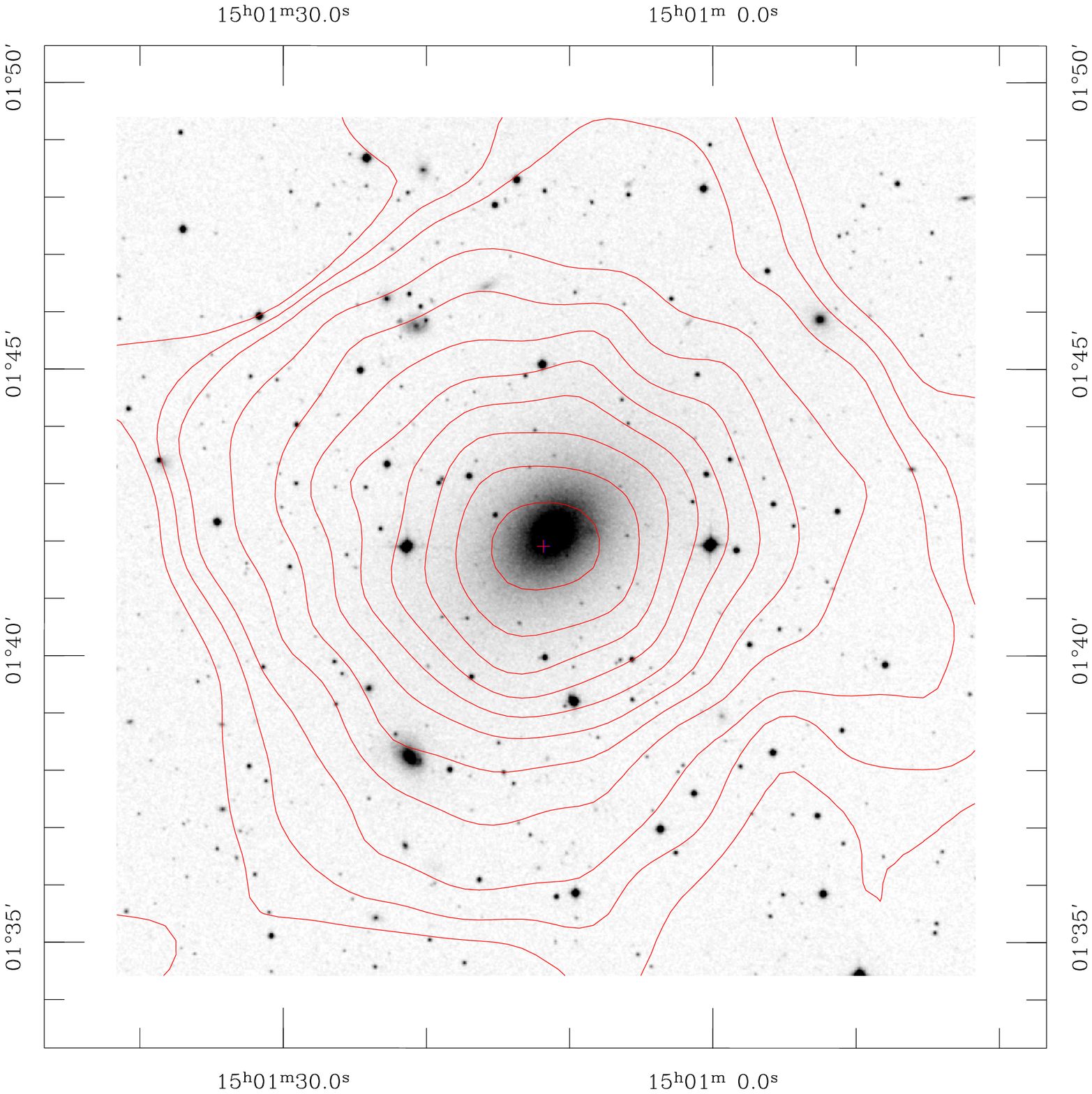}
}
\hbox{
\hspace{1cm}
   \includegraphics[width=7.5cm]{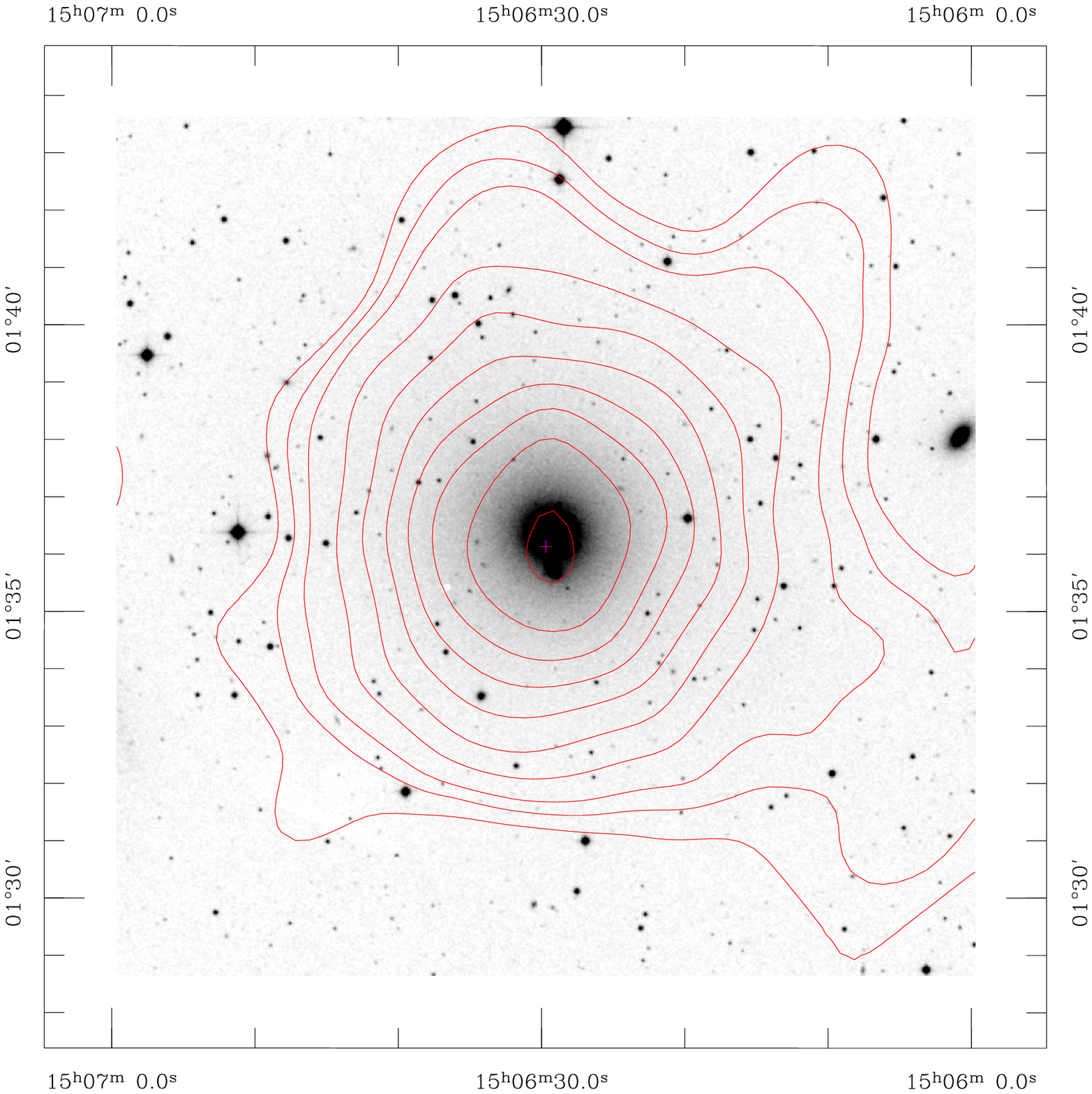}
}
\caption{Images of the Local Supercluster. Shown are contours of the X-ray surface brightness 
superposed on optical Digital Sky Survey images. All images in this and the following sections
are oriented such that north is up and east to the left.
{\bf Upper left:} RXCJ1226.2+1257, M86 
{\bf Upper right:}  RXCJ1230.7+1223, M87, 
{\bf Middle left:} RXCJ1242.8+0241, NGC 4636
{\bf Middle right:} RXCJ1501.1+0141,NGC 5813
{\bf Lower left:} RXCJ1506.4+0136, NGC 5846 
}\label{figA3}
\end{figure*}

\subsection{Images of the Centaurus Supercluster members}

Figs.~\ref{figA4} and \ref{figA5} provide images 
of the Centaurus SC with X-ray surface brightness 
contours from the RASS overlayed on optical Digital Sky Survey images. For two
clusters, RXCJ1349.3-3018 (A3574E) and RXCJ1403.5-3359 (NGC5328) we also
show optical images with X-ray surface brightness contours from XMM-Newton
observations. For the XMM-Newton data the exposures of the three detectors 
were combined with a scaling of the pn-images by a factor of 3.3 with respect
to the MOS images.

All members of the Centaurus SC shown here have significantly extended
X-ray emission in the RASS and no peculiar X-ray spectral properties.
An exception is the cluster RXCJ1349.3-3018 (A3574E) which harbours
an X-ray bright Sy 1.2 galaxy, IC4329A, which outshines
the cluster. In the RASS image, which is shown in Fig.~\ref{figA5}
upper right, we see mostly a point source and the additional cluster
emission is difficult to distinguish. Using, however, a pointed ROSAT
observation and even better an observation with XMM-Newton
(shown in Fig.~\ref{figA5} middle left) we could 
separate the soft point source emission from the AGN to
get approximate values for the cluster emission. This deblended 
X-ray luminosity is listed in Table~\ref{tab4}.

The two components of A3574, RXCJ1347.2-3025 and  RXCJ1349.3-3018,
appear as two distinct X-ray emission regions in the RASS.
One of the Centaurus SC members, RXCJ1321.2-4342, NGC 5090/5091
is located in the ZoA, at $b_{II} \sim 18.8^o$. 

\begin{figure*}[h]
\hbox{
\hspace{1cm}
   \includegraphics[width=7.5cm]{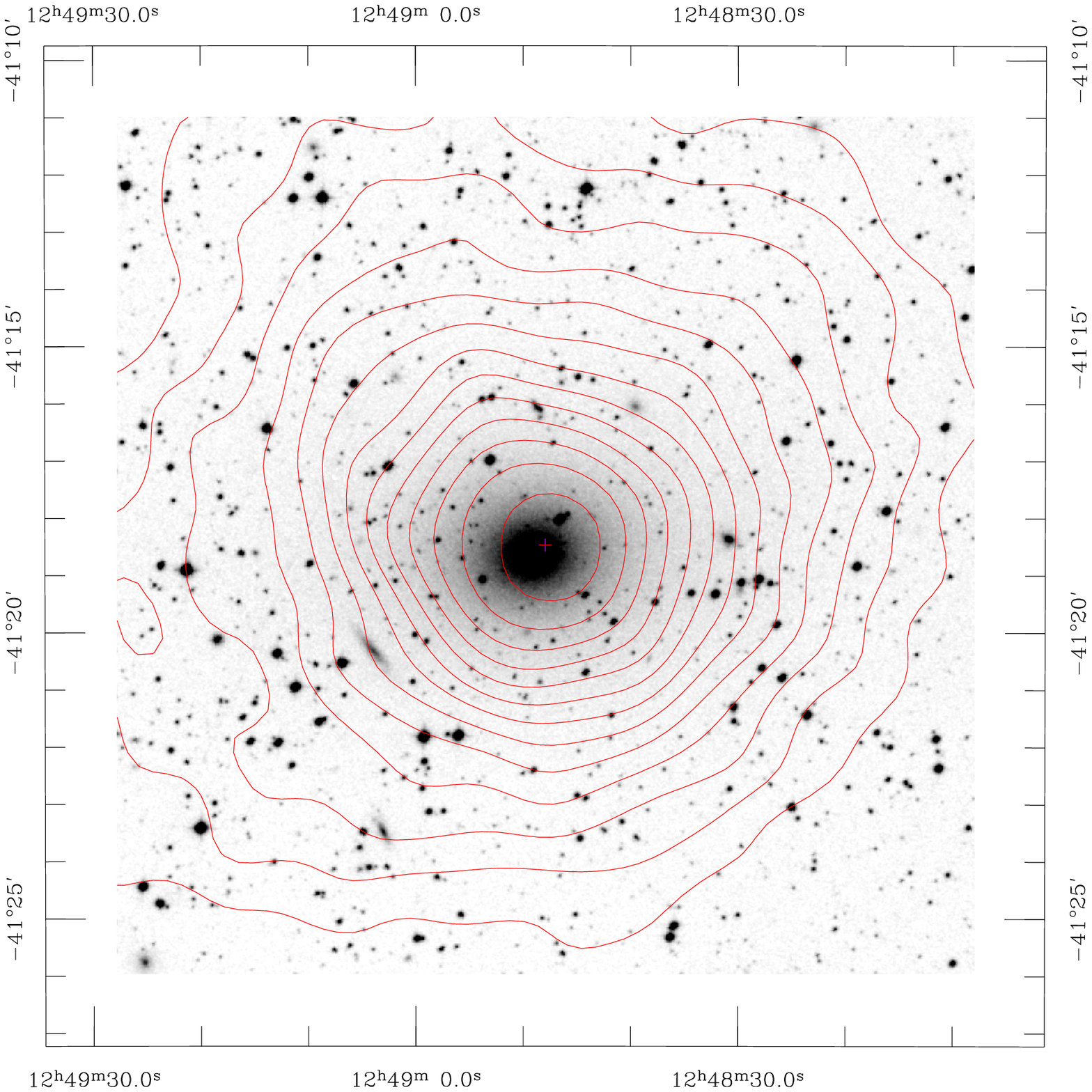}
\hspace{1cm}
   \includegraphics[width=7.5cm]{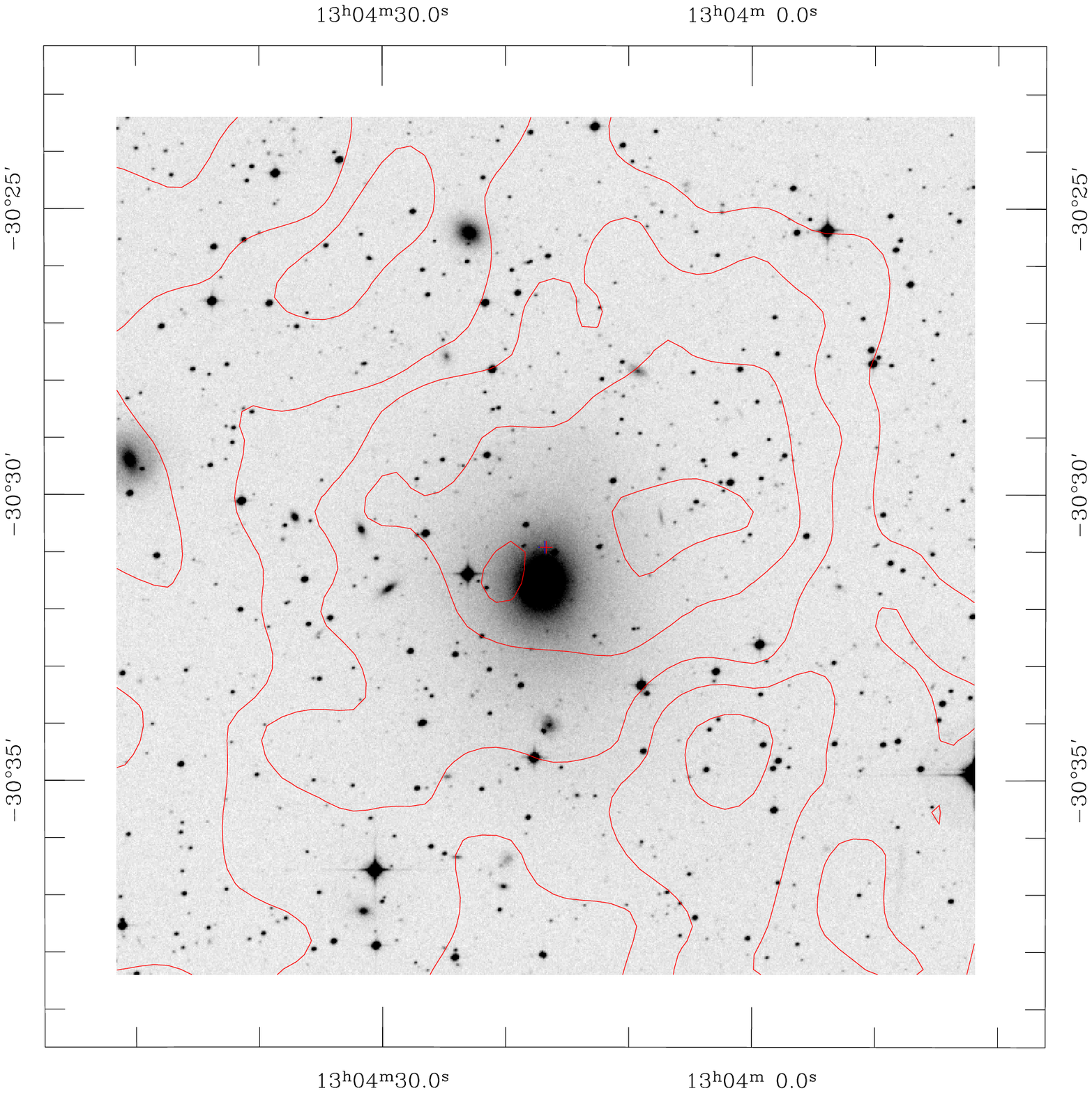}
}
\hbox{
\hspace{1cm}
   \includegraphics[width=7.5cm]{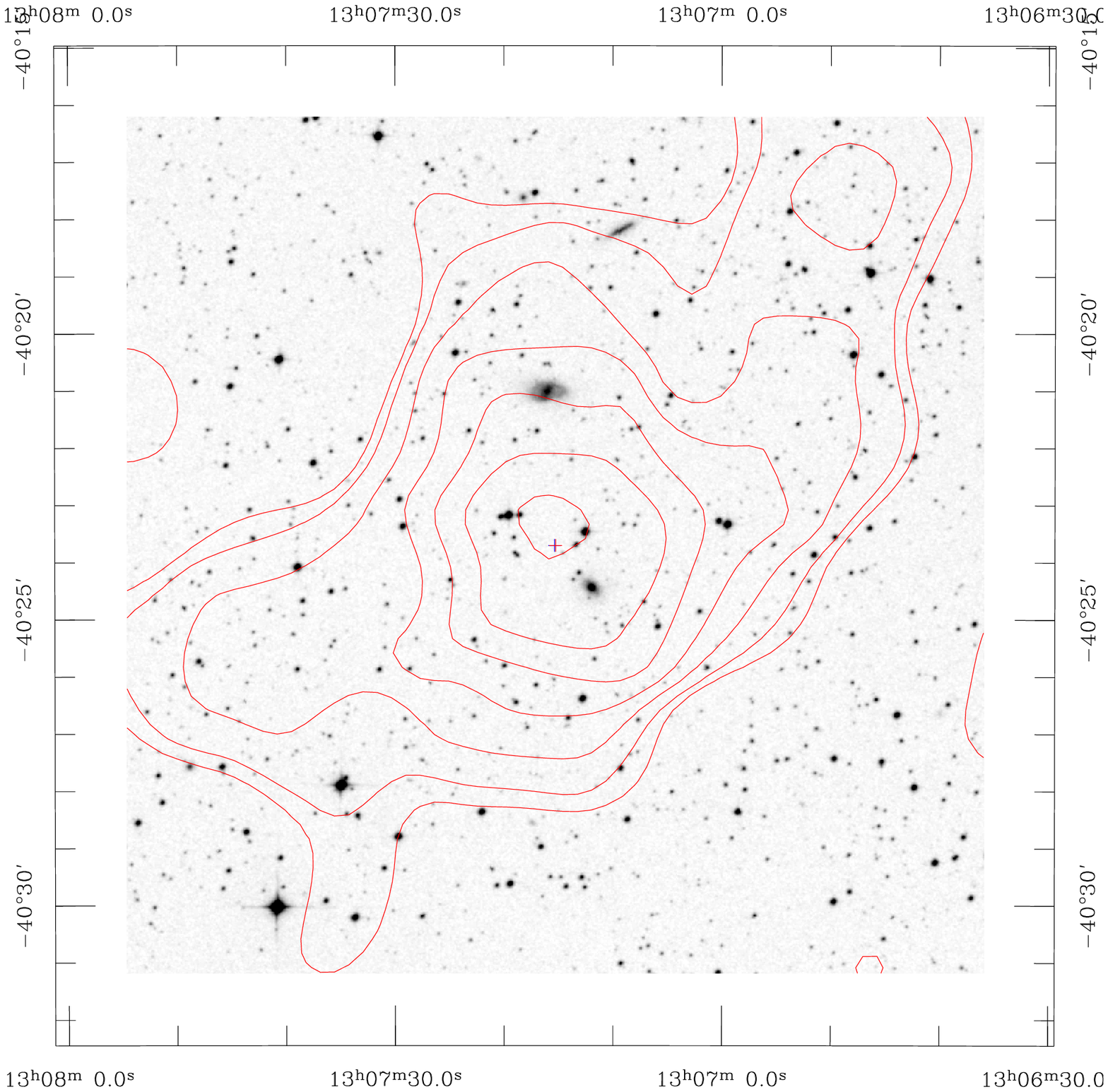}
\hspace{1cm}
   \includegraphics[width=7.5cm]{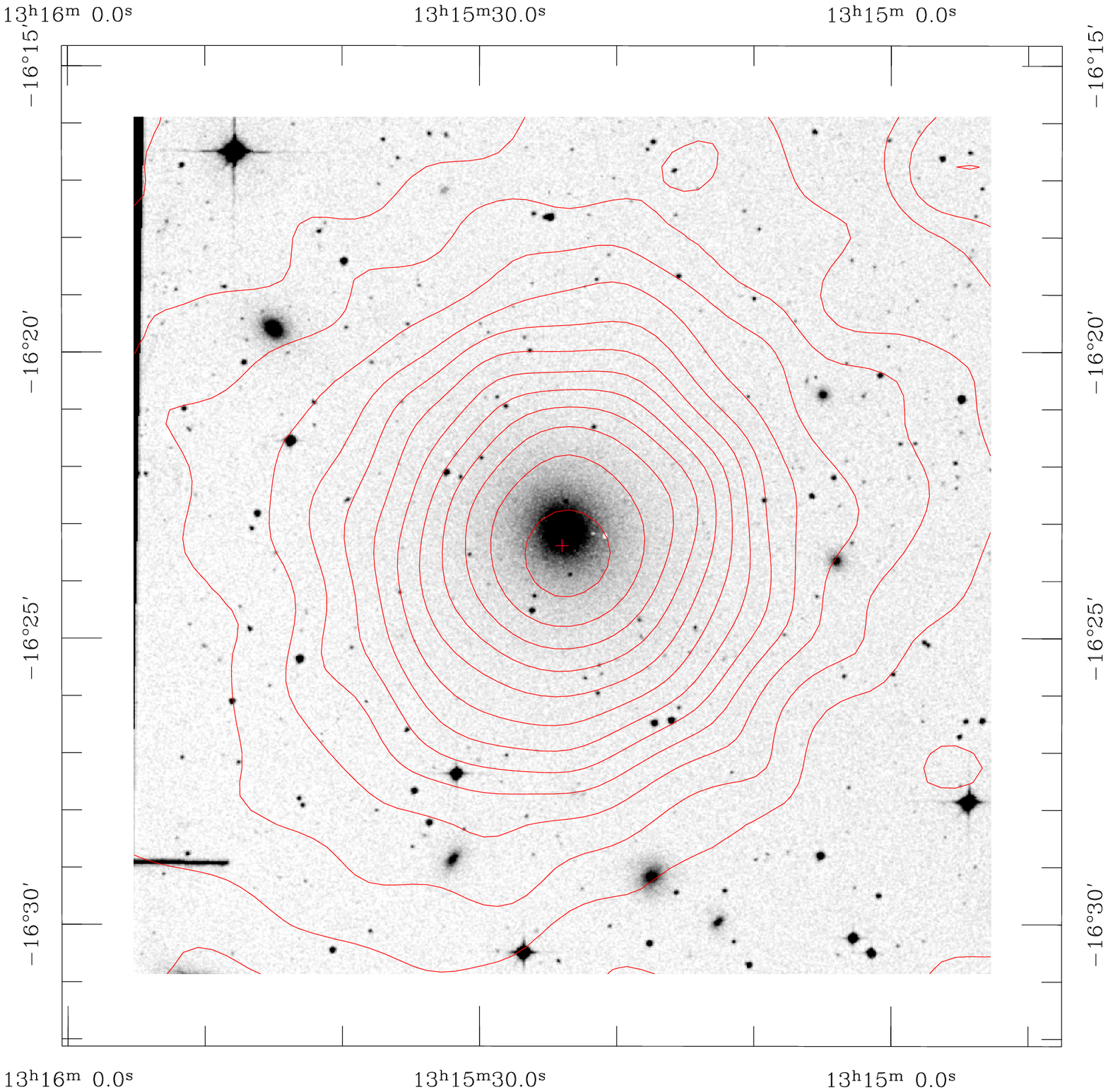}
}
\hbox{
\hspace{1cm}
   \includegraphics[width=7.5cm]{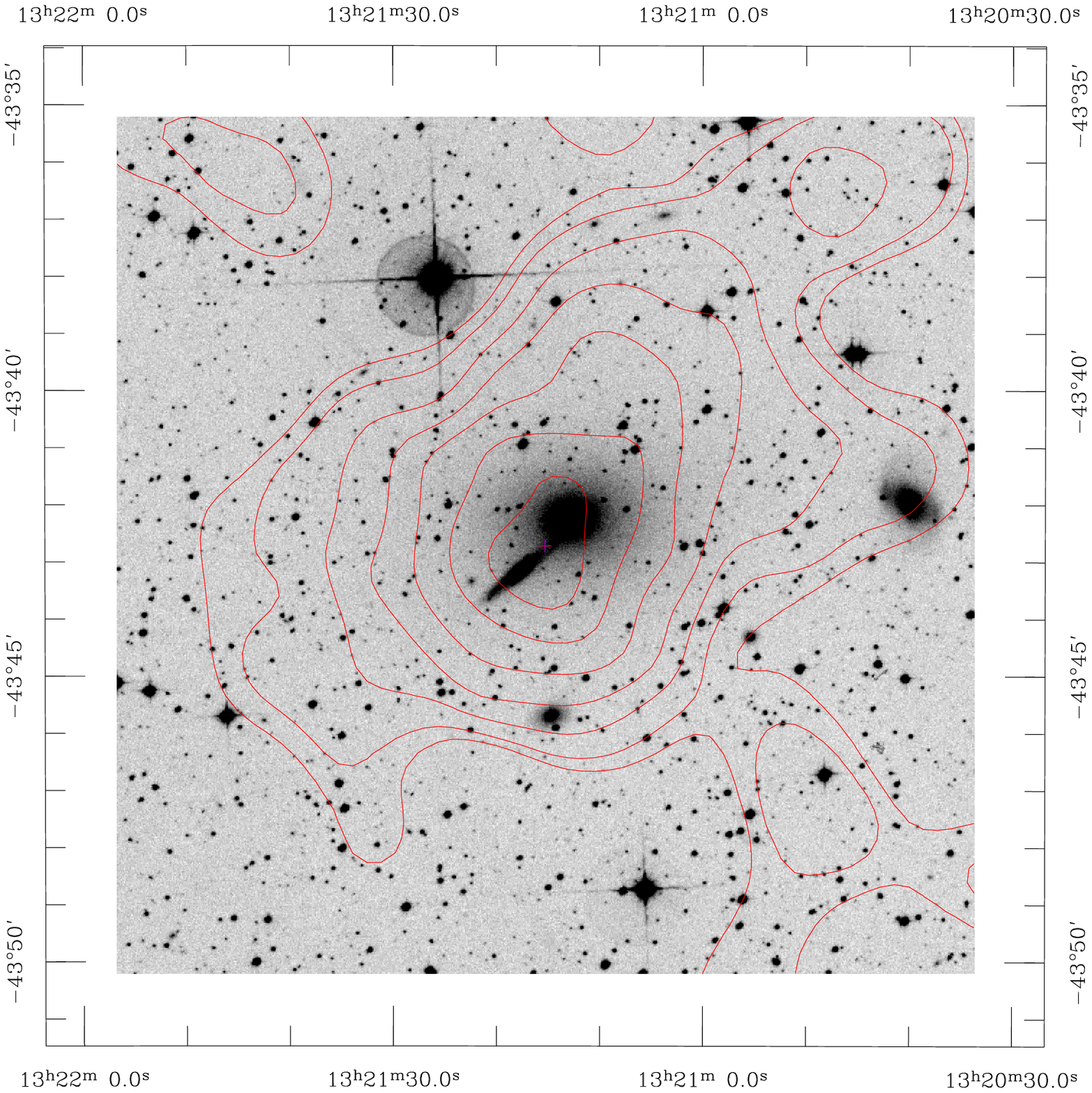}
\hspace{1cm}
   \includegraphics[width=7.5cm]{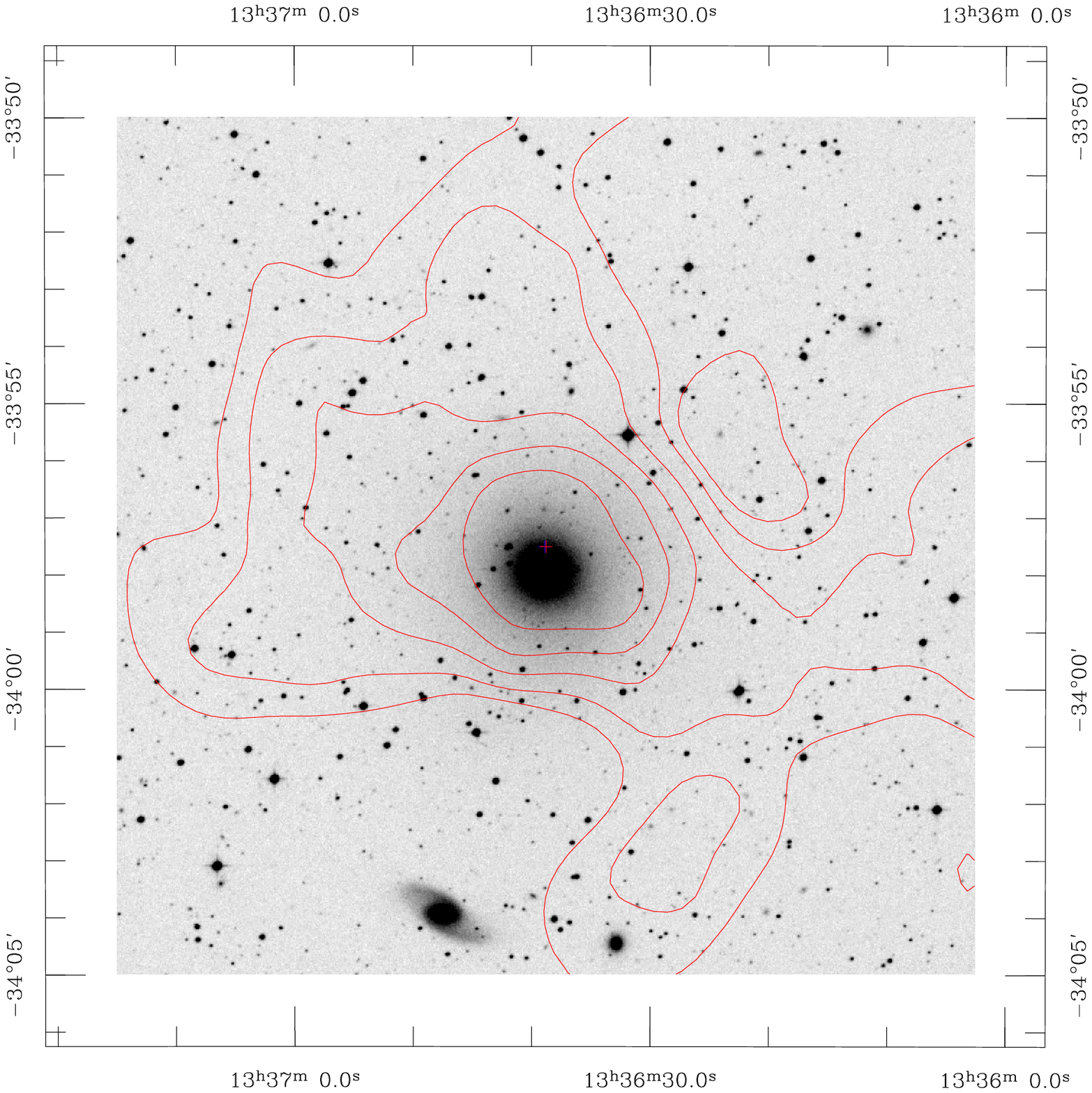}
}
\caption{Members of the Centaurus Supercluster.
Contours of the X-ray surface brightness superposed on optical images 
from the DSS database. The X-ray data are taken from the RASS or XMM-Newton observations.
{\bf Upper left:} RXCJ1248.7-4118, A3526, Centaurus cluster,
{\bf Upper right:} RXCJ1304.2-3030, NGC 4936,
{\bf Middle left:} RXCJ1307.2-4023, ESO-3230.0159,
{\bf Middle right:} RXCJ1315.3-1623, NGC 5044,
{\bf Lower left:} RXCJ1321.2-4342, NGC 5090/5091,
{\bf Lower right:} RXCJ1336.6-3357, A 3565.
}\label{figA4}
\end{figure*}

\begin{figure*}[h]
\hbox{
\hspace{1cm}
   \includegraphics[width=7.5cm]{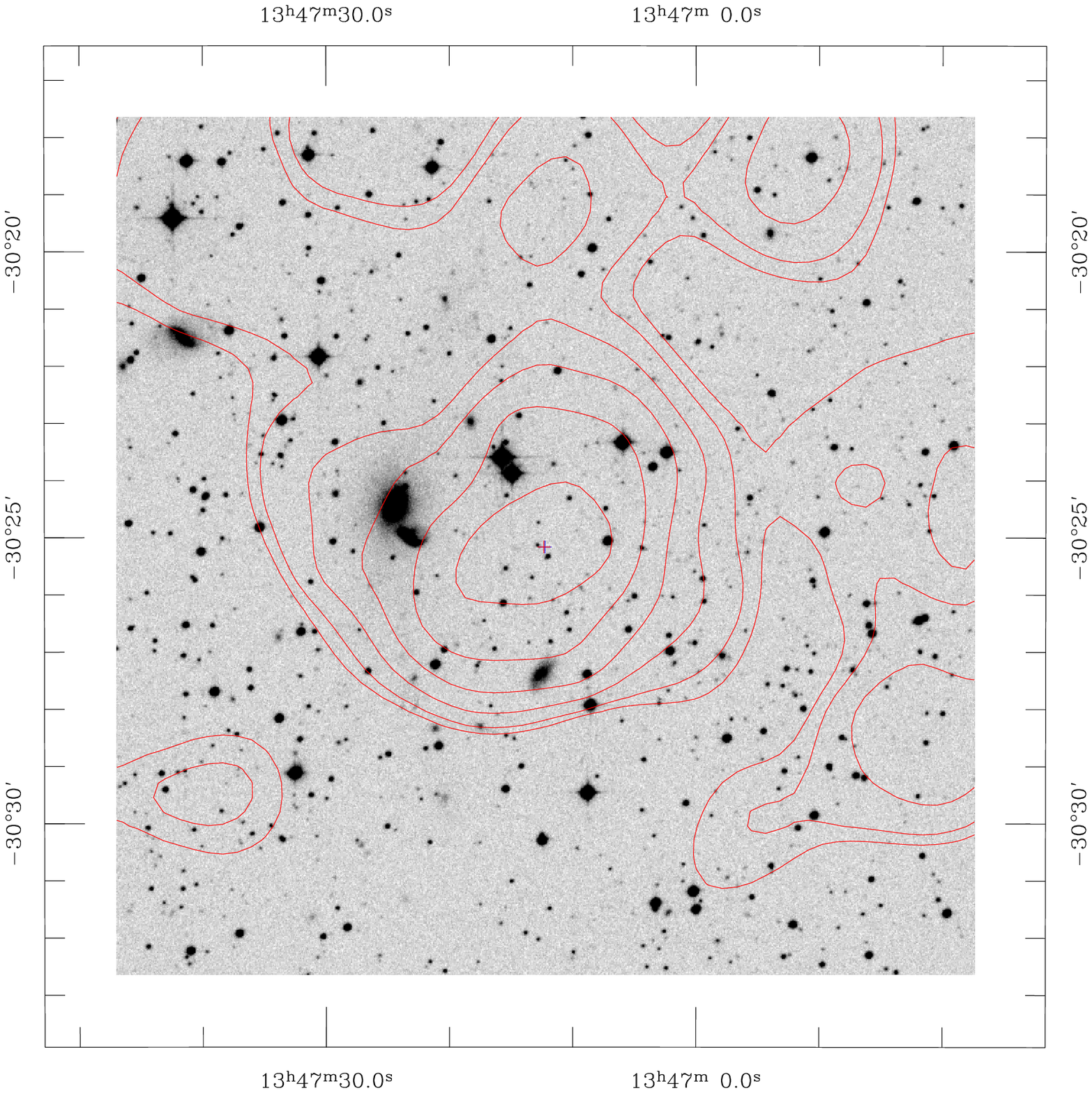}
\hspace{1cm}
   \includegraphics[width=7.5cm]{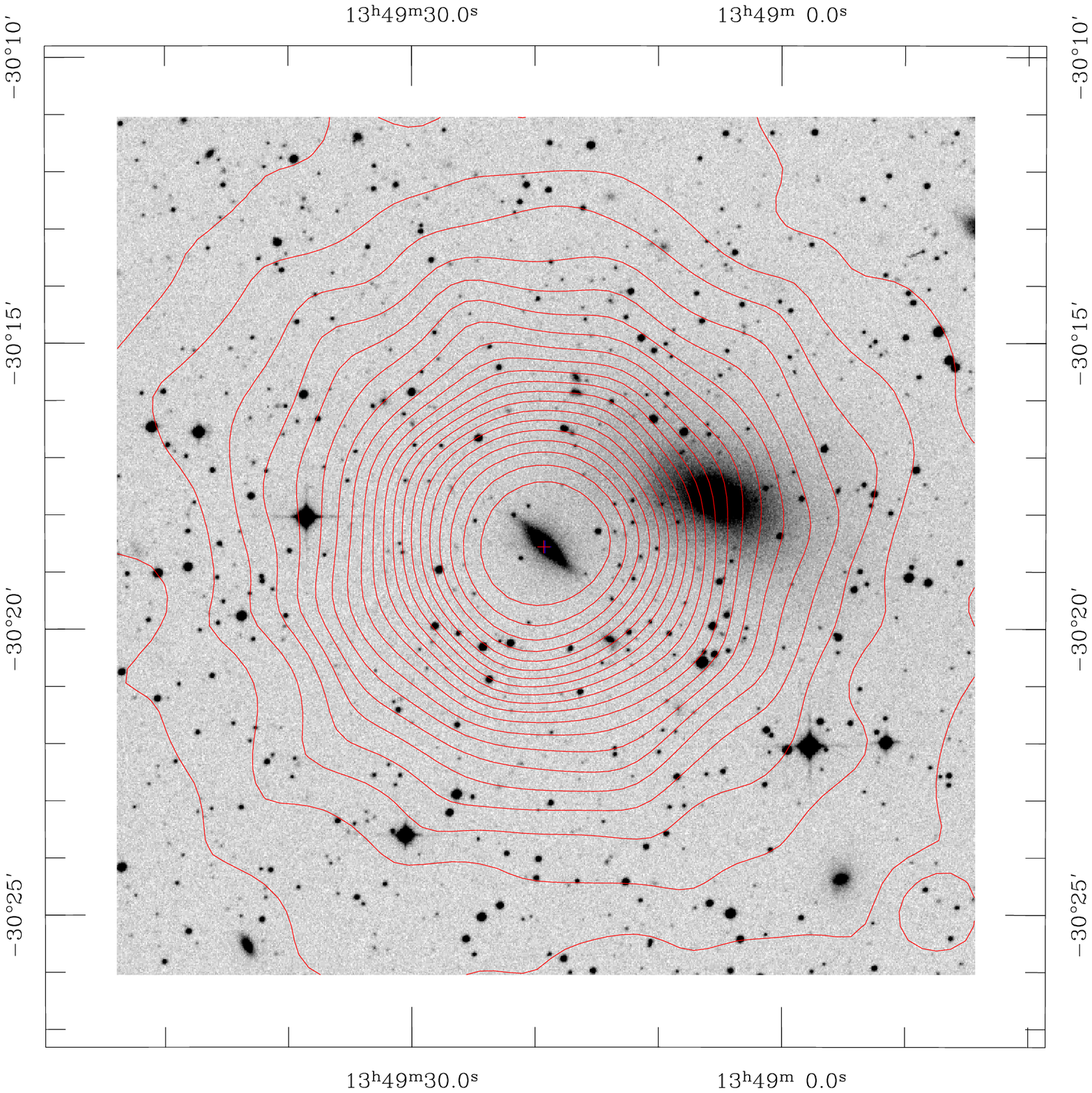}
}
\hbox{
\hspace{1cm}
   \includegraphics[width=7.5cm]{A3574E_OV.ps}
\hspace{1cm}
   \includegraphics[width=7.5cm]{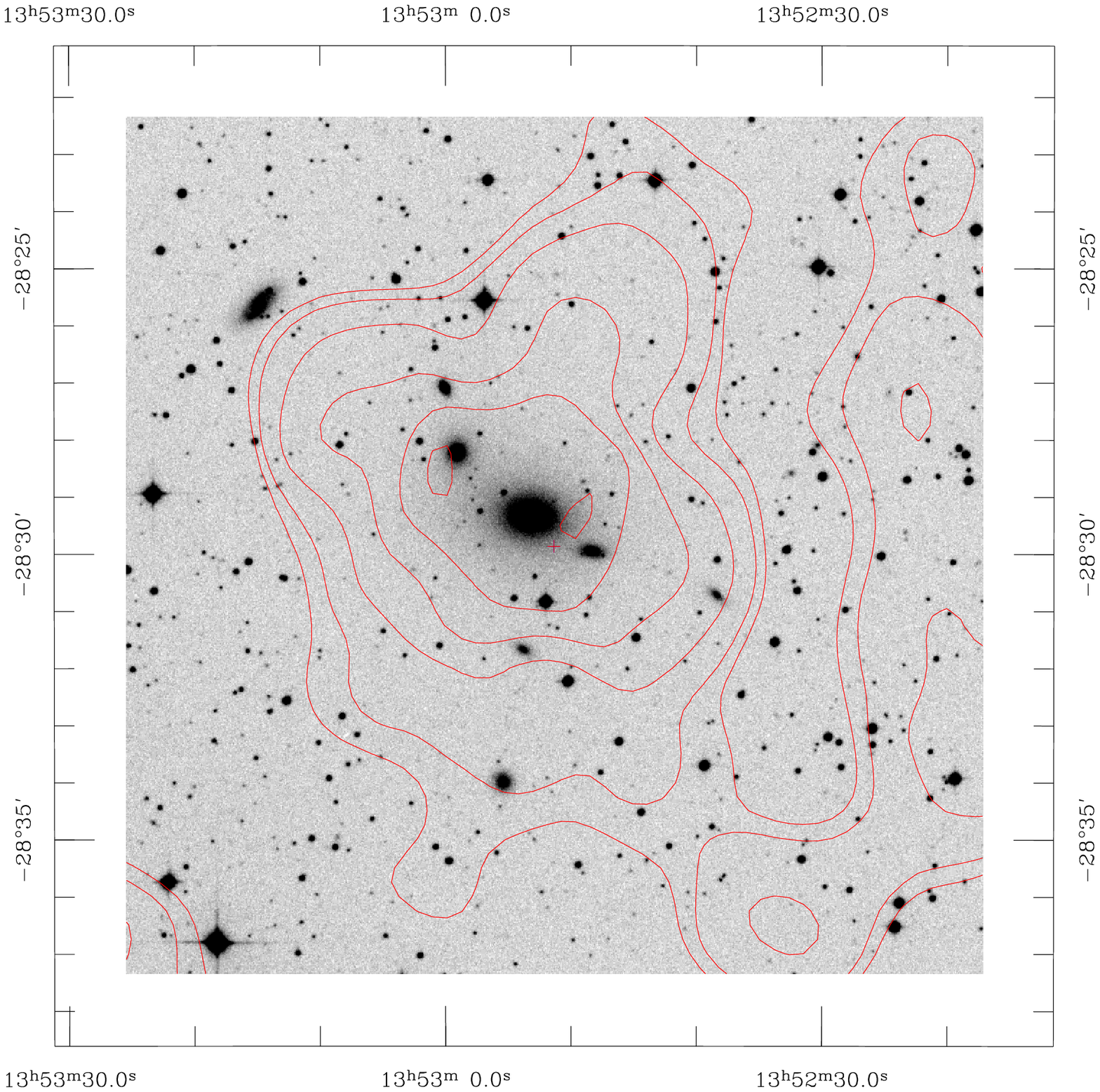}
}
\hbox{
\hspace{1cm}
   \includegraphics[width=7.5cm]{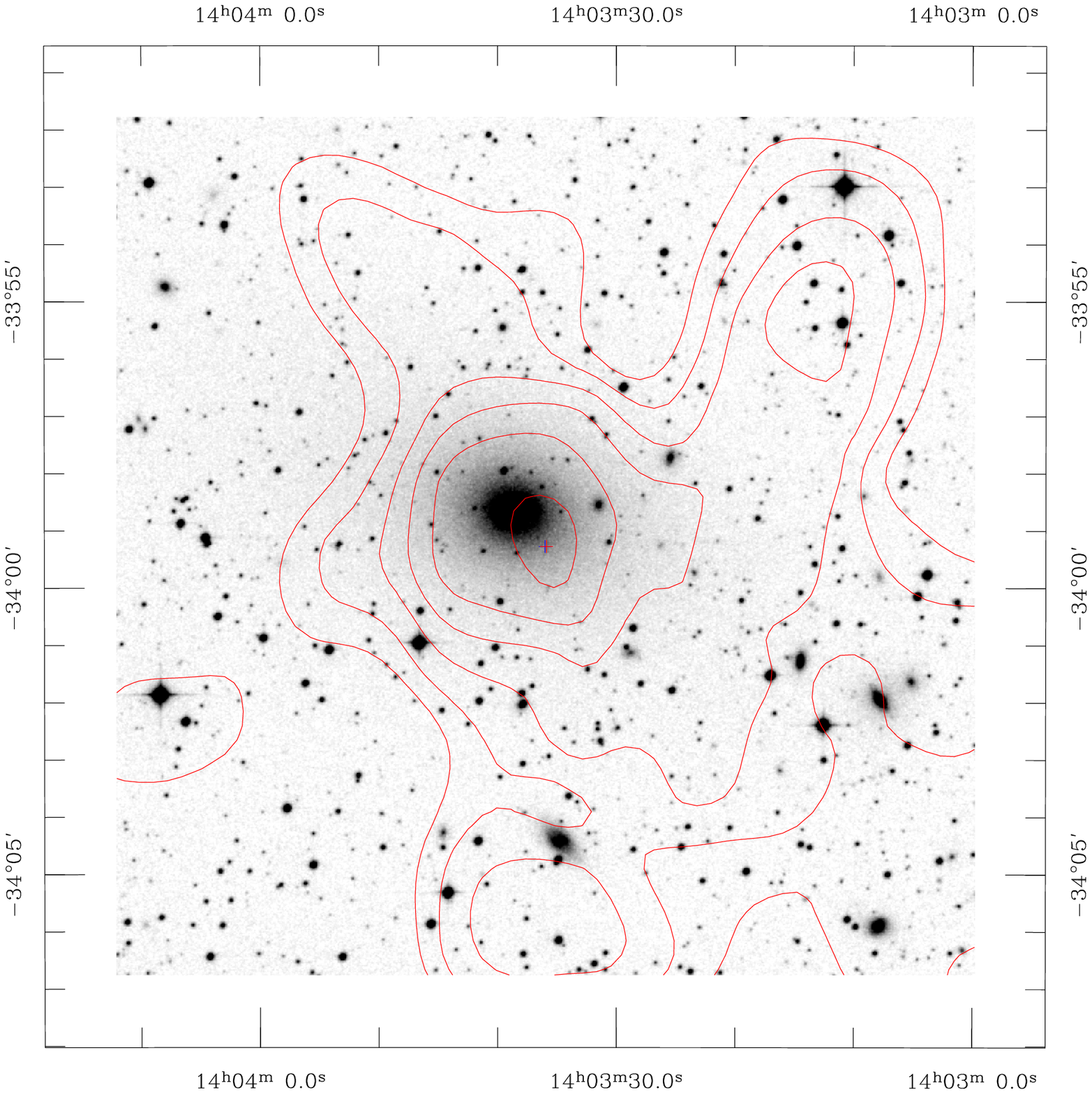}
\hspace{1cm}
   \includegraphics[width=7.5cm]{NGC5328_OV.ps}
}
\caption{Members of the Centaurus Supercluster continued.
{\bf Upper left:} RXCJ1347.2-3025, A 3574 W
{\bf Upper right:} RXCJ1349.3-3018, A 3574 E
{\bf Middle left:} XMM-Newton image of RXCJ1349.3-3018, A3574~E
{\bf Middle right:} RXCJ1352.8-2829, NGC 5328, 
{\bf Lower left:} RXCJ1403.5-3359, AS 753
{\bf Lower right:} XMM-Newton image of RXCJ1403.5-3359, AS 753.
}\label{figA5}
\end{figure*}

\subsection{Images of the Coma Supercluster members}

In this section we provide images of the member groups and 
clusters of the Coma SC (Figs.~\ref{figA6}, \ref{figA7} and \ref{figA8}). 
The images show overlays of X-ray 
contours from RASS on DSS images produced in the same way
as in the previous sections.
For two clusters with interesting internal structures,
A1185 (RXCJ1110.5+2843) and A1367 (RXCJ1145.0+1936), 
we also show images with X-ray contours from XMM-Newton
observations. We do not show an image of Coma because the
cluster is so large and there are plenty of detailed
images available in the literature, for example the new
image obtained with eROSITA by \citet{Chu2021}.
All groups and clusters of the Coma SC shown 
here have significantly extended X-ray emission in 
the RASS and no peculiar X-ray spectral properties.

\begin{figure*}[h]
\hbox{
\hspace{1cm}
   \includegraphics[width=7.5cm]{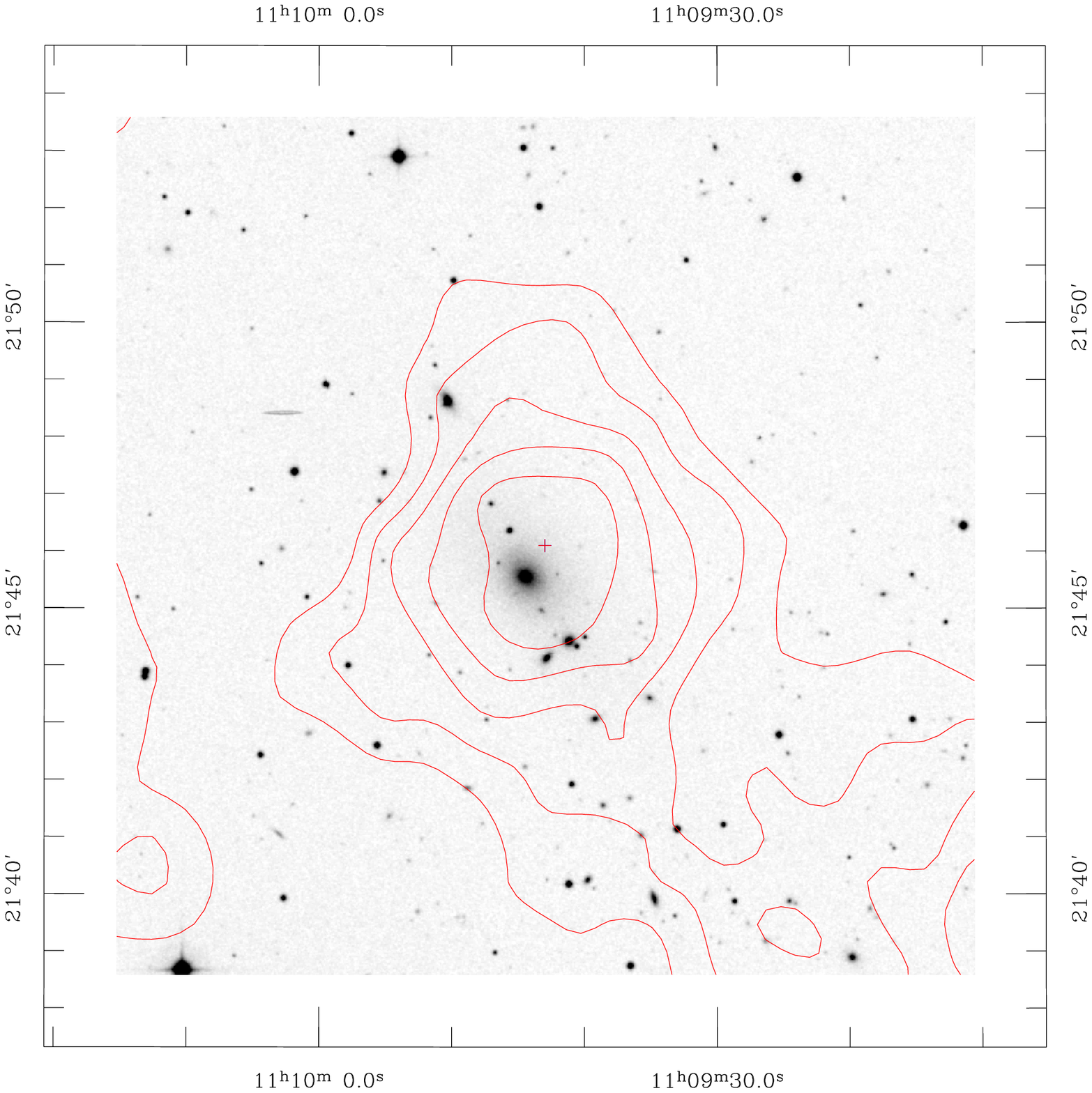}
\hspace{1cm}
   \includegraphics[width=7.5cm]{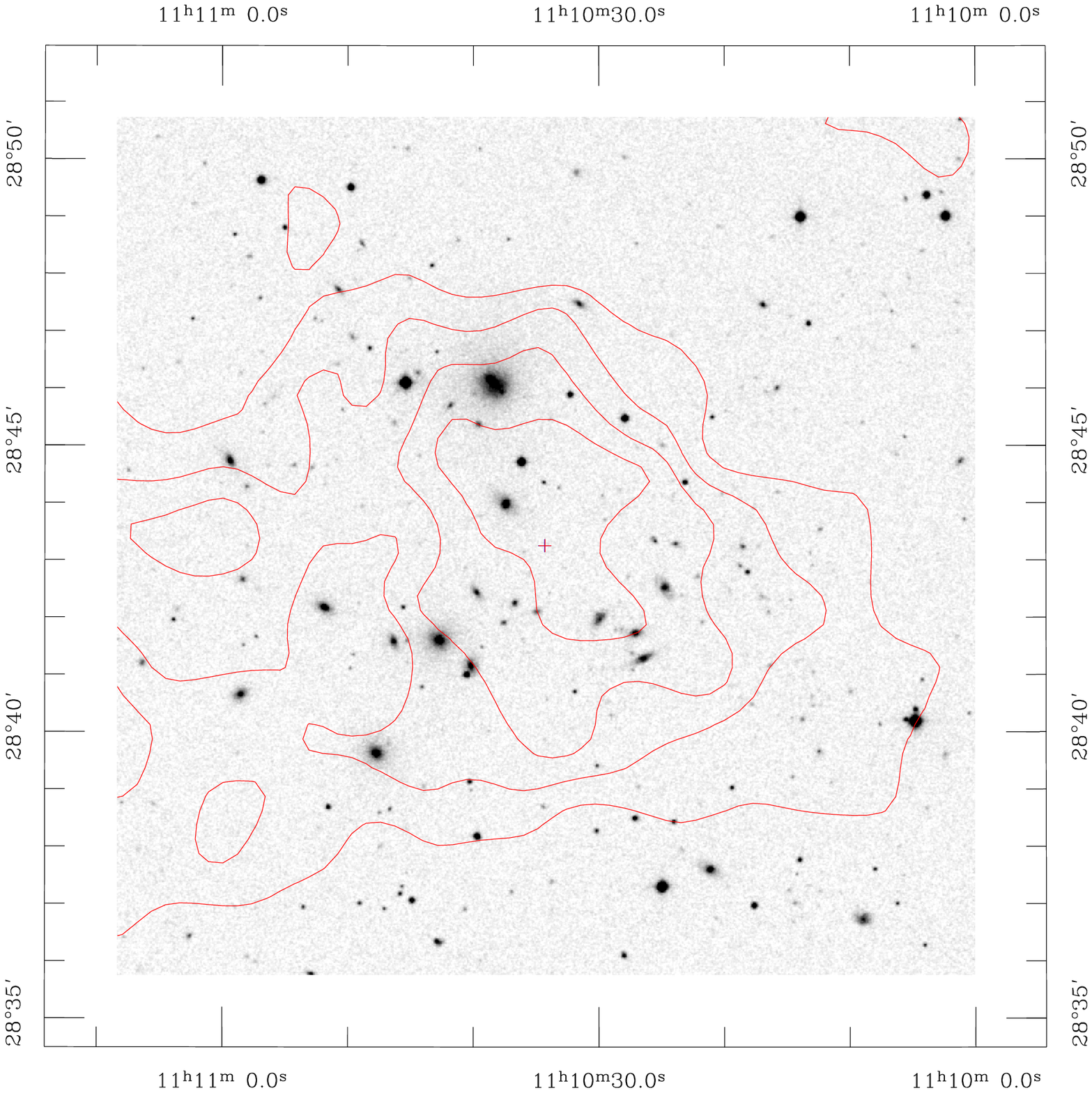}
}
\hbox{
\hspace{1cm}
   \includegraphics[width=7.5cm]{A1885_OV.ps}
\hspace{1cm}
   \includegraphics[width=7.5cm]{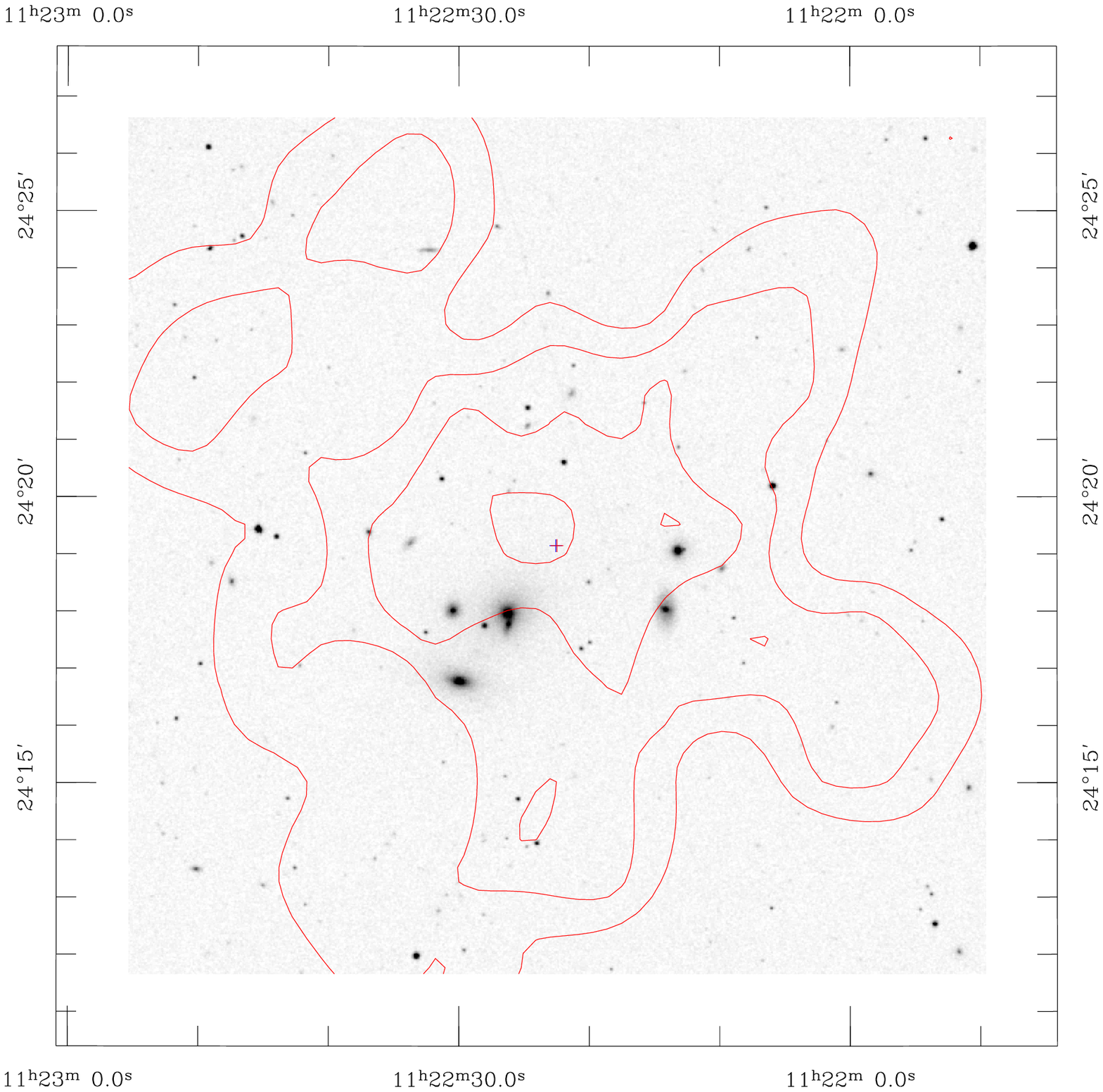}
}
\hbox{
\hspace{1cm}
   \includegraphics[width=7.5cm]{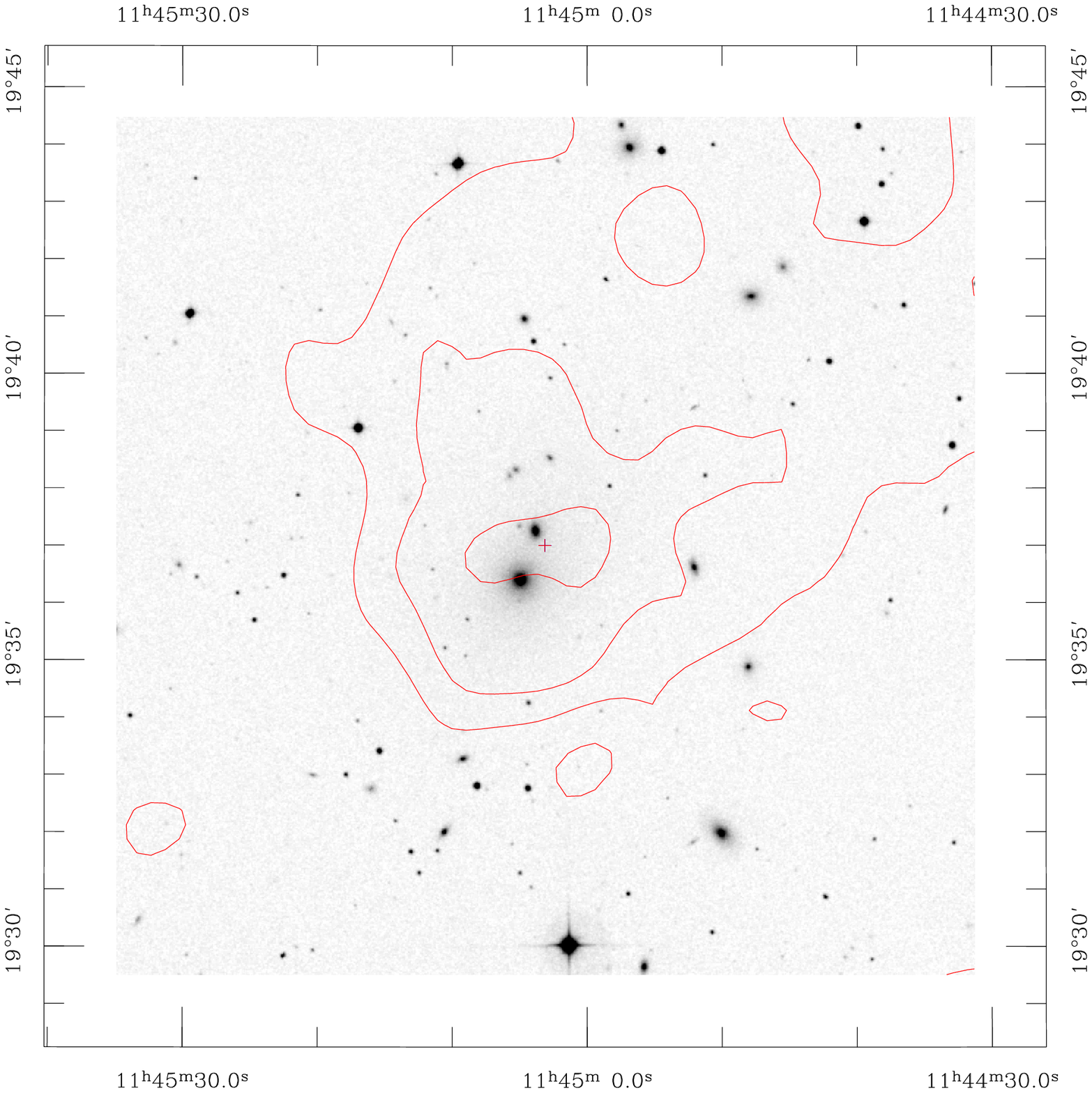}
\hspace{1cm}
   \includegraphics[width=7.5cm]{A1367_OV.ps}
}
\caption{Members of the Coma Supercluster.
{\bf Upper left:} RXCJ1109.7+2146, A 1177,
{\bf Upper right:} RXCJ1110.5+2843, A 1185,
{\bf Middle left:} XMM-Newton image of RXCJ1110.5+2843, A1885.
{\bf Middle right:} RXCJ1122.3+2419, HCG 51,
{\bf Lower left:} RXCJ1145.0+1936, A1367,
{\bf Lower right:} XMM-Newton image of RXCJ1145.0+1936, A1367.
}\label{figA6}
\end{figure*}

\begin{figure*}[h]
\hbox{
\hspace{1cm}
   \includegraphics[width=7.5cm]{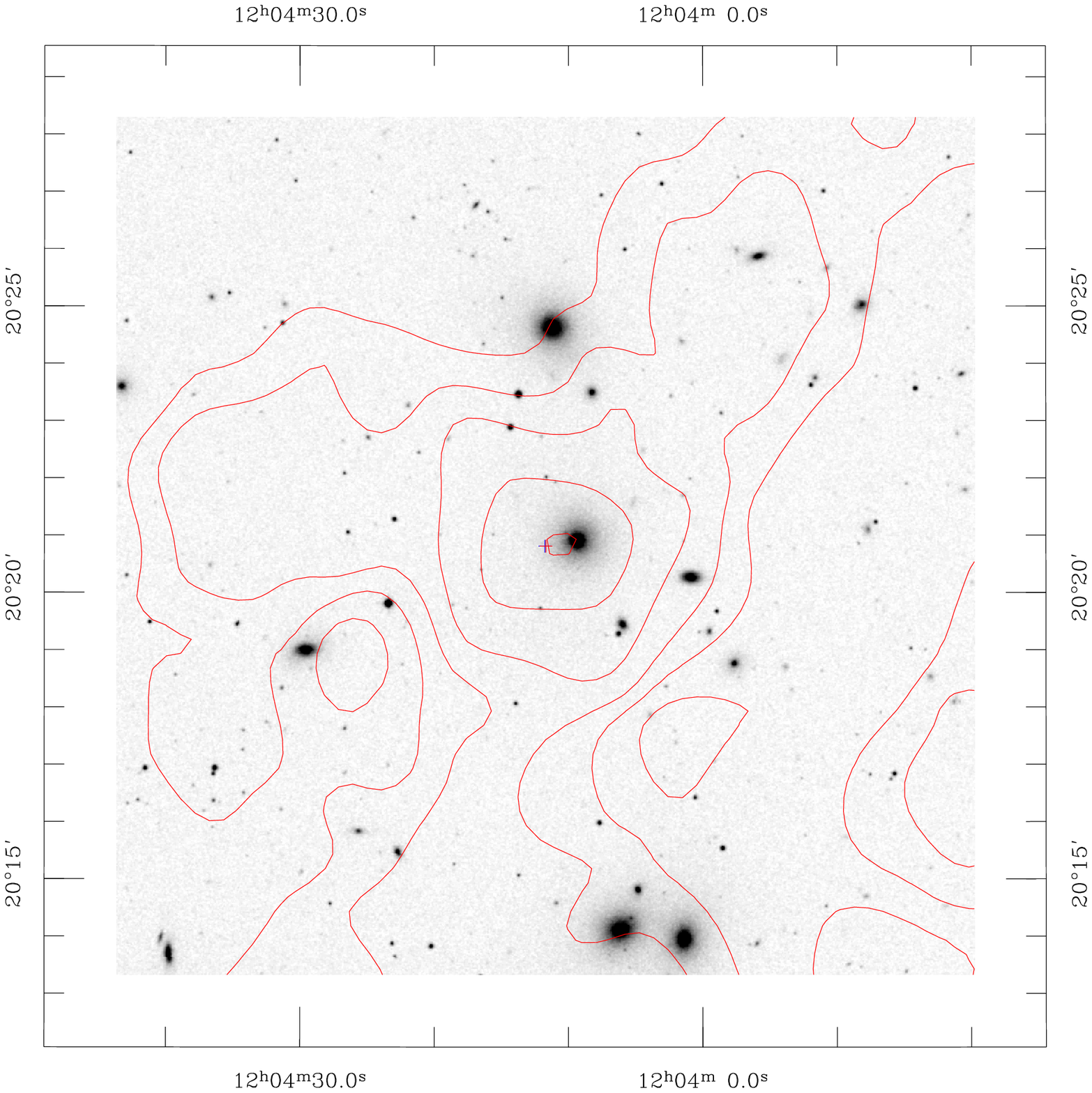}
\hspace{1cm}
   \includegraphics[width=7.5cm]{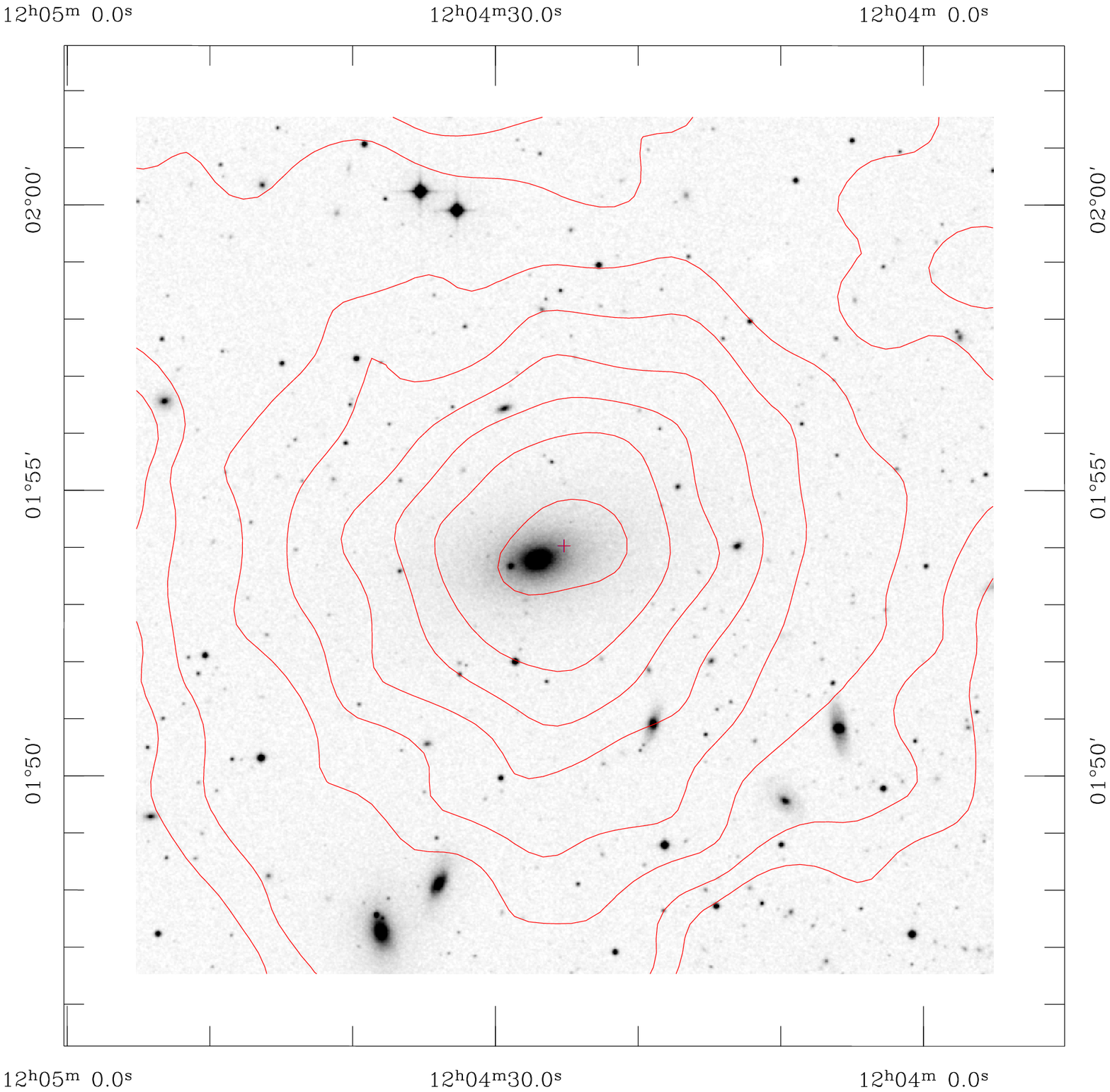}
}
\hbox{
\hspace{1cm}
   \includegraphics[width=7.5cm]{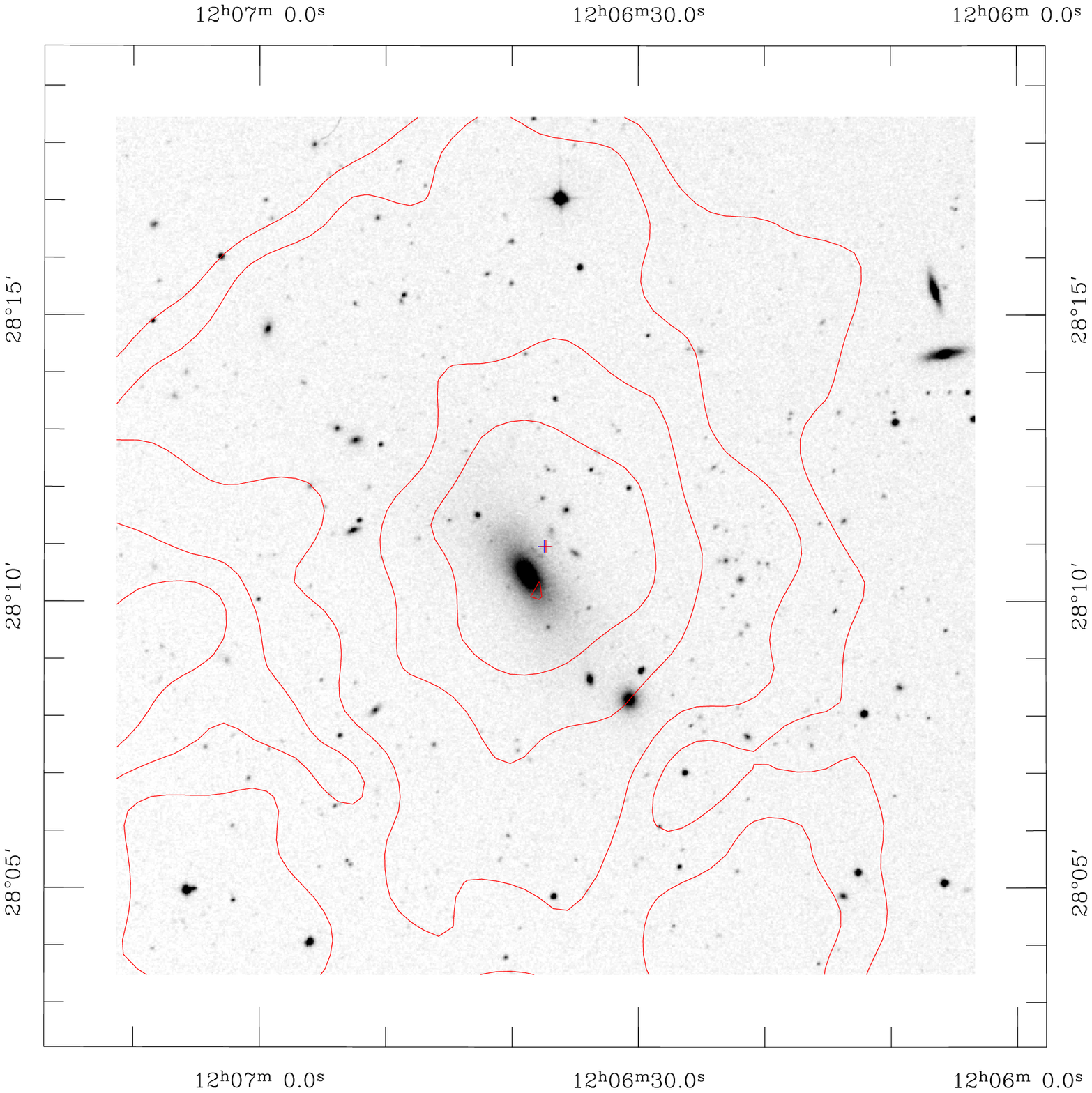}
\hspace{1cm}
   \includegraphics[width=7.5cm]{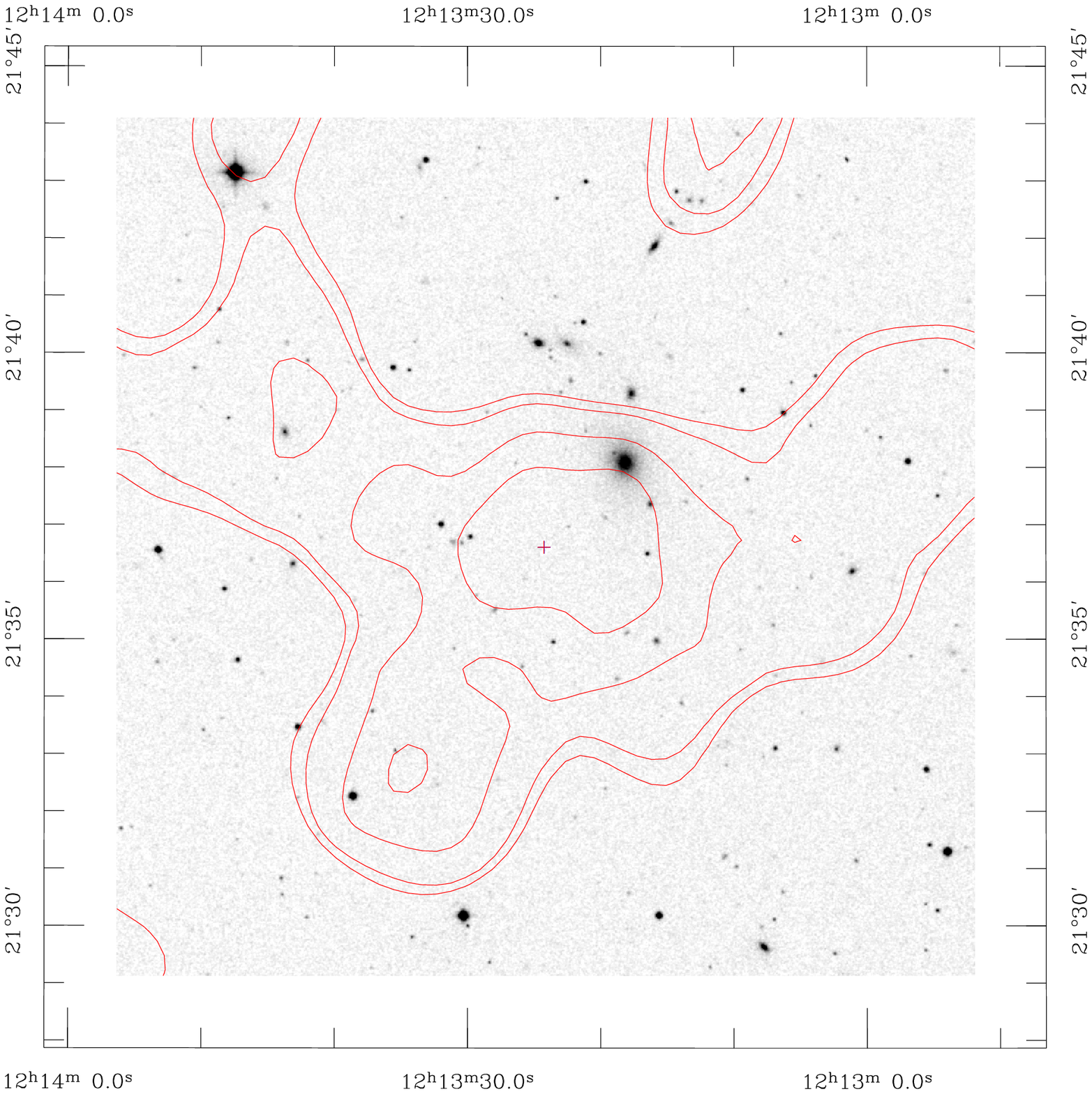}
}
\hbox{
\hspace{1cm}
   \includegraphics[width=7.5cm]{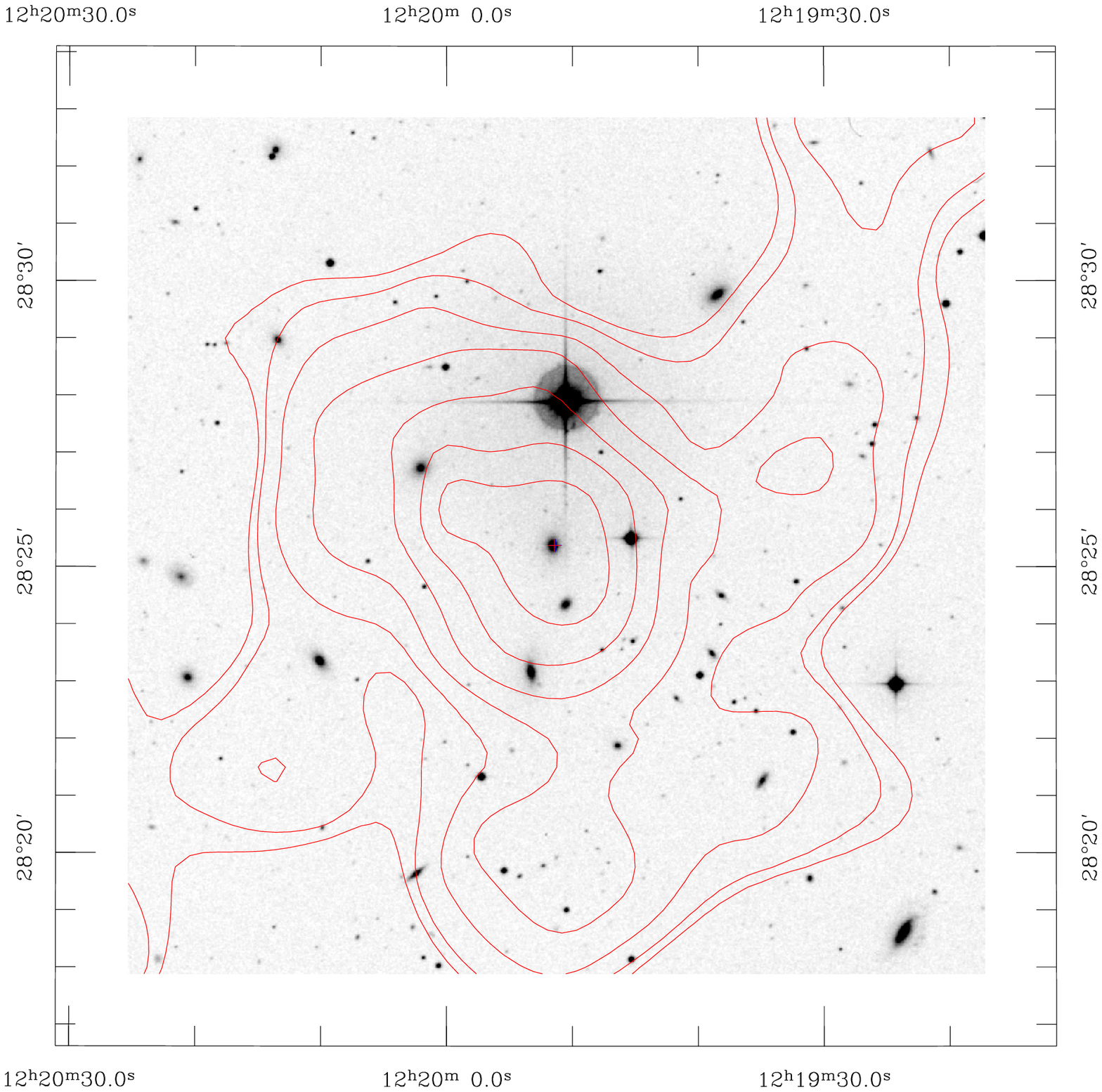}
\hspace{1cm}
   \includegraphics[width=7.5cm]{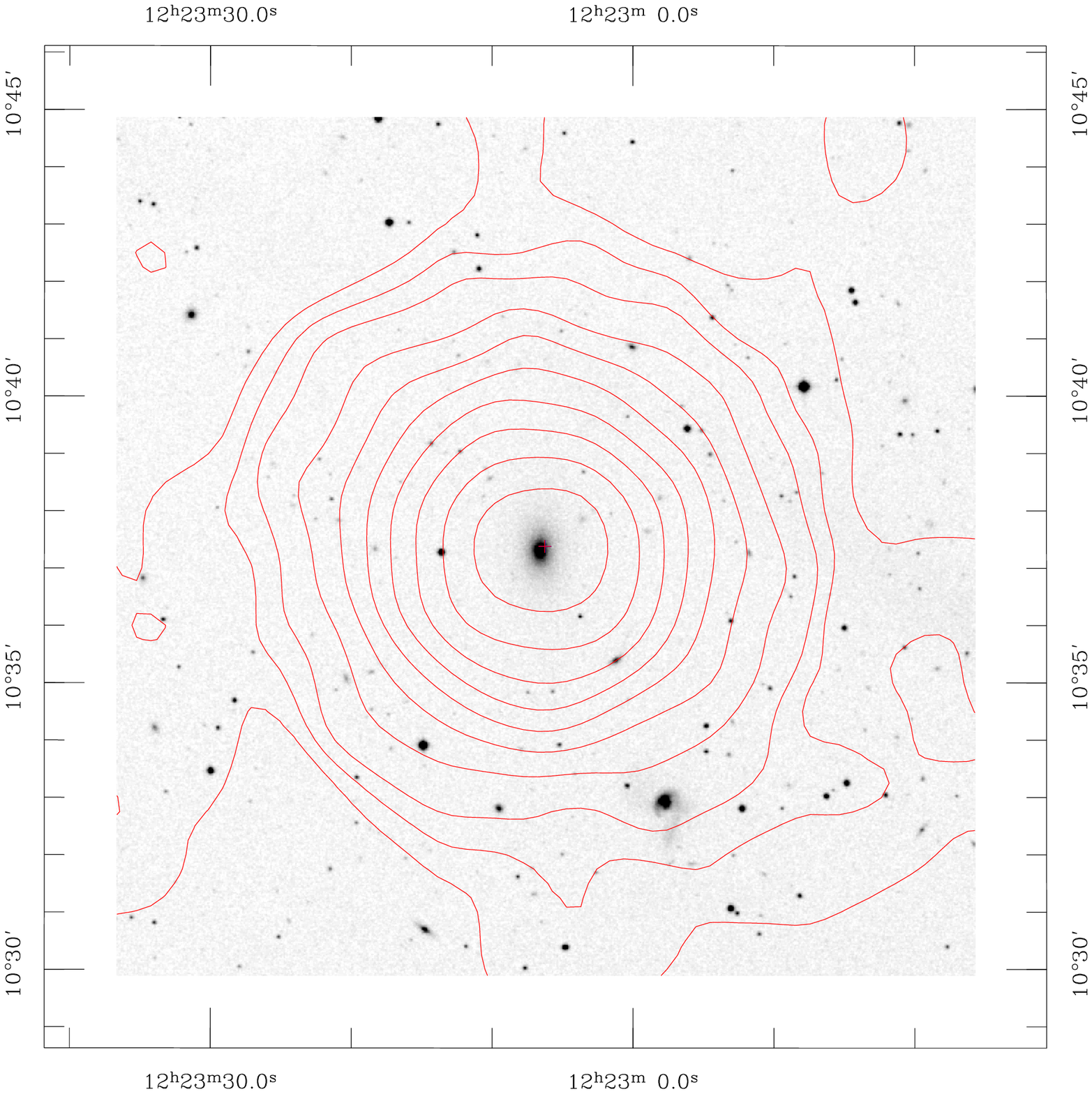}
}
\caption{Members of the Coma Supercluster continued.
{\bf Upper left:} RXCJ1204.1+2020, NGC 4066,
{\bf Upper right:} RXCJ1204.4+0154, MKW4,
{\bf Middle left:} RXCJ1206.6+2810, NGC 410
{\bf Middle right:} RXCJ1213.4+2136, UGC 7224,
{\bf Lower left:} RXCJ1219.8+2825,CGCG 185-075 
{\bf Lower right:} RXCJ1223.1+1037, NGC 4325.
}\label{figA7}
\end{figure*}

\begin{figure*}[h]
\hbox{
\hspace{1cm}
   \includegraphics[width=7.5cm]{NGC_4325_OV_XMM.ps}
\hspace{1cm}
   \includegraphics[width=7.5cm]{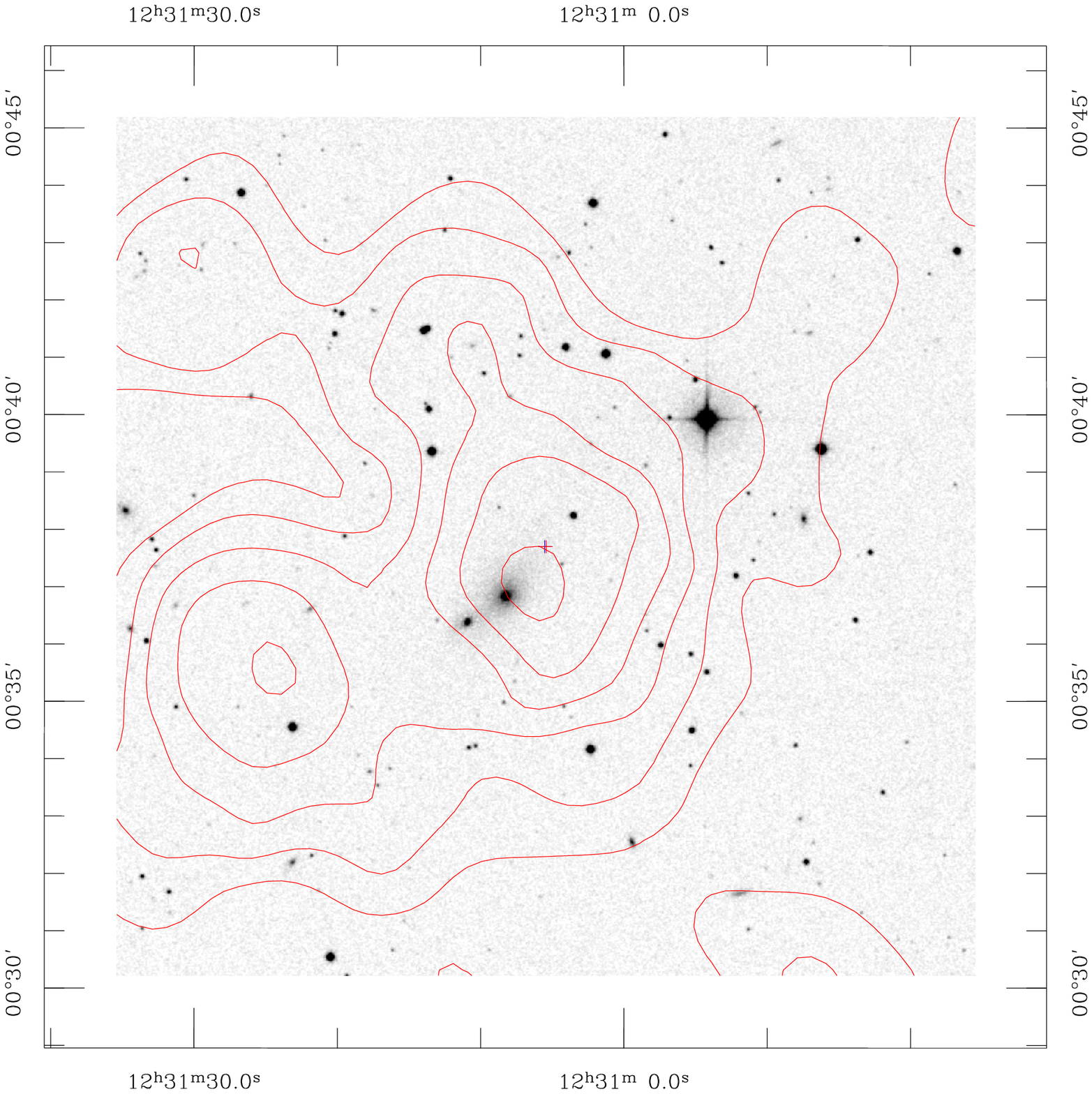}
}
\hbox{
\hspace{1cm}
   \includegraphics[width=7.5cm]{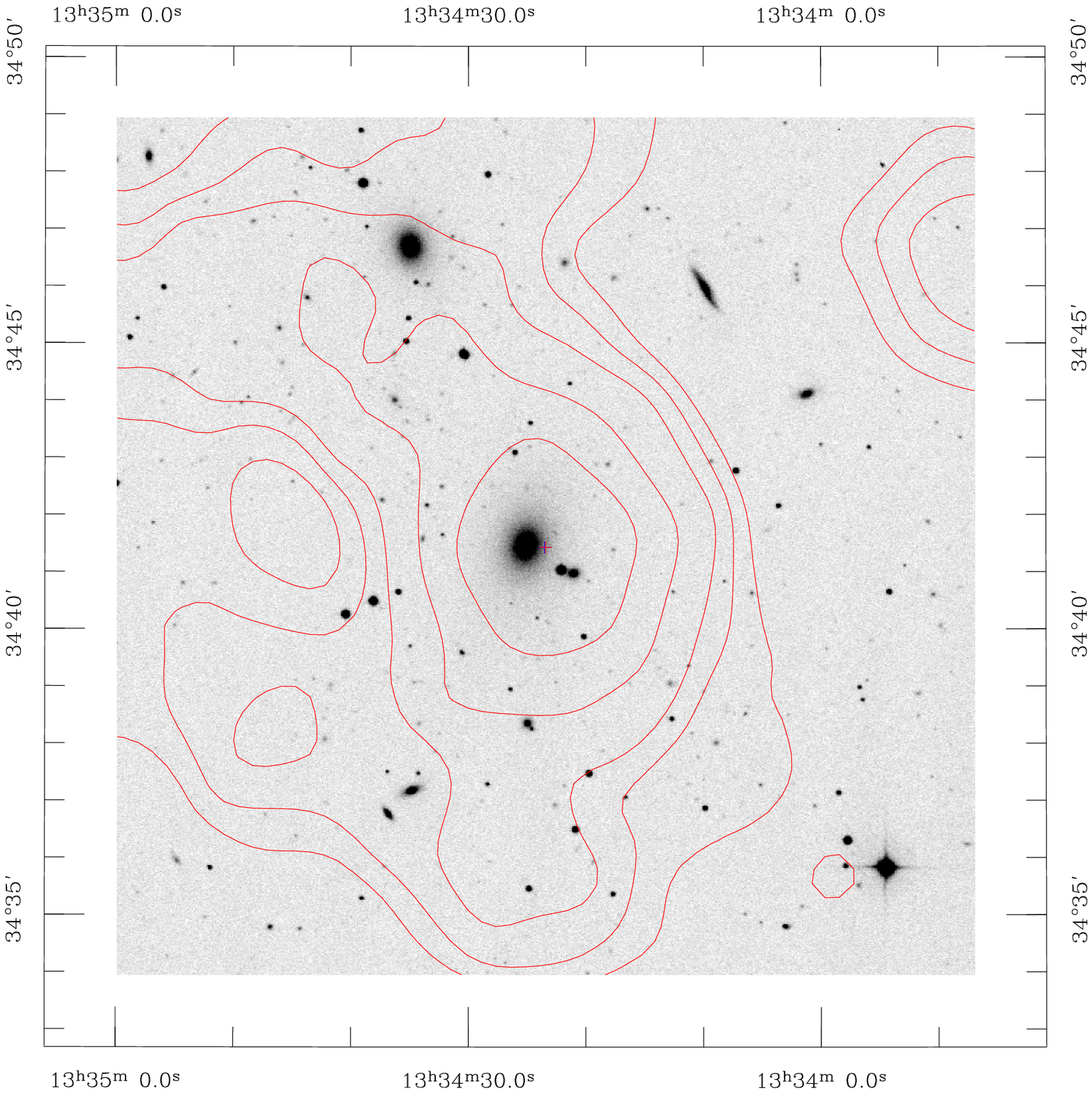}
\hspace{1cm}
}
\hbox{
\hspace{1cm}
\hspace{1cm}
}
\caption{Members of the Coma Supercluster continued.
{\bf Upper left:} XMM-Newton image of RXCJ1223.1+1037, NGC 4325
{\bf Upper right:} RXCJ1231.0+0037, NGC 4493,
{\bf Middle left:} RXCJ1334.3+3441, NGC 522.
}\label{figA8}
\end{figure*}

\clearpage

\subsection{Images of the Hercules Supercluster members}

This section provides images of the members of the
Hercules supercluster at $z \le 0.03$ (Figs.~\ref{figA9}, \ref{figA10}).
The images show overlays of X-ray 
contours from RASS on DSS images produced in the same way
as in the previous sections. 

Since this extension of the Hercules SC 
is less well known we remark on some of the cluster identifications. 
All groups and clusters have clearly extended X-ray emission in 
the RASS and no peculiar spectral properties. RXCJ1629.6+4049
is one of two parts of the cluster Abell 2197, which was found to have 
two clearly distinct X-ray emitting components, A 2197 W and A 2197 E, in the RASS
\citep{Mur1996}. The clusters
RXCJ1736.3+6803 and RXCJ1755.8+6236 have already been identified
in the RASS North Ecliptic Pole survey \citep{Hen1995} and they
are described in detail in this publication. RXCJ1714.3+4341
has also been found as a WBL group by \citet{Whi1999}.
RXCJ1723.4+5658
has been identified as a group of galaxies by \citet{Vdl2007},
and RXCJ1736.3+6803 by \citet{Lee2017}. 
One of the clusters, RXCJ1941.7+5037 lies in the ZoA at
$b_{II} \sim 13.3^o$.

\begin{figure*}[h]
\hbox{
\hspace{1cm}
   \includegraphics[width=7.5cm]{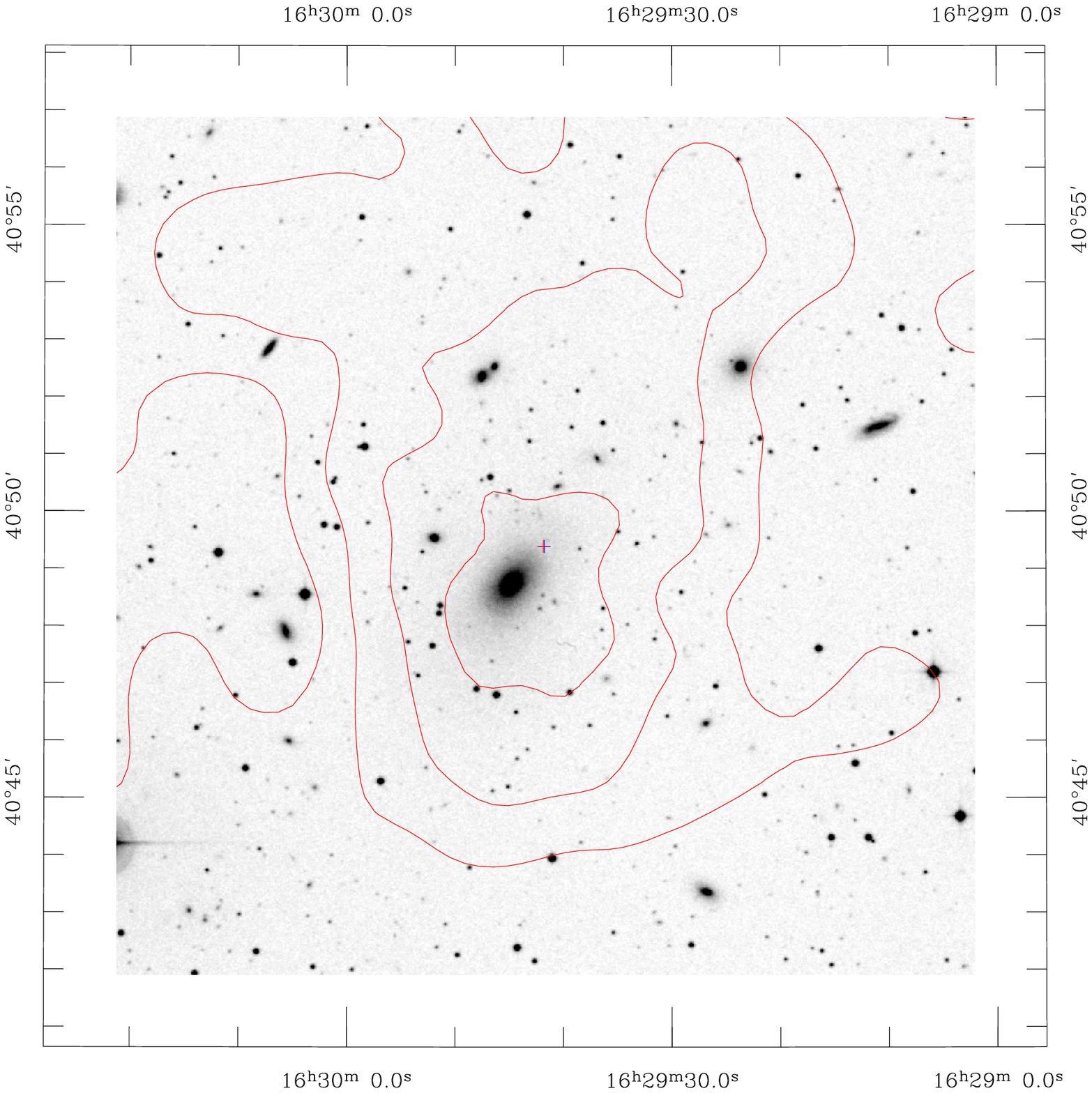}
\hspace{1cm}
   \includegraphics[width=7.5cm]{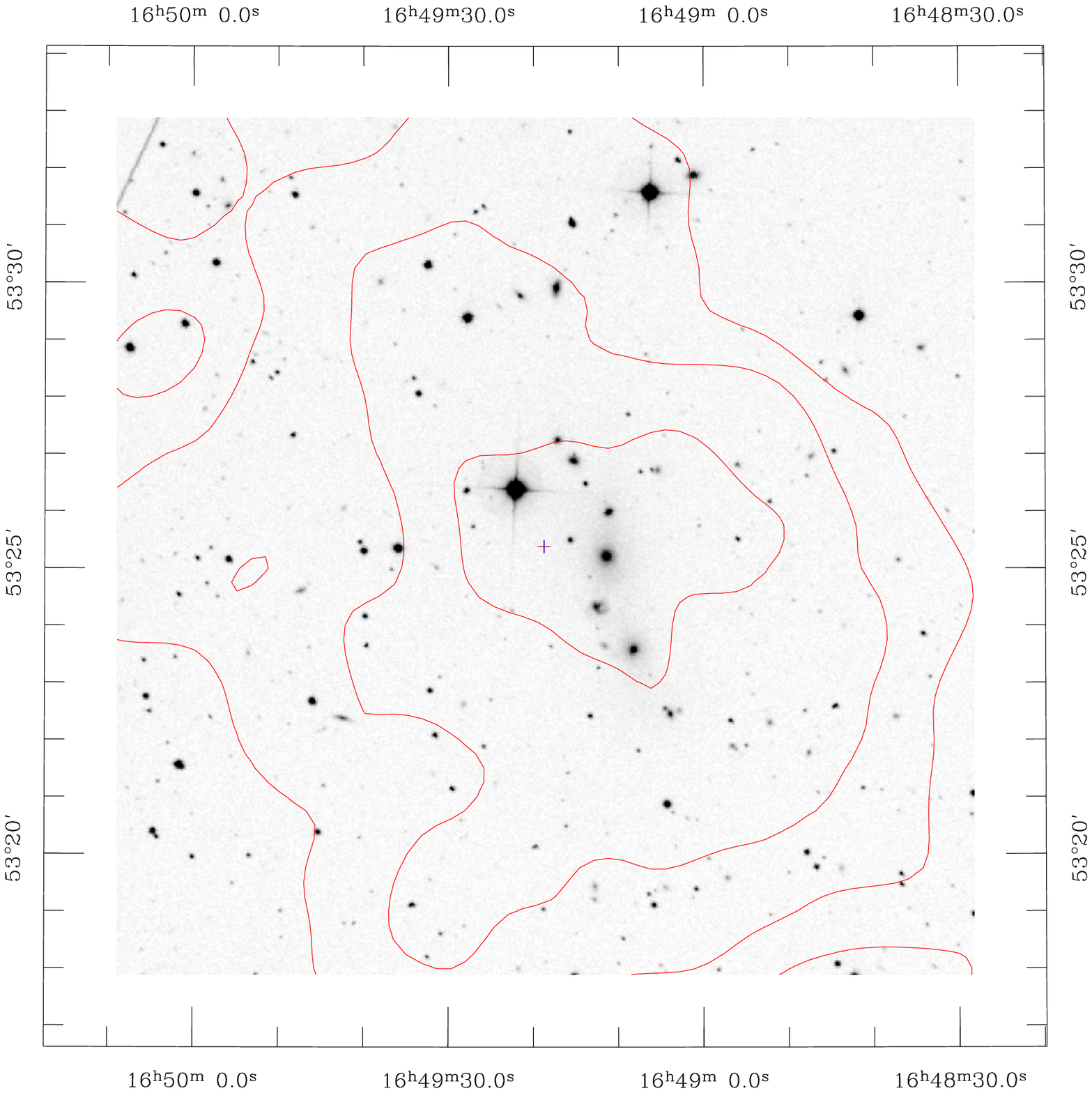}
}
\hbox{
\hspace{1cm}
   \includegraphics[width=7.5cm]{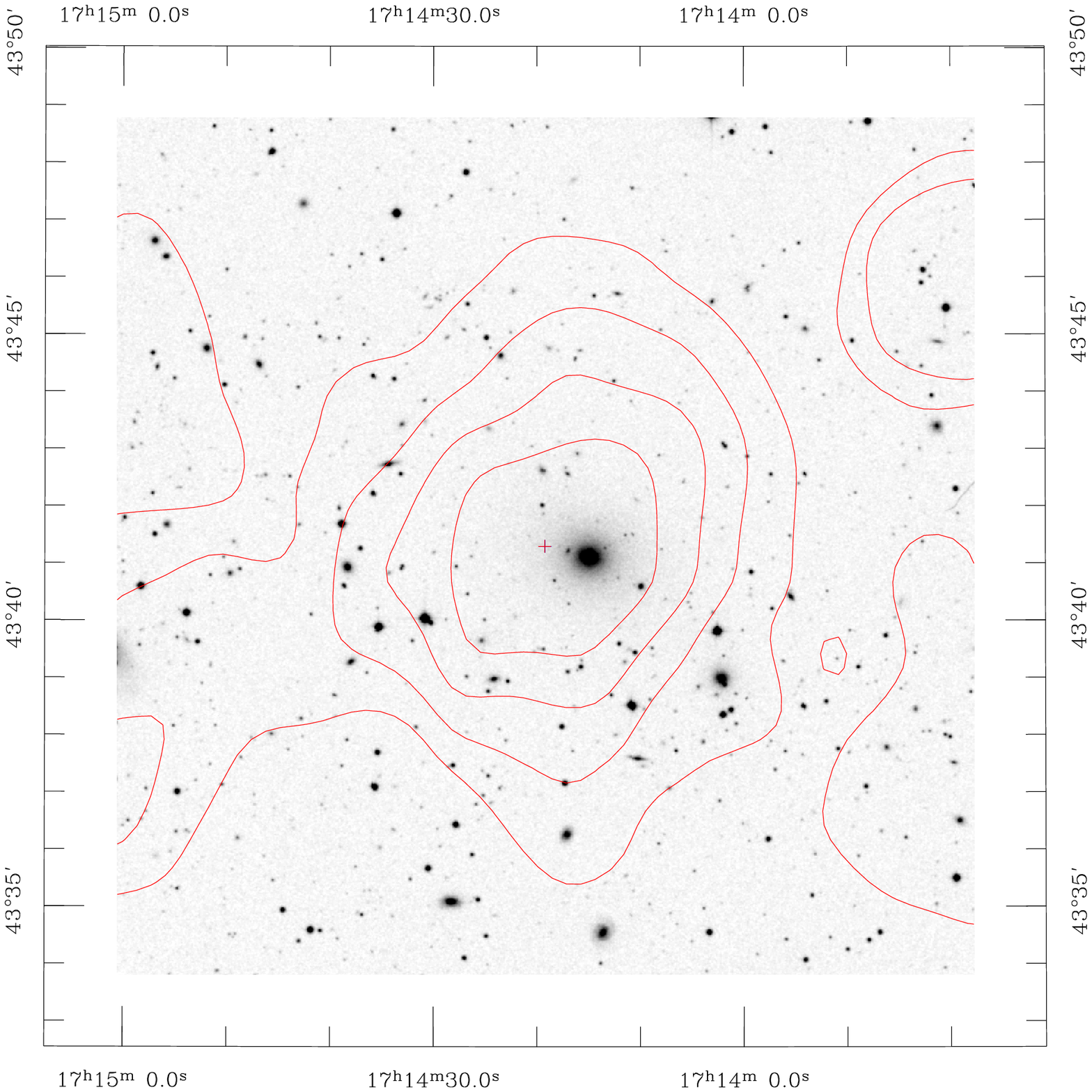}
\hspace{1cm}
   \includegraphics[width=7.5cm]{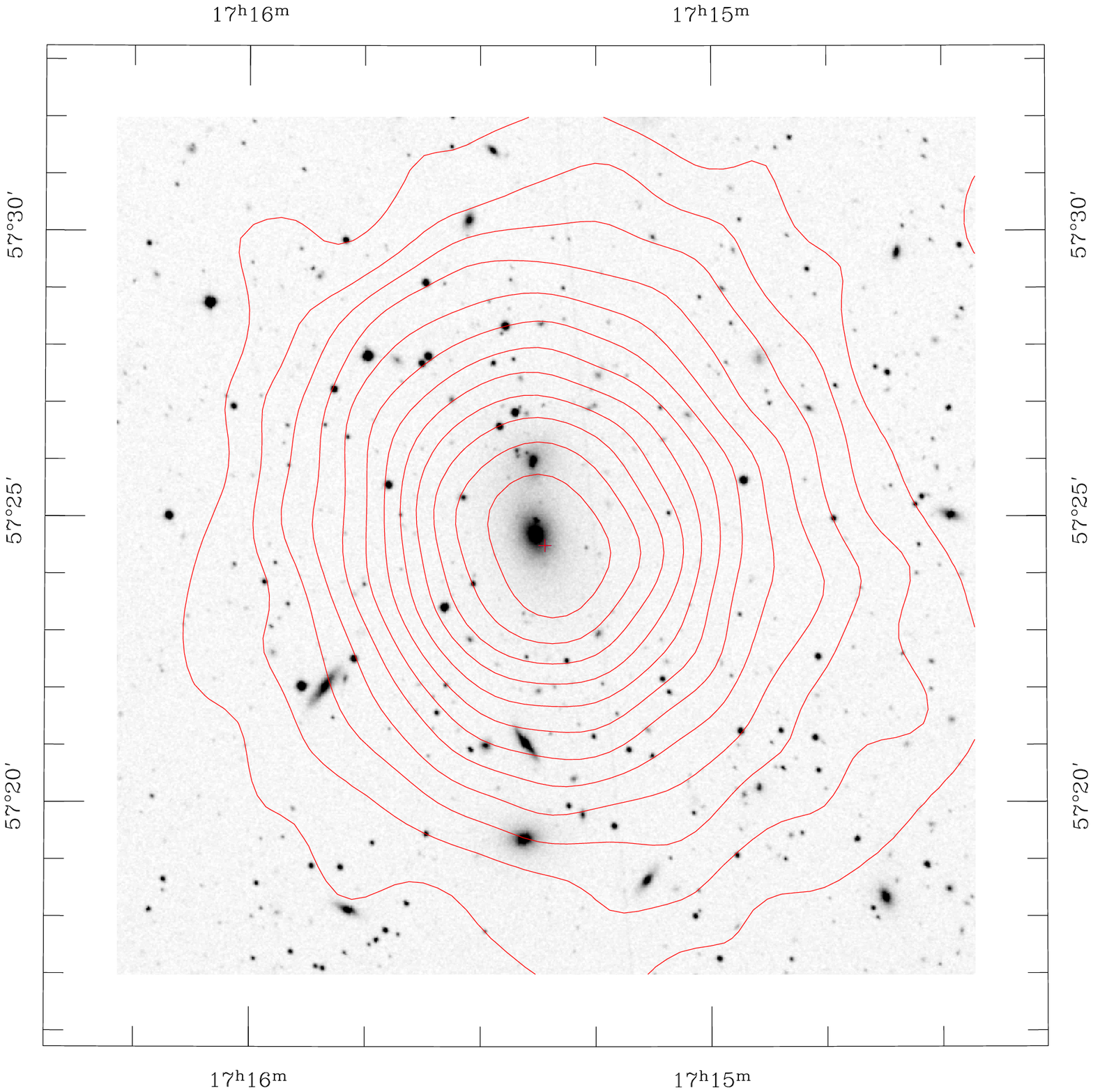}
}
\hbox{
\hspace{1cm}
   \includegraphics[width=7.5cm]{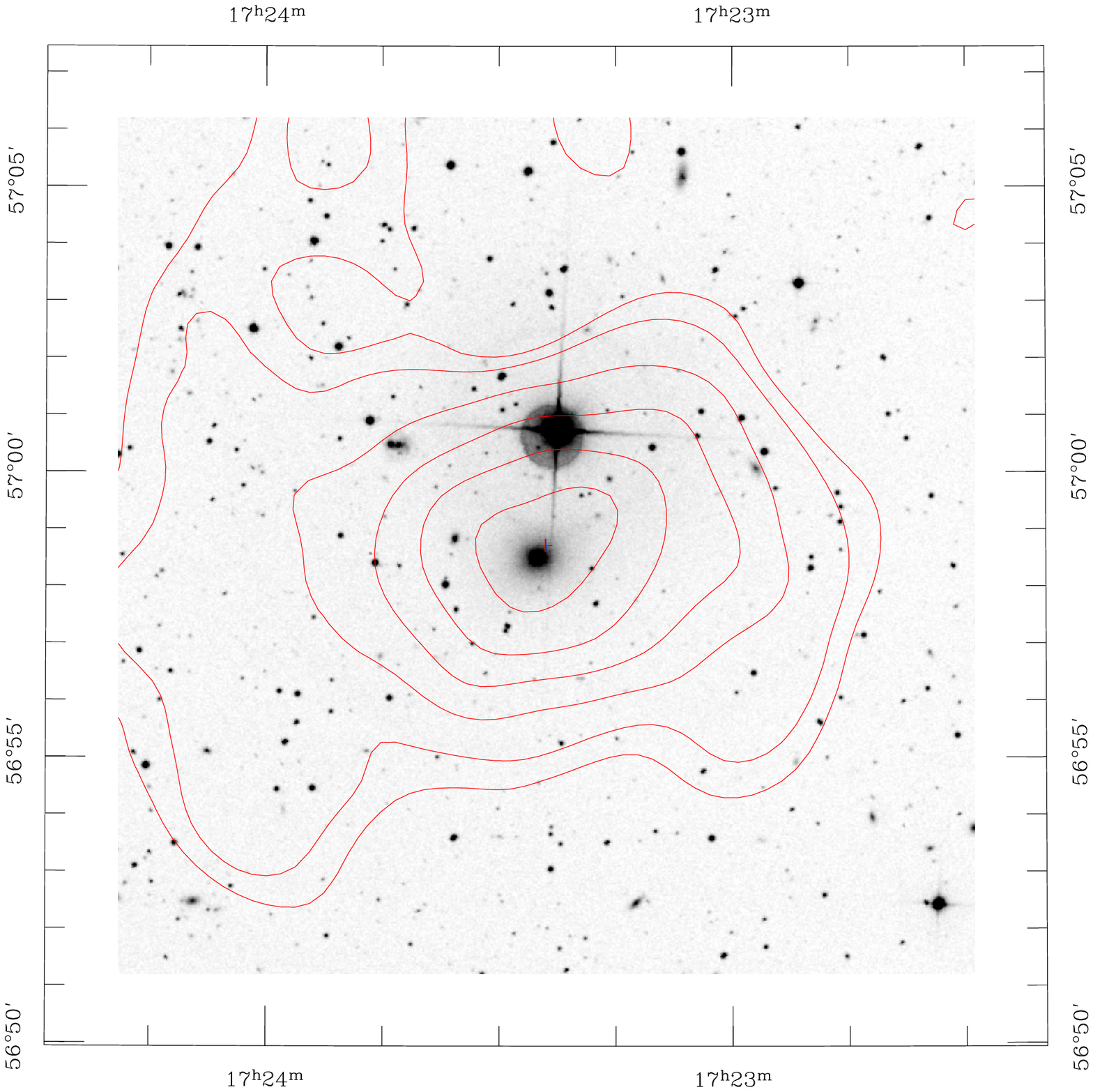}
\hspace{1cm}
   \includegraphics[width=7.5cm]{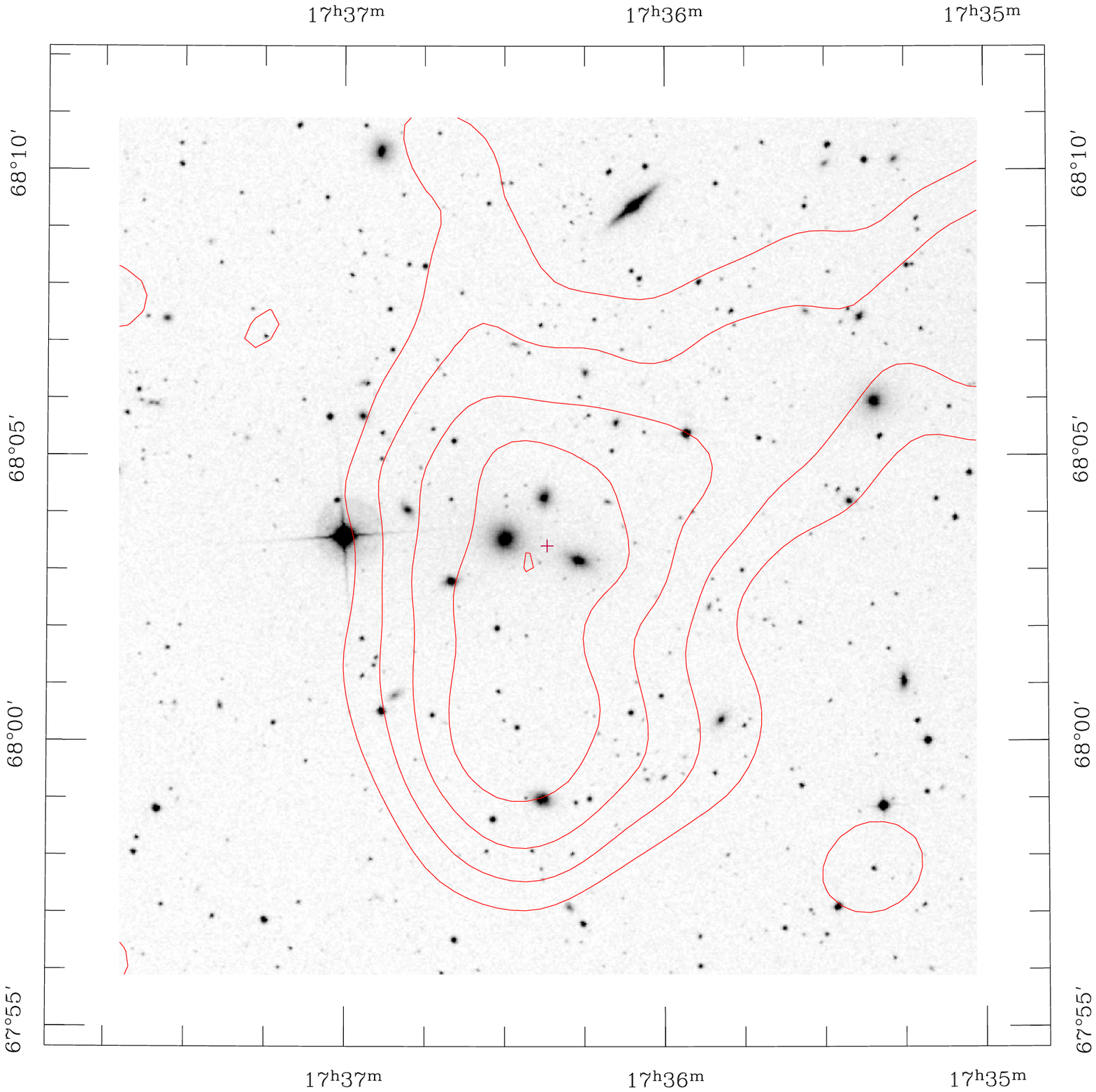}
}
\caption{Members of the Hercules supercluster.
{\bf Upper left:} RXCJ1629.6+4049, A2197~E,
{\bf Upper right:} RXCJ1649.3+5325, Arp 330,
{\bf Middle left:} RXCJ1714.3+4341, NGC 6329,
{\bf Middle right:} RXCJ1715.3+5724, NGC 6338,
{\bf Lower left:} RXCJ1723.4+5658, NGC 6370,
{\bf Lower right:} RXCJ1736.3+6803, NGC 6420.
}\label{figA9}
\end{figure*}

\begin{figure*}[h]
\hbox{
\hspace{1cm}
   \includegraphics[width=7.5cm]{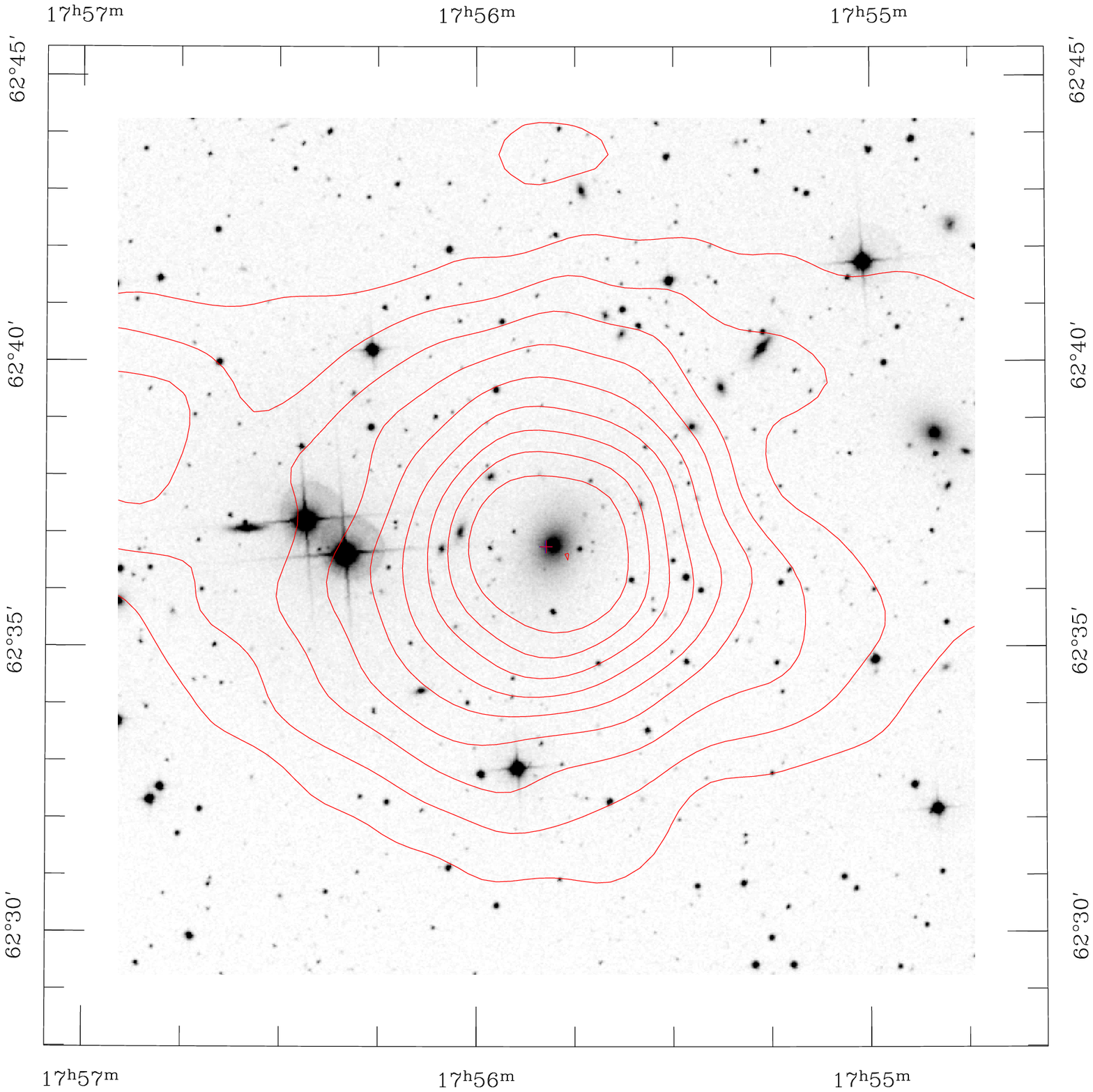}
\hspace{1cm}
   \includegraphics[width=7.5cm]{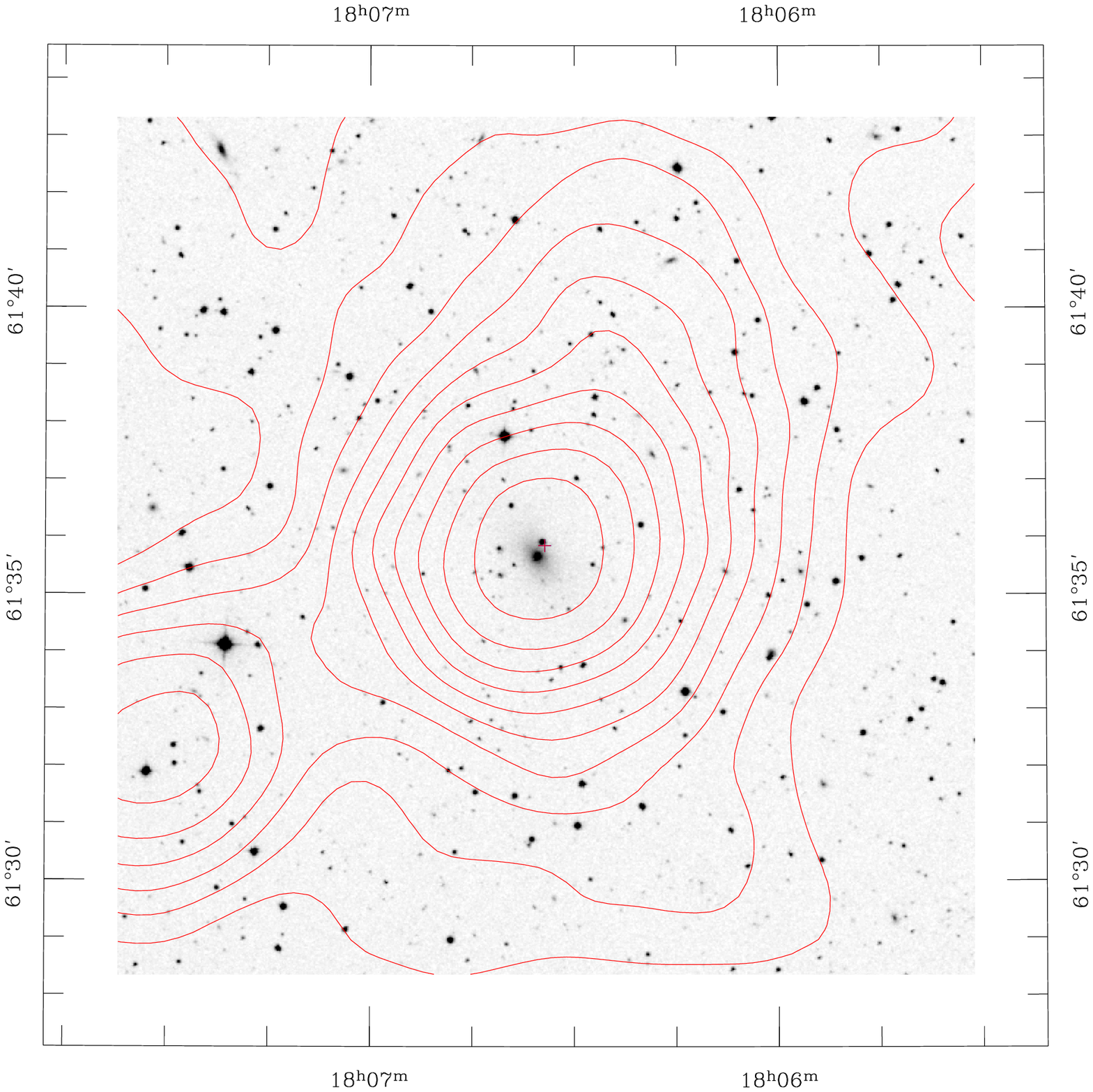}
}
\hbox{
\hspace{1cm}
   \includegraphics[width=7.5cm]{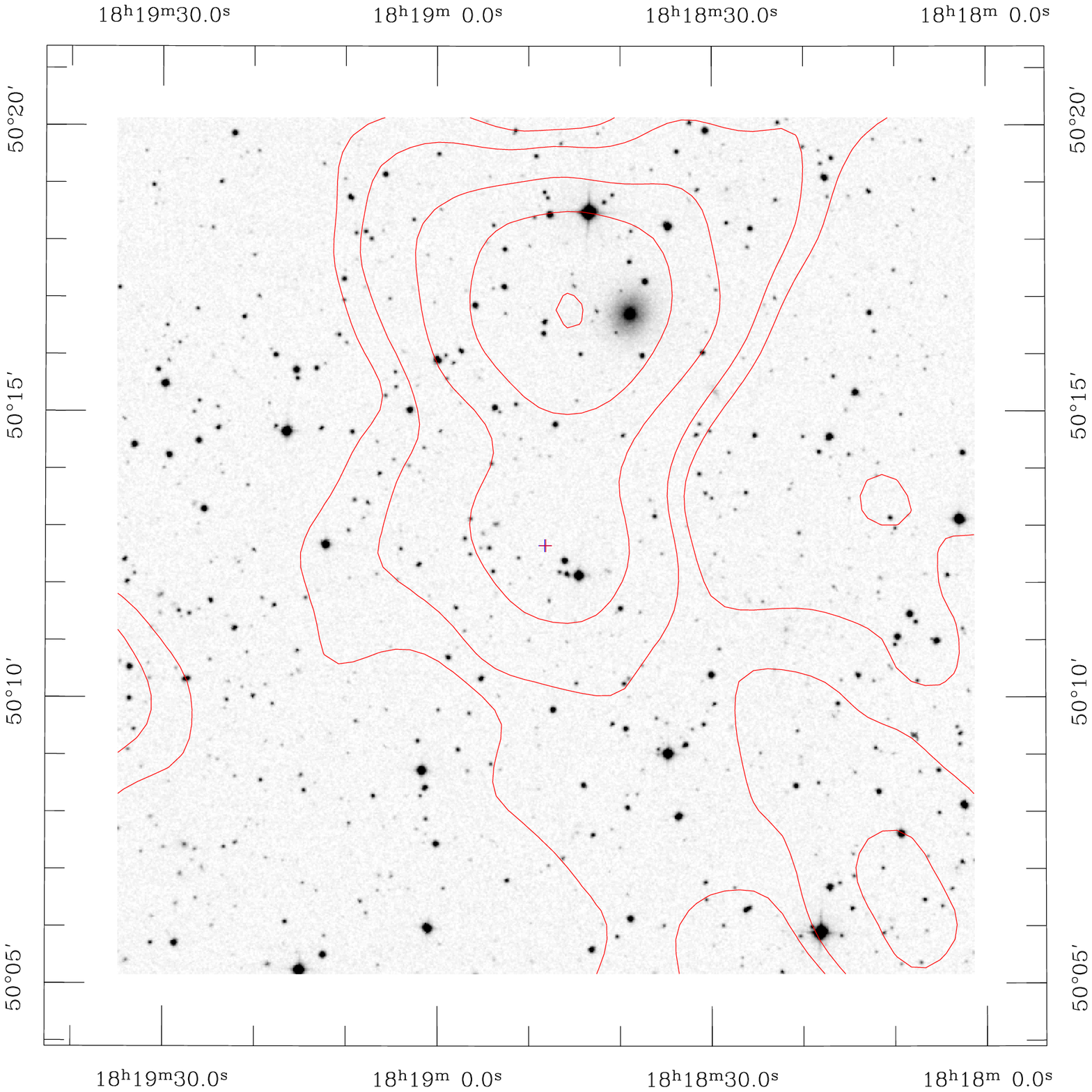}
\hspace{1cm}
   \includegraphics[width=7.5cm]{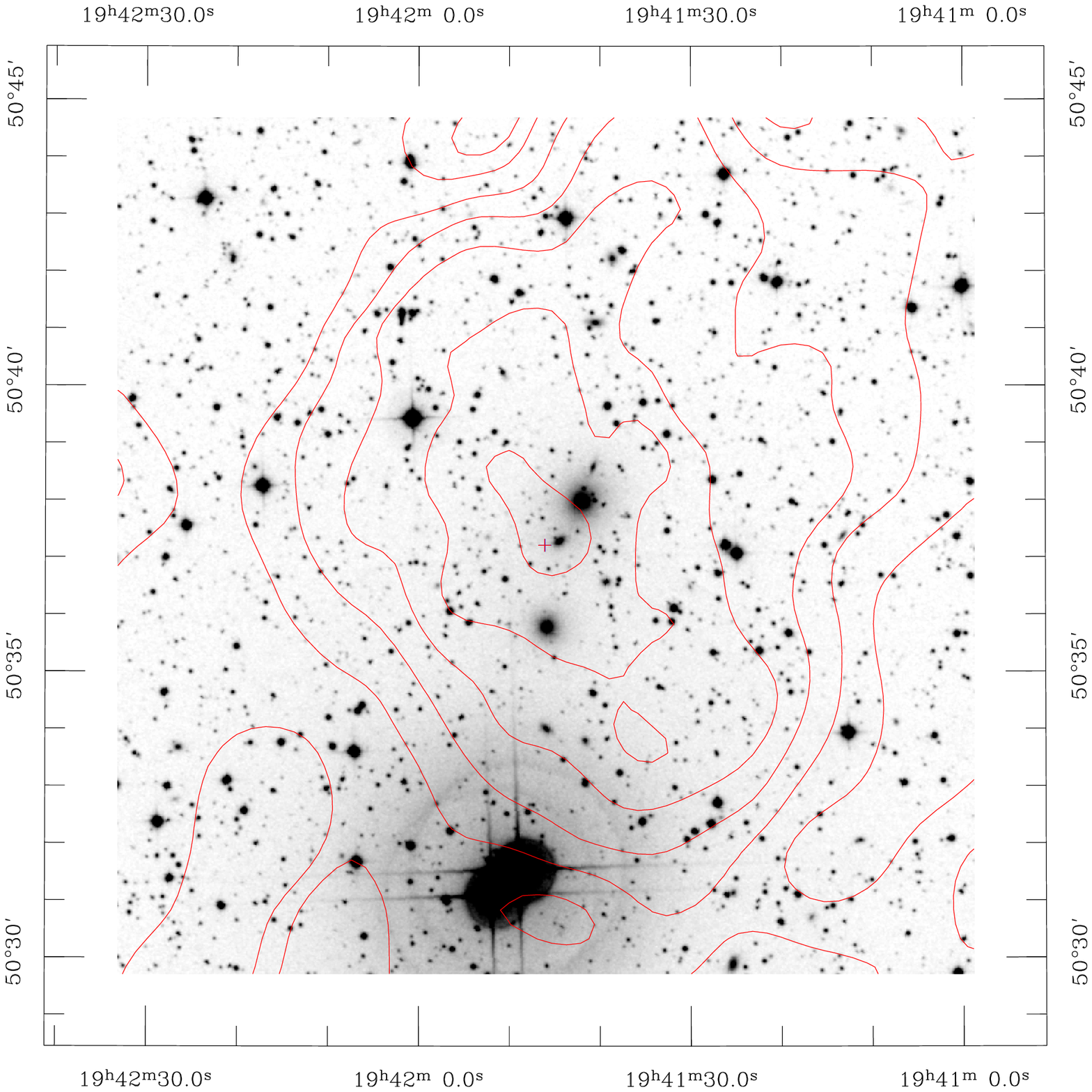}
}
\caption{Members of the Hercules supercluster continued.
{\bf Upper left:} RXCJ1755.8+6236 in the North Ecliptic Pole region,
{\bf Upper right:} RXCJ1806.5+6135, VII Zw 767,
{\bf Middle left:} RXCJ1818.7+5017, UGC 11202
{\bf Middle right:} RXCJ1941.7+5037, UGC 11465.
}\label{figA10}
\end{figure*}

\subsection{Images of the Sagittarius Supercluster members}

This section provides images of the groups and clusters of
the Sagittarius SC (Fig.~\ref{figA11}). The images show overlays of X-ray 
contours from RASS on DSS images produced in the same way
as in the previous sections. All objects have 
significantly extended X-ray emission in the RASS
and no unexpected spectral properties. 

\begin{figure*}[h]
\hbox{
\hspace{1cm}
   \includegraphics[width=7.5cm]{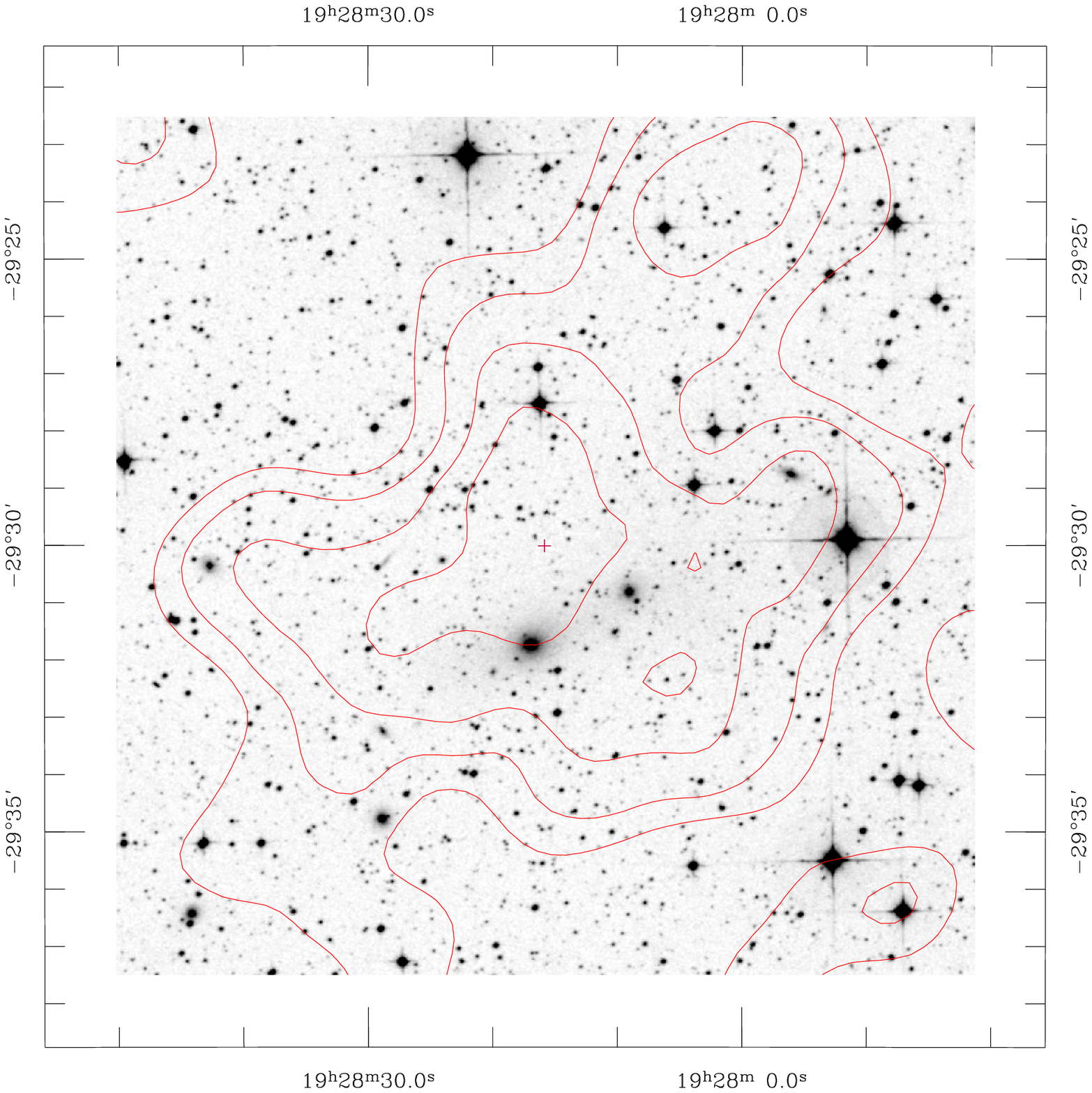}
\hspace{1cm}
   \includegraphics[width=7.5cm]{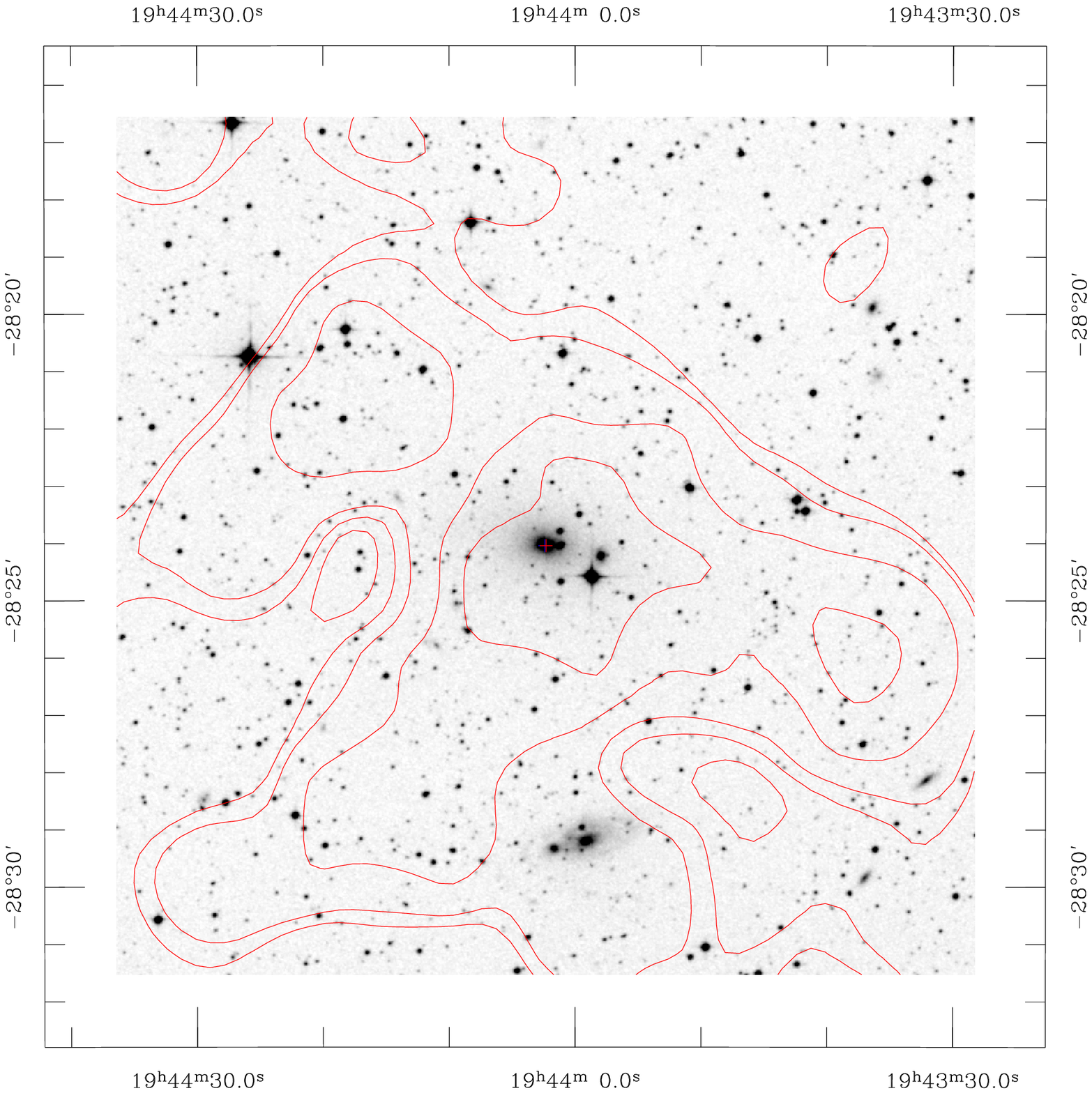}
}
\hbox{
\hspace{1cm}
   \includegraphics[width=7.5cm]{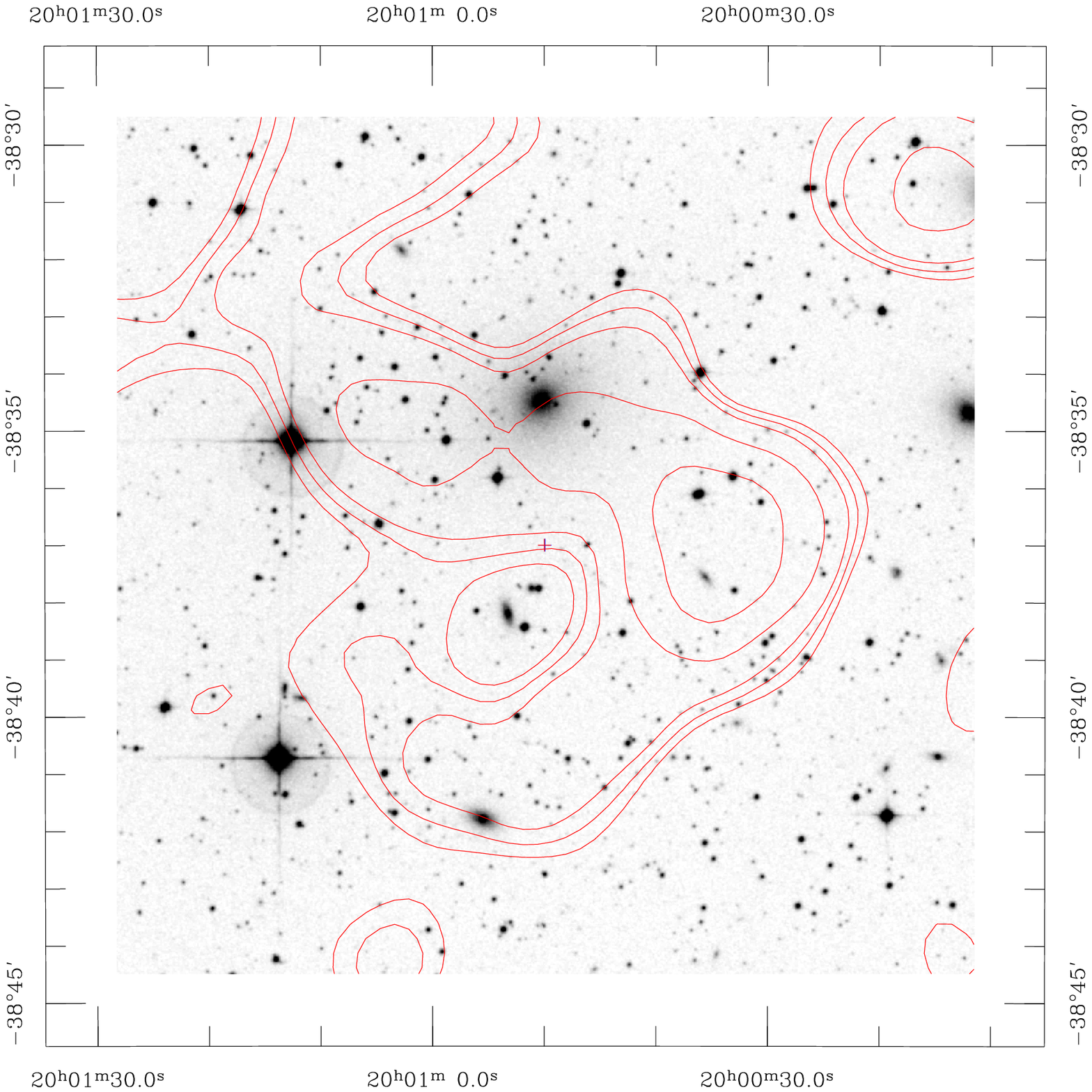}
\hspace{1cm}
   \includegraphics[width=7.5cm]{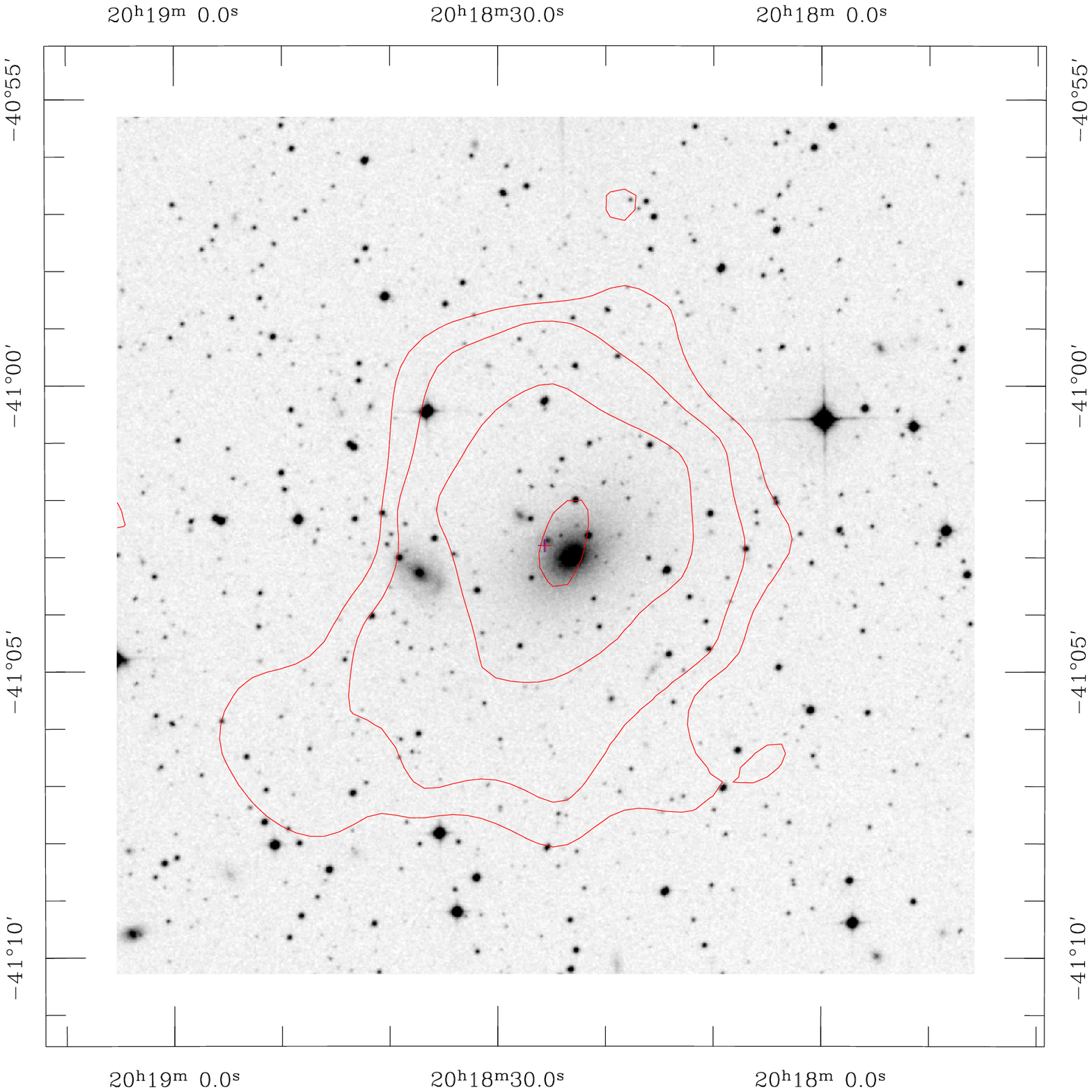}
}
\hbox{
\hspace{1cm}
   \includegraphics[width=7.5cm]{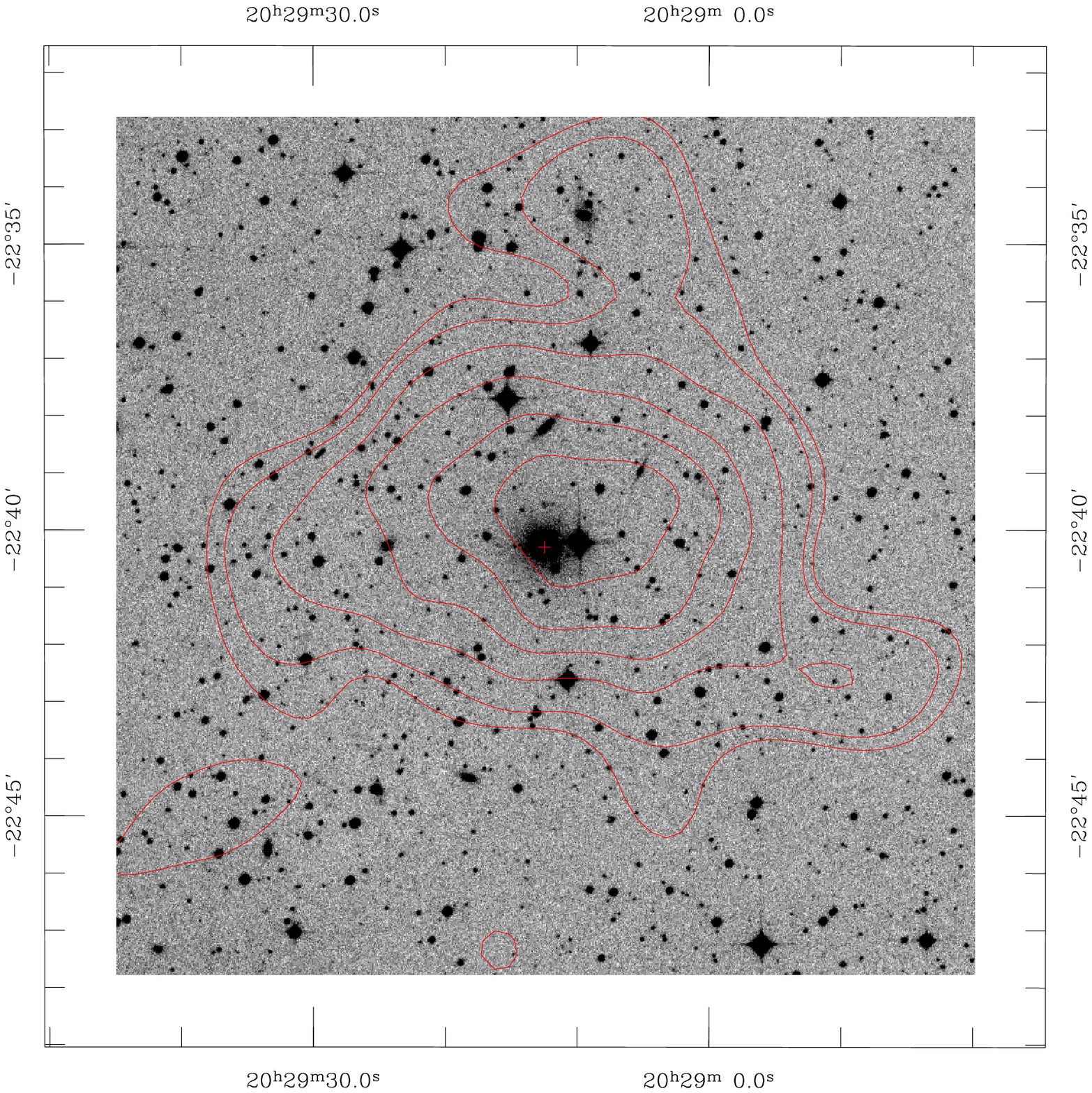}
\hspace{1cm}
   \includegraphics[width=7.5cm]{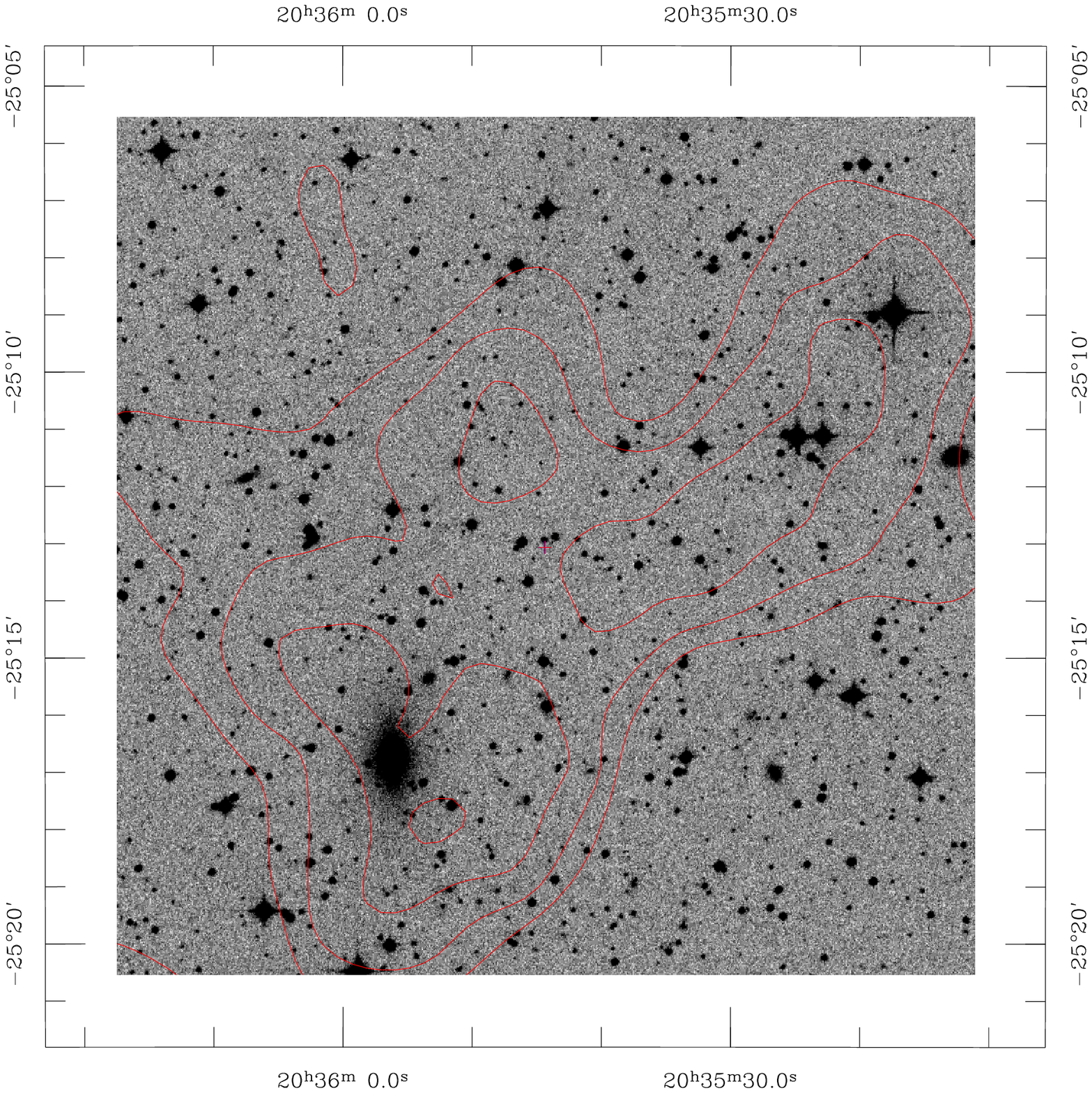}
}
\caption{Members of the Sagittarius Supercluster.
{\bf Upper left:} RXCJ1928.2-2930, ESO 460-G004,
{\bf Upper right:} RXCJ1944.0-2824, NGC6816
{\bf Middle left:} RXCJ2000.6-3837,
{\bf Middle right:} RXCJ2018.4-4102, IC 4991,
{\bf Lower left:} RXCJ2029.2-2240, ESO 528 - G008
{\bf Lower right:} RXCJ2035.7-2513, A3698
}\label{figA11}
\end{figure*}

\subsection{Images of the Lacerta Supercluster members}

This section provides images of the groups and clusters of
the Lacerta SC (Fig.~\ref{figA12}). The images show overlays of X-ray 
contours from RASS on DSS images produced in the same way
as in the previous sections. All objects have 
significantly extended X-ray emission in the RASS
and no unexpected spectral properties.

\begin{figure*}[h]
\hbox{
\hspace{1cm}
   \includegraphics[width=7.5cm]{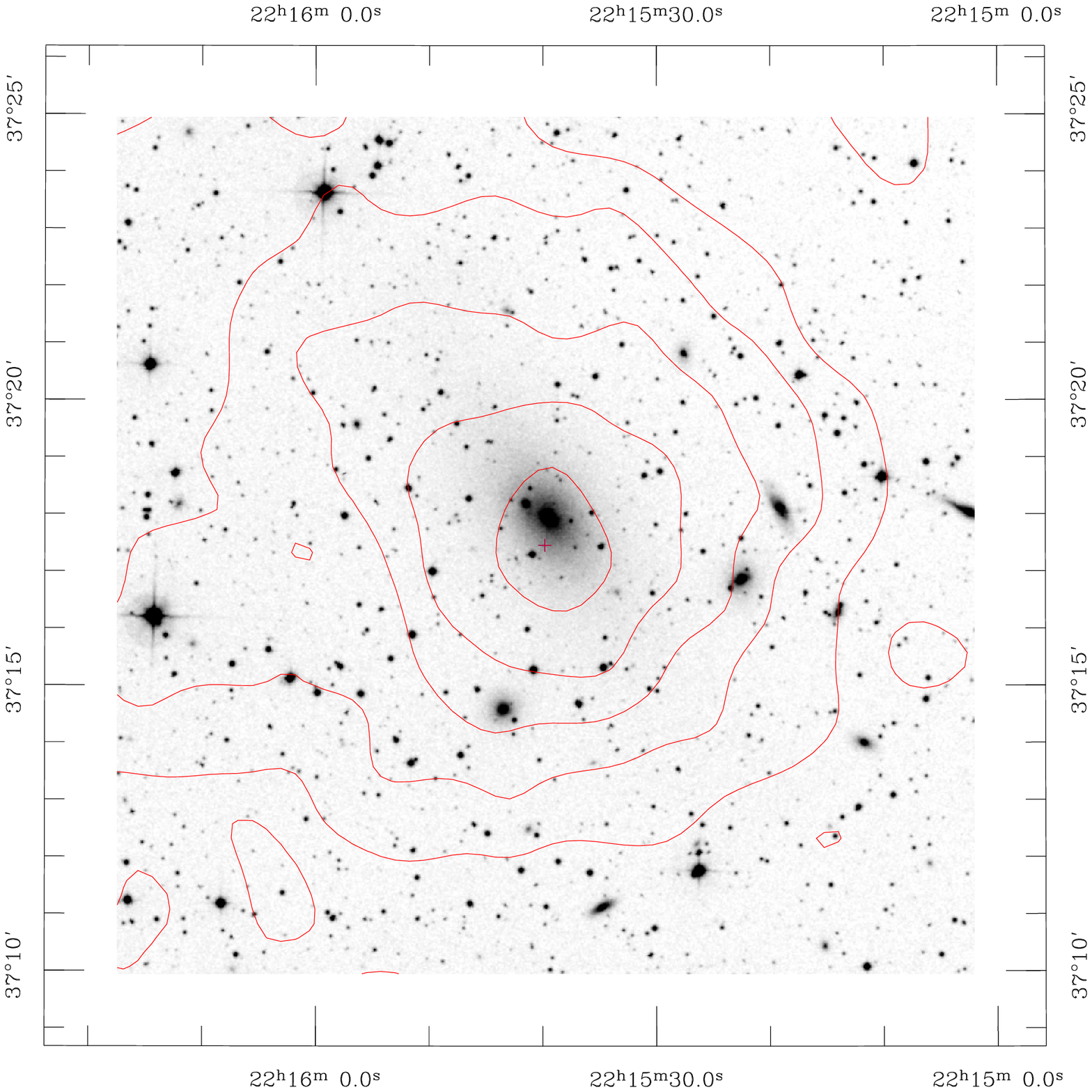}
\hspace{1cm}
   \includegraphics[width=7.5cm]{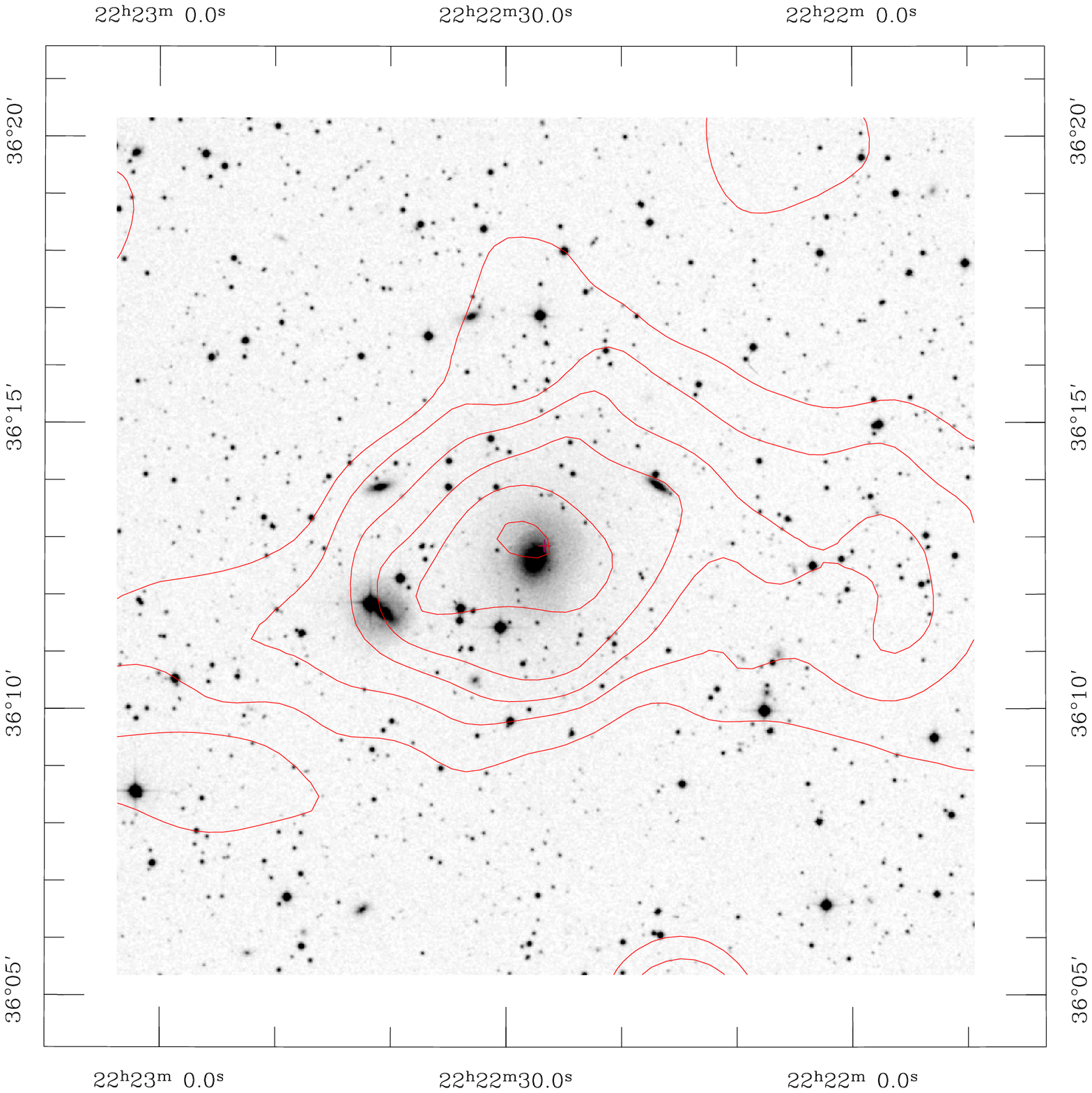}
}
\hbox{
\hspace{1cm}
   \includegraphics[width=7.5cm]{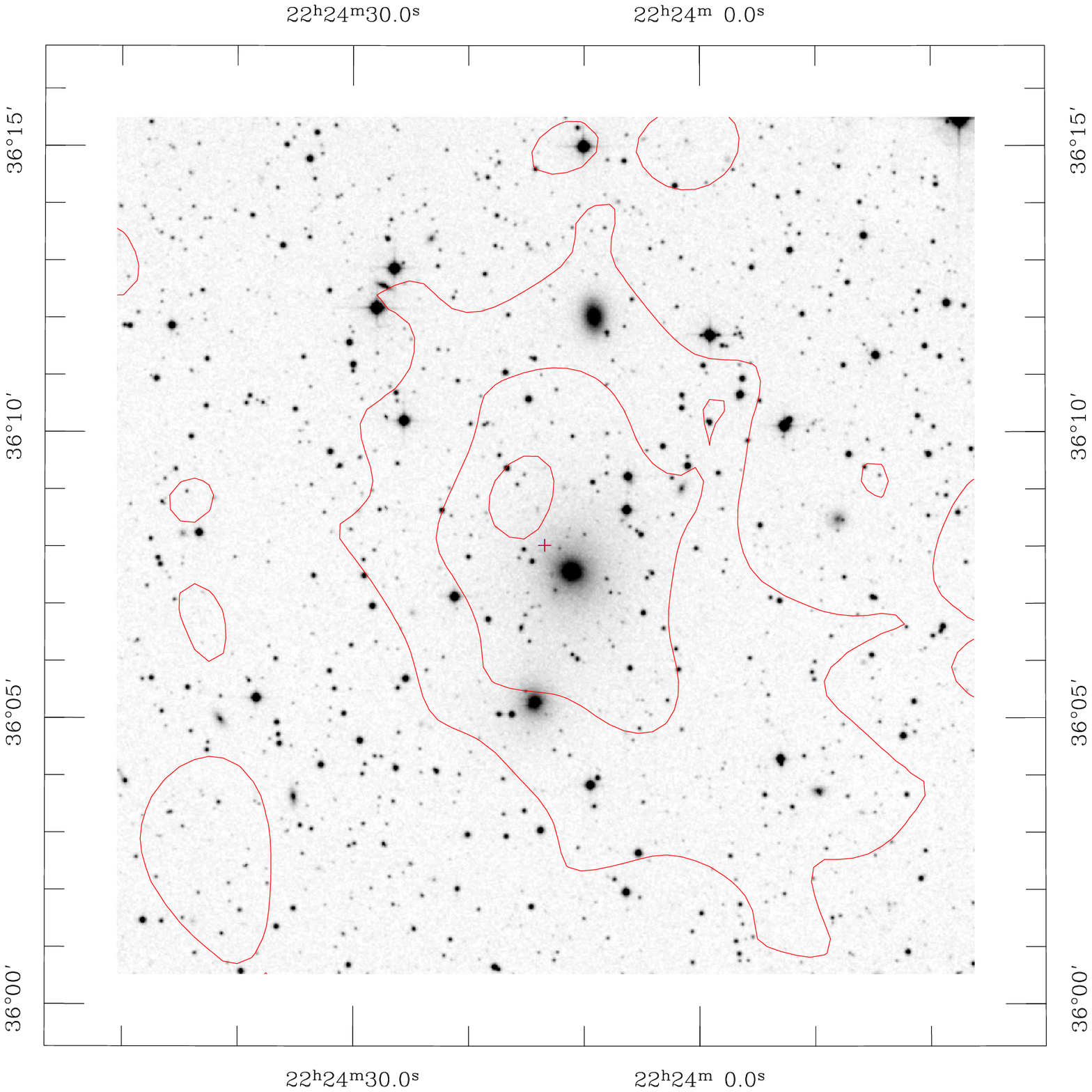}
\hspace{1cm}
   \includegraphics[width=7.5cm]{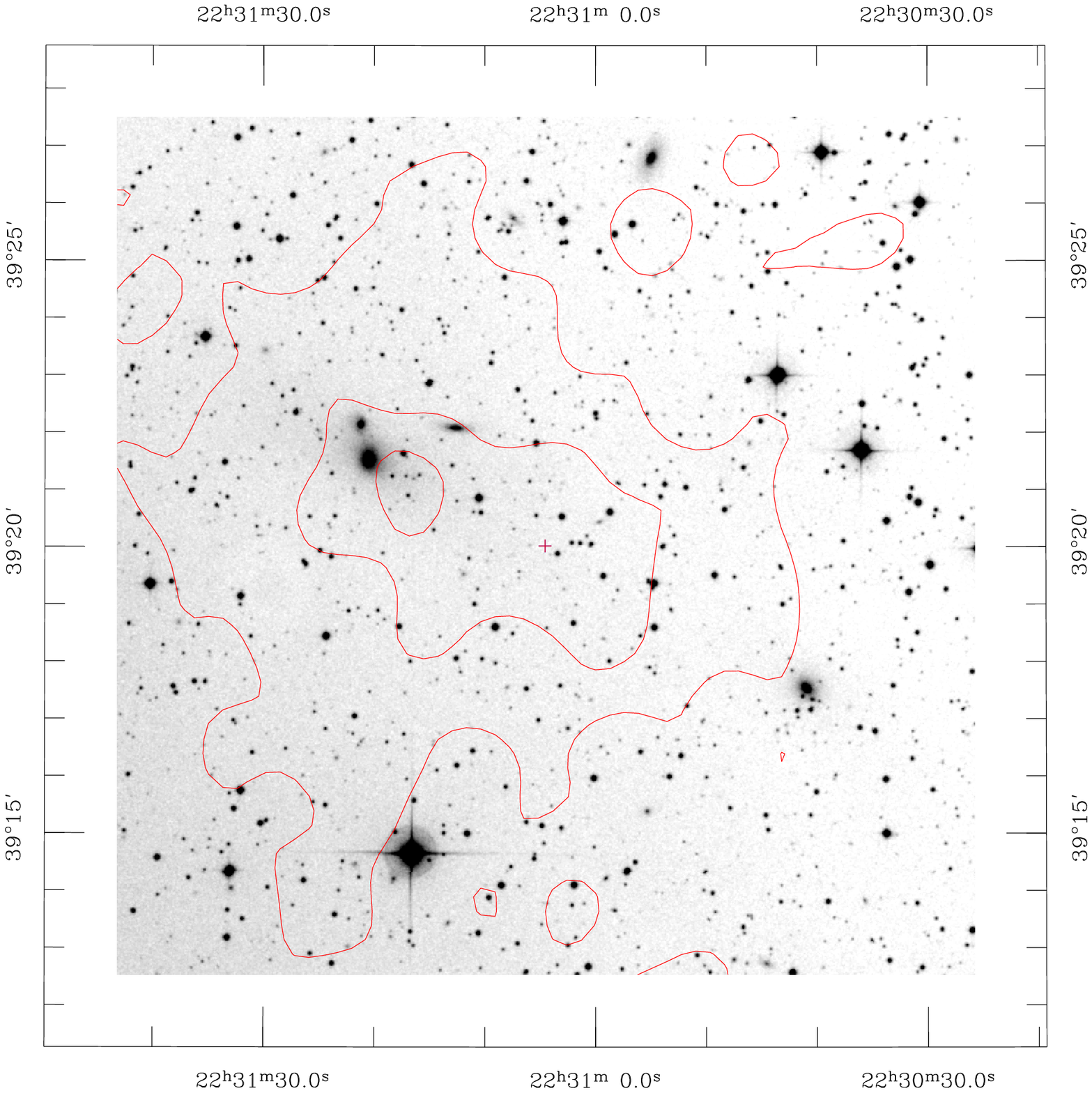}
}
\hbox{
\hspace{1cm}
   \includegraphics[width=7.5cm]{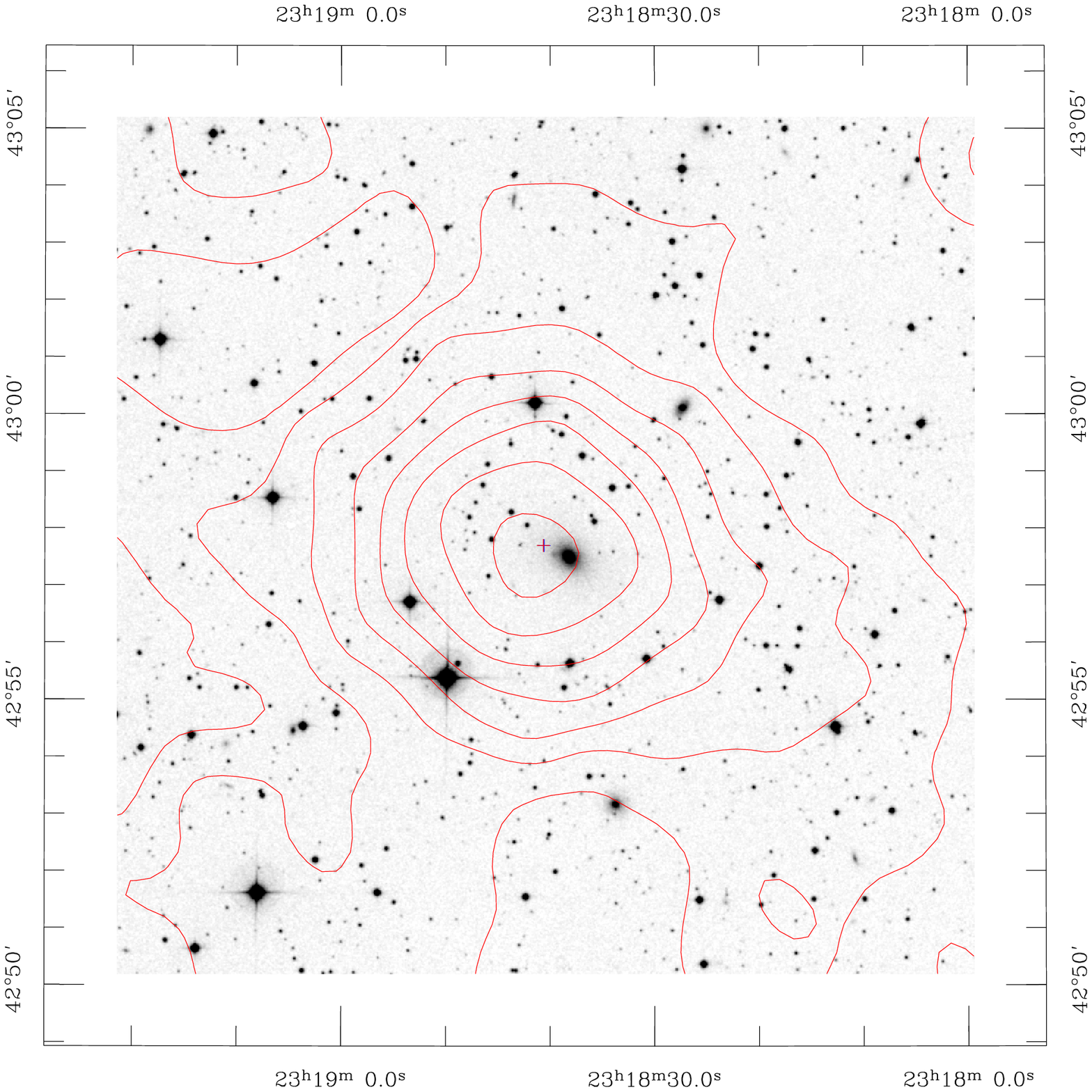}
\hspace{1cm}
   \includegraphics[width=7.5cm]{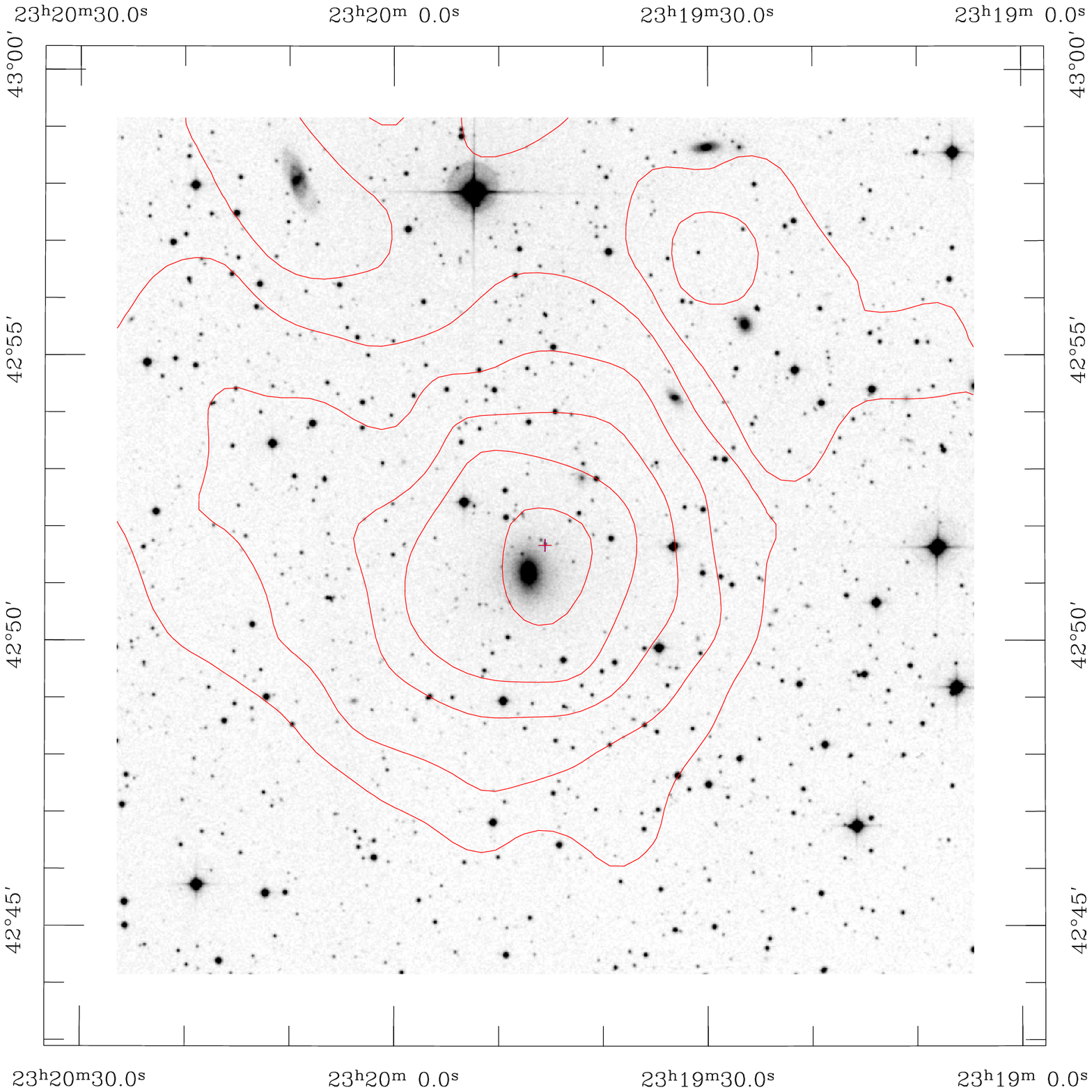}
}
\caption{Members of the Lacerta Supercluster.
{\bf Upper left:} RXCJ2215.6+3717, NGC 7242, 
{\bf Upper right:} RXCJ2222.4+3612, NGC 7265,
{\bf Middle left:} RXCJ2224.2+3608, NGC 7274,
{\bf Middle right:} RXCJ2231.0+3920,
{\bf Lower left:} RXCJ2318.6+4257, UGC 12491,
{\bf Lower right:} RXCJ2319.7+4251, NGC 7618.
}\label{figA12}
\end{figure*}

\subsection{Properties of supplementary clusters}

Table~\ref{tab10} provides data on those additional clusters, which were linked to Centaurus SC 
and the Coma SC with the alternative linking schemes discussed in section 5.

   \begin{table*}
      \caption{Groups and clusters which were found with the alternative linking schemes to
  be SC members.
  The meaning of the columns is the same as in Table 3, except for the column
  labelled 'memb.', which provides the name of the SC of which the cluster is a member: 
  C = Centaurus, G = Coma Supercluster.}
         \label{Tab8}
      \[
         \begin{array}{lrrrrrrrrrl}
            \hline
            \noalign{\smallskip}
{\rm name}&{\rm RA}&{\rm DEC}&{\rm redshift}&{\rm flux}& {\rm err.}&L_X&m_{200}&n_H & {\rm memb.} & {\rm alt. name} \\
            \noalign{\smallskip}
            \hline
            \noalign{\smallskip}
{\rm RXCJ1030.0-3521}& 157.5009 & -35.3515 & 0.0093 &  21.1936 &   6.20 &   0.0521 &   0.598 &   6.4& C & {\rm Antlia}\\
{\rm RXCJ1036.6-2731}& 159.1740 & -27.5243 & 0.0126 &  84.7802 &   6.50 &   0.3203 &   1.841 &   4.9& C &{\rm Hydra} \\
{\rm RXCJ1253.0-0912}& 193.2730 &  -9.2004 & 0.0146 &   7.5690 &  13.20 &   0.0438 &   0.536 &   2.9& C &{\rm HCG~62}\\
{\rm RXCJ1324.2+1358}& 201.0517 &  13.9811 & 0.0236 &   2.8567 &  22.00 &   0.0427 &   0.525 &   1.8& G &{\rm NGC~5129} \\
{\rm RXCJ1329.5+1147}& 202.3861 &  11.7888 & 0.0229 &   6.7478 &  15.00 &   0.0851 &   0.806 &   2.0& G &{\rm NGC~5171} \\
{\rm RXCJ1407.4-2700}& 211.8670 & -27.0152 & 0.0230 &  26.7265 &   5.80 &   0.3697 &   2.003 &   4.3& C &{\rm A~3581}\\
{\rm RXCJ1324.7-5736}& 201.1800 & -57.6144 & 0.0190 &  52.1226 &   9.30 &   0.5207 &   2.481 &  41.0& C &{\rm CIZA J1324.7-5736}\\
{\rm RXCJ1615.3-6055}& 243.8368 & -60.9230 & 0.0157 &  97.2861 &   4.10 &   0.6927 &   2.966 &  20.8& C &{\rm Norma}\\
            \noalign{\smallskip}
            \hline
            \noalign{\smallskip}
         \end{array}
      \]
\label{tab10}
   \end{table*}


\begin{thebibliography}{}

\bibitem[Abell (1958)]{Abe1958}
Abell, G.O., 1958, ApJS, 3, 211

\bibitem[Abell (1961)]{Abe1961}
Abell, G.O., 1961, AJ, 66, 607

\bibitem[Abell et al. (1989)]{Abe1989}
Abell, G.O., Corvin, H.G., Olowin, R.P., 1989, ApJS, 70, 1

\bibitem[Alpaslan et al. (2015)]{Alp2015}
Alpaslan, M., Driver, S., Robotham, A.S.G., et al., 2015, MNRAS, 451, 3249 

\bibitem[Bahcall \& Soneira (1983)]{Bah1983}
Bahcall, N.A. \& Soneira, R.M., 1983, ApJ, 270, 20

\bibitem[Bahcall \& Soneira (1984)]{Bah1984}
Bahcall, N.A. \& Soneira, R.M., 1984, ApJ, 277, 27

\bibitem[Bahcall (1988)]{Bah1988}
Bahcall, N., 1988, ARA\&A, 26, 631

\bibitem[Balaguera-Antolinez et al. (2011)]{Bal2011}
Balaguera-Antolinez, A., Sanchez, A., B\"ohringer, H., et al., 2011, MNRAS, 413, 386

\bibitem[Balaguera-Antolinez et al. (2012)]{Bal2012}
Balaguera-Antolinez, A., Sanchez, A., B\"ohringer, H., et al., 2012, MNRAS, 425, 2244

\bibitem[Bardeen et al. (1986)]{Bar1986}
Bardeen, J.M., Bond, J.R., Kaiser, N., et al., 1986, ApJ, 304, 15

\bibitem[Barmby \& Huchra (1998)]{Bar1998}
Barmby, P. \& Huchra, J.P., 1998, AJ, 115, 6

\bibitem[Basilakos et al. (2001)]{Bas2001}
Basilakos, S., Plionis, M., Rowan-Robinson, M., 2001,  MNRAS, 323, 47

\bibitem[Batuski \& Burns (1985)]{Bat1985}
Batuski, D.J. \& Burns, J.O., 1985, ApJ, 299, 5

\bibitem[Binggeli  et al. (1987)]{Bin1987}
Binggeli, B., Tammann, G.A., Sandage, A., 1987, AJ, 94, 251 

\bibitem[B\"ohringer et al. (1994)]{Boe1994}
B\"ohringer, H., Briel, U.G., Schwarz, R.A., et al., 1994, Nat, 368, 828

\bibitem[B\"ohringer et al. (2004)]{Boe2004}
B\"ohringer, H., Schuecker, P., Guzzo, L., et al., 2004, A\&A, 425, 367

\bibitem[B\"ohringer et al. (2013)]{Boe2013}
B\"ohringer, H., Chon, G., Collins, C.A., et al., 2013, A\&A, 555, A30

\bibitem[B\"ohringer et al. (2014)]{Boe2014}
B\"ohringer, H., Chon, G., Collins, C.A., et al., 2014, A\&A, 570, A31

\bibitem[B\"ohringer et al. (2015)]{Boe2015}
B\"ohringer, H., Chon, G., Bristow, M., et al., 2015,  A\&A, 574, A26

\bibitem[B\"ohringer et al. (2017)]{Boe2017}
B\"ohringer, H., Chon, G., Retzlaff, J., et al., 2017, AJ, 153, 220 

\bibitem[B\"ohringer et al. (2020)]{Boe2020}
B\"ohringer, H., Chon, G., Collins, C.A., 2020,  A\&A, 633, 19

\bibitem[B\"ohringer et al. (2021a)]{Boe2021a}
B\"ohringer, H., Chon, G., Tr\"umper, J., 2021a,  A\&A, in press

\bibitem[B\"ohringer et al. (2021b)]{Boe2021b}
B\"ohringer, H. Chon, G., Tr\"umper, J., 2021b,  A\&A, in press


\bibitem[Cautun et al. (2014)]{Cau2014}
Cautun, M., van de Weygaert, R., Jones, B.J.T., et al., 2014, MNRAS, 441, 2923

\bibitem[Chamaraux et al. (1990)]{Cha1990}
Chamaraux, P., Cayatte, V., Balkoski, C., et al., 1990, A\&A, 229, 340

\bibitem[Chincarini et al. (1983)]{Chi1983}
Chincarini, G.L., Giovanelli, R., Haynes, M.P., 1983, A\&A, 121, 5

\bibitem[Chon \& B\"ohringer (2013)]{Cho2013}
Chon, G., \& B\"ohringer, H.,  2013, MNRAS, 429, 3272

\bibitem[Chon et al. (2014)]{Cho2014}
Chon, G., B\"ohringer, H., Collins, C.A., et al., 2014   A\&A, 567, A144

\bibitem[Chon et al. (2015)]{Cho2015}
Chon, G., \& B\"ohringer, H. \& Zaroubi, S., 2015, A\&A, 575, L14

\bibitem[Churazov et al. (2021)]{Chu2021}
Churazov, E., Khabibullin, I., Lyskova, N., et al., 2020, arXiv201211627

\bibitem[Collins et al. (2000)]{Col2000}
Collins, C.A., Guzzo, L., B\"ohringer, H., et al., 2000, MNRAS, 319, 939

\bibitem[Costa-Duarte et al. (2011)]{Cos2011}
Costa-Duarte M.V., Sodre, L. jr., Durret, F., 2011, MNRAS, 411, 1716 

\bibitem[Courtois et al. (2013)]{Cou2013}
Courtois, H.M., Pom\`arede, D., Tully, R.B., et al., 2013, AJ, 146, 69

\bibitem[Crook et al. (2007)]{Cro2007}
Crook, A.C., Huchra, J.P., Martimbeau, N., et al., 2007, ApJ, 655, 790

\bibitem[Croton et al. (2006)]{Cro2006}
Croton, D.J., Springel, V., White, S.D.M., et al., 2006, MNRAS, 365, 11

\bibitem[DeVaucouleurs (1953)]{Dev1953}
de Vaucouleurs, G., 1953, AJ, 58, 30

\bibitem[DeVaucouleurs (1956)]{Dev1956}
de Vaucouleurs, G., 1956, VA, 2, 1584

\bibitem[DeVaucouleurs (1958)]{Dev1958}
de Vaucouleurs G., 1958, ApJ, 63, 223

\bibitem[de Vaucouleurs (1991)]{Dev1991}
de Vaucouleurs G., de Vaucouleurs A., Corwin H. G. Jr, Buta R., Paturel
G., Fenque P., 1991, The Third Catalogue of Bright Galaxies (RC3).
University of Texas Press, Austin

\bibitem[Dickey  \& Lockman (1990)]{Dic1990}
Dickey, J.M. \& Lockman, F.J., 1990, ARA\&A, 28, 215

\bibitem[Dupuy et al. (2019)]{Dup2019}
Dupuy, A., Courtois, H.M., Dupont, F., et al., 2019, MNRAS, 489,L1

\bibitem[Dupuy et al. (2020)]{Dup2020}
Dupuy, A., Courtois, H.M., Libeskind, N.I., et al., 2020, MNRAS, 493, 3513

\bibitem[Einasto et al. (1997)]{Ein1997}
Einasto, M., Tago, E., Jaaniste, J., et al., 1997, A\&AS, 123, 119

\bibitem[Einasto et al. (2001)]{Ein2001} 
Einasto, M., Einasto, J., Tago, E., et al., 2001, AJ, 122, 2222 

\bibitem[Einasto et al. (2003a)]{Ein2003a} 
Einasto, J., H\"utsi, G., Einasto, M., et al., 2003, A\&A, 405, 425 

\bibitem[Einasto et al. (2003b)]{Ein2003b} 
Einasto, J., Einasto, M., H\"utsi, G., et al., 2003, A\&A, 410, 425 

\bibitem[Einasto et al. (2006)]{Ein2006} 
inasto, J., Einasto, M., Saar, E., et al., 2006, A\&A, 459, 1 

\bibitem[Einasto et al. (2007a)]{Ein2007a}
Einasto, J., Einasto, M., Saar, E., et al., 2007a, A\&A, 462, 397

\bibitem[Einasto et al. (2007b)]{Ein2007b}
Einasto, J., Einasto, M., Tago, E., et al., 2007b, A\&A, 462, 811

\bibitem[Einasto et al. (2016)]{Ein2016}
Einasto, M., Lietzen, H., Gramann, M., et al., 2016, A\&A, 595. A70.

\bibitem[Einasto et al. (2018)]{Ein2018}
Einasto, J., Suhhonenko, I., Liivam\"agi, L.J., et al., 2018, A\&A, 616, A141

\bibitem[Einasto et al. (2019)]{Ein2019}
Einasto, J., Suhhonenko, I., Liivam\"agi, L.J., et al., 2019, A\&A, 623, A97

\bibitem[Einasto et al. (2020)]{Ein2020}
Einasto, M., Deshev, B., Tenjes, P., et al., 2020, A\&A, 641, A172

\bibitem[Einasto et al. (2021)]{Ein2021}
Einasto, J., H\"utsi, G., Suhhonenko, I., et al., 2021, A\&A, 647, A17

\bibitem[Giovanelli (1983)]{Gio1983}
Giovanelli, R., 1983, in  {\it Early Evolution of the Universe and its Present Structure},
IAU Symp. No. 104, Chincarini, G., Abell, G., eds., p. 273


\bibitem[Giovanelli et al. (1986)]{Gio1986}
Giovanelli, R., Haynes, M.P., Chincarini, G.L., 1986, ApJ, 300, 77

\bibitem[Giovanelli et al. (1997)]{Gio1997}
Giovanelli, R., Haynes, M.P., Herter, T., et al., 1997, AJ, 113, 53

\bibitem[Giovanelli et al. (1999)]{Gio1999}
Giovanelli, R., Dale, D.A., Haynes, M.P., et al., 1999, ApJ, 525, 25

\bibitem[Gottl\"ober et al. (2010)]{Got2010}
Gottl\"ober, S., Hoofman, Y., Yepes, G., 2010, High Performance Computing
in Science and Engineering, Gaching, 2009, p. 309

\bibitem[Gregory et al. (1981)]{Gre1981}
Gregory, S.A., Thompson, L.A., Tifft, W.G., 1981, ApJ, 243, 411

\bibitem[Gregory et al. (1981)]{Gre1981}              
Gregory, S.A. \& Thompson, L.A., 1984, ApJ, 286, 422 

\bibitem[Guzzo et al. (2009)]{Guz2009}
Guzzo, L., Schuecker, P., B\"ohringer, H., et al., 2009, A\&A, 499, 357

\bibitem[Hauser \& Peebles (1973)]{Hau1973}
Hauser, M.G. \& Peebles, P.E.J., 1973, ApJ, 185, 757

\bibitem[Henry et al. (1995)]{Hen1995}
Henry, J.P., Gioia, I.M., Huchra, J.P., et al., 1995, ApJ, 449, 422

\bibitem[Hauschildt (1987)]{Hau1987}
Hauschildt, M., 1987, A\&A, 184, 43

\bibitem[Joeveer \& Einasto (1978)]{Joe1978}
Joeveer, M.  \& Einasto, J., 1978, IAU Symp. 79, 241

\bibitem[Joeveer et al. (1978)]{Joe1978b}
Joeveer, M., Einasto, J., Tago, E., 1978, MNRAS, 185, 357

\bibitem[Kaiser (1986)]{Kai1986}
Kaiser, N., 1986, MNRAS, 222, 323

\bibitem[Kalberla et al. (2005)]{Kal2005}
Kalberla, P.M.W., Burton, W.B., Hartmann, D., 2005, A\&A, 440, 775

\bibitem[Kerscher et al. (2001)]{Ker2001}
Kerscher, M., Mecke, K., Schuecker, P., et al., 2001, A\&A, 377, 1 

\bibitem[Kraan-Korteweg et al. (2018)]{Kra2018}
Kraan-Korteweg, R.C., van Driel, W., Schr\"oder, A.C., et al., 2018, MNRAS, 481, 1262

\bibitem[Kraft et al. (2006)]{Kra2006}
Kraft, R.P., Jones, C., Nulsen, P.E.J., 2006, ApJ, 640, 762

\bibitem[Lahav et al. (2000)]{Lah2000}
Lahav, O., Santiago, B.X., Webster, A.M., et al., 2000, MNRAS, 312, 166

\bibitem[Lee et al. (2017)]{Lee2017}
Lee, G.H., Hwang, H.S., Sohn, J., et al., 2017, ApJ, 835, 280.

\bibitem[Liebeskind et al. (2018)]{Lie2018}
Liebeskind, N.I., van de Weygaert, R., Cautun, M., et al., 2018, MNRAS, 473, 1195

\bibitem[Lietzen et al. (2007)]{Lie2007}
Lietzen, H., Tempel, E., Hein\"am\"aki, P., et al., 2012

\bibitem[Liivamaegi et al. (2012)]{Lii2012}
Liivamaegi, L.J., Tempel, E., Saar, E., 2012, A\&A, 539. A80

\bibitem[Luparello et al. (2011)]{Lup2011}
Luparello, H., Lares, M., Lambas, D.G., Padilla, N., 2011, MNRAS, 415, 964

\bibitem[Machacek et al. (2011)]{Mac2011}
Machacek, M.E., Diab, J., Kraft, R., et al., 2011, ApJ, 743, 15

\bibitem[Mahdavi et al. (2000)]{Mah2000}
Mahdavi, A., B\"ohringer, H., Geller, M.J., et al., 2000, ApJ, 534, 114

\bibitem[Muriel et al. (1996)]{Mur1996}
Muriel, H., B\"ohringer, H., Voges, W., 1996, Int. Conf. X-ray 
Astronomy and Astrophyics: R\"ontgenstrahlung from the Universe, p. 601 

\bibitem[Oort (1983)]{Oor1983}
Oort, J.H., 1983, ARAA, 21, 373

\bibitem[O'Sullivan et al. (2017)]{Osu2017}
O'Sullivan, E., Ponman, T.J., Kolokythas, K., et al., 2017, MNRAS, 472, 1482

\bibitem[O'Sullivan et al. (2019)]{Osu2019}
O'Sullivan, E., Schellenberger, G., Burke, D.J., et al., 2019, MNRAS, 488, 2925

\bibitem[Mo  \& White (1996)]{Mo1996}
Mo, H.J. \& White, S.D.M., 1996, MNRAS, 282, 347 

\bibitem[Pandage et al. (2012)]{Pan2012}
Pandage,M.B., Vagshette, N.D., David, L.P., et al., 2012, MNRAS, 421, 808 

\bibitem[Park et al. (2007)]{Par2007}
Park, C., Choi, Y.-Y., Vogeley, M.S., 2007, ApJ, 658, 898

\bibitem[Pratt et al. (2009)]{Pra2009}
Pratt, G.W., Croston, J.H., Arnaud, M., B\"ohringer, H., 2009,
A\&A, 498, 361

\bibitem[Ramatsoku et al. (2016)]{Ram2016}
Ramatsoku, M., Verheijen, M.A.W., Kraan-Korteweg, R.C., et al., 2016, MNRAS, 460, 923

\bibitem[Randall et al. (2015)]{Ran2015}
Randall, S.W., Nulsen, P.E.J., Jones, C., et al., 2015, ApJ, 805, 112

\bibitem[Shapley (1932)]{Sha1932}
Shapley, H., 1932, Ann. Harvard College Observatory, 88, 41 

\bibitem[Schuecker et al. (2001)]{Sch2001}
Schuecker, P., B\"ohringer, H., Guzzo, L., et al., 
2001, A\&A,368, 86

\bibitem[Schuecker et al. (2002)]{Sch2002}
Schuecker, P., Guzzo, L., Collins, C.A., et al., 2002, MNRAS, 335, 807

\bibitem[Schuecker et al. (2003a)]{Sch2003a}
Schuecker, P., B\"ohringer, H., Collins, C.A. et al., 2003a, 
A\&A, 398, 867

\bibitem[Schuecker et al. (2003b)]{Sch2003b}
Schuecker, P., Caldwell, R.R., B\"ohringer, H., et al., 2003b, A\&A, 402, 53

\bibitem[Sheth \& Tormen (1999)]{She1999}
Sheth, R.K. \& Tormen, G., 1999, MNRAS, 308,  119 

\bibitem[Springel et al. (2005)]{Spr2005}
Springel, V., White, S.D.M., Jenkins, A., et al. 2005, Nature, 435, 629

\bibitem[Tarenghi et al. (1979)]{Tar1979}
Tarenghi, M., Tifft, W.G., Chincarini, G., 1979, ApJ, 234, 793

\bibitem[Tarenghi et al. (1980)]{Tar1980}
Tarenghi, M., Chincarini G., Rood, H.J., et al., 1980, ApJ, 235, 724

\bibitem[Tinker et al. (2010)]{Tin2010}
Tinker, J.L., Robertson, B.E., Kravtsov. A.V., 2010, ApJ, 724, 878
 
\bibitem[Trasarti-Battistani (1998)]{Tra1998}
Trasart-Battistani, R., 1998, A\&AS, 130, 341

\bibitem[Tr\"umper (1993)]{Tru1993}
Tr\"umper, J., 1993, Science, 260, 1769

\bibitem[Tully et al. (2014)]{Tul2014}
Tully, R.B., Courtois, H., Hoffman, Y., et al., 2014, Nat, 531, 71

\bibitem[Tully et al. (2019)]{Tul2019}
Tully, R.B., Pomarede, D., Graziani, R., et al., 2019, ApJ, 880, 24 

\bibitem[van der Linden et al. (2007)]{Vdl2007}
van der Linden, A., Best, P.N., Kauffmann, G., et al., 2007, MNRAS, 379, 867

\bibitem[Voges et al. (1999)]{Vog1999}
Voges, W., Aschenbach, B., Boller, T., et al. 1999,  A\&A, 349, 389

\bibitem[Wen et al. (2009)]{Wen2009}
Wen, Z.L., Han, J.L., Liu, F.S., 2009, ApJS, 183, 197

\bibitem[White et al. (1999)]{Whi1999}
White, R.A., Bliton, M., Bhavsar, S., et al., 1999, AJ, 118, 2014

\bibitem[Yepes et al. (2009)]{Yep2009}
Yepes, G., Martinez-Vaquero, L.A., Gottl\"ober, S., et al., 2009, in
Balazs, C., Wang, F. eds. AIP Conf. Proc. Vol. 1178, Am Inst. Phys, New York, p.64 

\bibitem[Zucca et al. (1993)]{Zuc1993}
Zucca, E. Zamorani, G., Scaramella,R., et al., 1993, ApJ, 407, 470

\end{thebibliography}
\end{document}